\documentclass[aps,prd,twocolumn,longbibliography,nofootinbib,floatfix]{revtex4-2}

\usepackage{graphicx}
\usepackage{dcolumn}
\usepackage{bm}
\usepackage{mathtools}

\usepackage{amsmath,amsfonts,amssymb}
\usepackage{mathrsfs,bbm}
\usepackage{orcidlink}
\usepackage{cleveref}
\usepackage[normalem]{ulem} 
\usepackage[dvipsnames]{xcolor}

\def\p{\partial} 
\def\non{\nonumber}
\def\Lie{\mathcal{L}}

\begin{document}


\title{3d Summation-by-Parts scheme for Linear Wave Equations on Hyperboloidal Slices}

\author{Anuraag Reddy$^{1,2}$\orcidlink{0009-0008-6390-0678}}
\author{Shalabh Gautam$^{3,4}$\orcidlink{0000-0003-2230-3988}}
\author{Prayush Kumar$^2$~\orcidlink{0000-0001-5523-4603}}

\affiliation{${}^1$Department of Physics, Indian Institute of Science Education and Research (IISER) Pune, Dr. Homi Bhabha Road, Pashan, Pune - 411008, India \\
${}^2$International Centre for Theoretical Sciences (ICTS), Tata Institute of Fundamental Research (TIFR), Survey No. 151, Shivakote, Hesaraghatta Hobli, Bengaluru - 560 089, India \\
${}^3$Beijing Institute of Mathematical Sciences and Applications (BIMSA), \\
No. 544, Hefangkou Village Huaibei Town, Huairou District, Beijing 101408, China \\
${}^4$  Yau Mathematical Sciences Center~(YMSC), Jingzhai, Tsinghua University, Haidian District, Beijing 100084, China
}

\date{\today}

\begin{abstract}
We derive a fully 3-dimensional Summation-By-Parts scheme for a class of linear wave equations on hyperboloidal slices that meet future null infinity on a Minkowski background. The scheme is derived in spherical polar coordinates, with a major strength being that it is provably stable and allows having grid points at the origin and on the $z$-axis, despite coordinate singularities, and at infinity, by introducing compactification followed by rescaling. Reducing it to the standard Cauchy problem, or on finite spacelike slices with an outer boundary, will follow a similar procedure. Interesting relations are obtained between the rescaling and compactification factors that simplify the equations, and the conditions on constraint addition terms are discovered to maintain symmetric hyperbolicity. Numerical implementation is achieved using finite-difference methods at second-order accuracy, which can be generalized to higher-order or spectral accuracies as well. Dissipation operators are given a more abstract treatment, which makes it possible to define them everywhere in the domain, including at the boundary points, in curvilinear coordinates, such that they satisfy the dissipative property~(DP) in our energy norms. These generalizations reduce to the well-known Kreiss-Oliger dissipation operators whenever defined on a Cartesian grid in the bulk and satisfy the~DP in the standard~$L^2$-norms. We also propose new norm convergence tests that produce more accurate outputs. Promising results are obtained, giving hope for application to fully nonlinear systems, like the Einstein Field Equations, and extracting the resulting gravitational waves free of systematic errors or gauge ambiguities.
\end{abstract}


\maketitle

\section{Introduction}\label{sec:intro}

Numerical relativity~(NR) concerns itself with the goal of numerically solving the Einstein field equations~(EFEs) to study diverse phenomena of interest. These phenomena range from studying the consequences of general relativity~(GR) to the nature of gravity, like the existence of critical phenomena discovered by Choptuik in~$1993$~\cite{Cho93}, strong and weak cosmic censorship conjectures~\cite{Chr99, Pen69}, to probing its implications in astronomy. One facet of the latter goal is to extract the gravitational wave~(GW) waveforms from different progenitors. One class of these progenitors is the binary systems, such as binary black holes, binary neutron stars, and black hole-neutron star mergers. Another class consists of isolated systems like supernova explosions, rapidly rotating deformed neutron stars, etc. These waveforms then act as reference signals in various domains of~GW research, such as calibrating semi-analytic waveform models such as~IMR-Phenom and~EOB,~GW searches and parameter estimations, testing~GR, and studying population channels. Therefore, generating highly accurate~GW waveforms is undoubtedly of primary importance.

A crucial point to note is that the~GW degrees of freedom from any system of interest are defined unambiguously only at infinite distances. One such region is future null infinity, denoted by~$\mathscr{I}^+$, where these waves reach in their infinite future. As described by Penrose~\cite{Pen63},~$\mathscr{I}^+$ is the set of endpoints of all future-directed null geodesics in the exterior in a given spacetime. Accurate extraction of these~GW-degrees of freedom without any gauge ambiguity, therefore, requires inclusion of~$\mathscr{I}^+$ in the computational domain with a well-posed formalism of the~EFEs in~NR simulations. The state-of-the-art method for this inclusion is Cauchy-Characteristic Matching~(CCM), which was proposed by Winicour~\cite{Win12} and recently implemented in the~NR simulations by Ma et al~\cite{MaMoxSch23, MaSchMox24}. However, Cauchy-Characteristic Extraction~(CCE) \cite{BisGomLeh96, BisGomLeh97, MoxSchTeu21} still wears the crown for producing the most accurate waveforms numerically. Another popular, and simpler, method is extrapolation \cite{BoyMro09}, which is not as accurate as~CCE and~CCM, as it contains systematic errors that depend on the choice of extrapolation method.

Although the GW-degrees of freedom are well-defined in the characteristic region for both CCE and CCM, defining them in the extraction region remains gauge dependent, making it one of the primary limitations of these methods. One of the recent claims regarding~CCM is that, even if both the Cauchy and the characteristic formulations of the~EFEs are separately well posed, being in tandem with each other might not lead to the same situation~\cite{GiaHilZil20, GiaBisHil22, GiaBisHil23}. CCE has an additional problem that the characteristic evolution does not feedback the backreaction effects from the wavezone to the Cauchy domain, making it an additional source of inaccuracy that is purely an artefact of this method. Both~CCE and extrapolation methods suffer from the time limit up to which the evolutions could be performed. This limit depends on how far the boundary of the Cauchy domain is fixed, which determines when the boundary effects start affecting the extraction region.

One promising method that overcomes all these limitations consists of foliating the entire spacetime domain using hyperboloidal slices, as they are spacelike everywhere but meet null infinity, denoted by~$\mathscr{I}$. Waveform extraction requires the ones that meet~$\mathscr{I}^+$. This approach was first proposed by Friedrich~\cite{Fri81, Fri81a, Fri83} in~1981-83, together with the conformal~EFEs~(cEFEs), which use conformal compactification suggested by Penrose~\cite{Pen63} in~$1963$. The initial value problems associated with this formulation were explored more in detail by Andersson et al in~$1992-94$~\cite{AndChrFri92, AndChr93, AndChr94}. The first numerical implementation was achieved by H\"ubner~\cite{Hub94, Hub99, Hub99a, Hub01} and Frauendiener~\cite{Fra97, Fra97a, Fra98}. This formulation offers a completely regular system of evolution equations and has been used to study the global properties of spacetime. However, the potential weakness of this method is that~$\mathscr{I}^+$ here moves along the incoming null direction and collapses to a single grid point, corresponding to the future timelike infinity~$i^+$, in a finite conformal time. For this reason, it is not computationally feasible to attain highly accurate binary black hole simulations and the~GW extraction even from perturbative environments.

To solve this problem, Zengino{\u{g}}lu~\cite{Zen07, Zen08} proposed fixing~$\mathscr{I}^+$ to a finite compactified radial coordinate on these slices, so-called scri-fixing, or~$\mathscr{I}^+$-fixing. The time coordinate in this approach remains uncompactified. The strength of this method is that one can attain long-time evolutions without losing computational resolution, as has been demonstrated, for example, in 3+1 simulations of test fields with given backgrounds~\cite{ZenKid10, Zen11, ZenGal12, YanZimZen13, Rin25}.

There have been several formulations of the conformally rescaled~EFEs with~$\mathscr{I}^+$-fixing \cite{MonRin08, Rin09, Rin10, RinMon13, Rin14, BaaRin16, MalRin18, VanHusHil14, VanHus14, Van15, VanHus16, VanHus17, Van23, Van23a, VanVal24, AlvVan25, AlvVan25a}, but none of them gives a completely regular system of equations. Attaining a fully regular system for the fully nonlinear~EFEs is still an open problem, even in the Dual-Foliation~(DF) approach proposed by Hilditch~\cite{Hil15, HilHarBug16}. However, a lot of progress has been made in this direction \cite{GasHil19, GasGauHil20, DuaFenGas21, DuaFenGas22, DuaFenGas22a, PetGauRai23, PetGauVan24, NakNakRac17, CsuRac19, CsuRac23, CsuRac25}, and it is hoped that this problem will be solved in all the above formulations in the near future.
 
In this paper, we address the numerical challenges of this approach by deriving a suitable discretization scheme for a regularized system on hyperboloidal slices. While the numerical aspect requires exploration even for the simplest cases, such as the linear wave equation~(LWE), our framework is designed with the fully nonlinear regime in mind. The~EFEs in generalized harmonic gauge (GHG)~\cite{Fri85,Gar02} take the form of a quasilinear system of wave equations where the principal part is identical to the~LWE operator. By providing a provably stable~$3D$ summation-by-parts~(SBP) scheme for the~LWE, we establish the mathematical foundation necessary for the numerical stability of the~GHG system's principal part. Furthermore, this~SBP framework specifically simplifies the implementation of~EFEs in~GHG by:
\begin{enumerate}
\item {\it Handling Coordinate Singularities:} It allows for grid points directly on the origin and the~$z$-axis without requiring complex coordinate transformations or manual regularization of the~GHG variables at these points.
\item {\it Stable Boundary Inclusion:} It provides an energy-stable treatment of~$\mathscr{I}^+$, allowing for the unambiguous extraction of~GW within the computational domain, free from the systematic errors of extrapolation.
\item {\it Constraint Damping Integration:} The first-order reduction and constraint damping techniques developed here for the~LWE map directly onto the constraint damping mechanisms essential for the long-term stability of~GHG simulations.
\item {\it Artificial Dissipation:} It proposes a covariant way to derive artificial dissipation operators and to implement them in the associated discrete system for effectively eliminating the numerical noise in our coordinate basis.
\end{enumerate}

The~SBP scheme~\cite{Str94} has been proved to be a very powerful tool in various fields of computational physics and mathematics~\cite{HalHarNch26, ManMalNch26, SteDur26, WorFerZin26, GlaIskLam26, RicLeeDur24, SteLeeDur24, GlaRanHen25, RanWinSch23}. Working in Minkowski spacetime in spherical polar coordinates, the~LWE encounters coordinate singularities at the origin and on the~$z$-axis. In a general~$d+2$ dimensional spacetime, consisting of time and radial directions along with a~$d$-dimensional sphere, the standard~LWE has polar singularities along~$d$ axes. The problem of these polar singularities was circumvented in~\cite{GunGarGar10, CsiLasRac12}, where they reduced the entire problem to an effective spherically symmetric one by writing the general solution in terms of~$d$-dimensional spherical harmonics, and gave ways to overcome the singularity at the origin. These spherical modes were then superposed analytically to obtain the general solution. This spherically symmetric system was then generalized to hyperboloidal slices in~\cite{GauVanHil21}.

This paper is a generalization of these previous works~\cite{GunGarGar10, CsiLasRac12, GauVanHil21}, which introduces discretization along the angular directions, producing a fully spatial discretization of the regularized~LWE on hyperboloidal slices. Generalizing to a~$d+2$ dimensional Minkowski spacetime, and on any kind of foliation consisting of spacelike slices, will follow similarly. For the time being, we simply use the standard~RK4 time integration, assuming that it satisfies certain required properties of the continuum time integration. More thorough studies on discrete time integrators have been performed in~\cite{Tad23, MarBoyBru14, MarBoyGle19, BoyMarSil22, MarBraZen23, BoyMar23, CorCerIba12}. Similarly, we confine ourselves to second-order accurate finite-difference~(FD) methods. Similar schemes with higher-order accurate~FD operators, and for pseudo-spectral methods could also be constructed.

We will denote the standard Cauchy coordinates using uppercase Latin letters with primed indices, represented as~$X^{\mu'} = (T,R,\theta,\phi)$, and the hyperboloidal coordinates in lowercase with unprimed indices,~$x^\mu = (t,r,\theta,\phi)$. All lowercase Greek indices, such as~$\mu, \nu, \ldots$, will represent spacetime indices~$(0,1,2,3)$, while the latter half of the Latin ones,~$i, j, k, \ldots$ will denote spatial components~$(1,2,3)$. We will use the first half of lowercase Latin indices~$a, b, c, \ldots$ to denote the abstract indices, and adopt the signature~$(-,+,+,+)$ for our spacetime metric.

The remainder of this paper is organized as follows. Section~\ref{sec:continuum_setup} sets the stage by deriving the continuum set of equations and studying some of its properties. This setup will then be discretized in Sec.~\ref{sec:Discretization}, and then implemented numerically for some special cases in Sec.~\ref{sec:numerics}. We finally conclude in Sec.~\ref{sec:conclusions} by summarising overall results and outlook.

\section{Continuum Setup}\label{sec:continuum_setup}

In this section, we define our analytical setup, which starts by introducing a foliation on the background spacetime, which we confine to Minkowski. We then rewrite our evolution equations in terms of the coordinates adapted to the foliation and regularize them at the points of coordinate singularities, which include the origin, the~$z$-axis, and~$\mathscr{I}^+$ using l'H{\^o}pital's rule and rescaling. Then, we will rewrite this regular system in the form we will discretize and derive the associated conserved quantities that we will use to achieve numerical stability. Finally, we will analyse the hyperbolicity of the resulting system we will discretize in the next section.

With the signature~$(-,+,+,+)$, and in standard spherical polar coordinates~$X^{\mu'} = (T,R,\theta,\phi)$, the Minkowski metric can be described by the following line element
\begin{align}\label{eq:MK_metric}
ds^2 = -dT^2 + dR^2 + R^2 \, d\theta^2 + R^2 \, \sin^2 \theta \, d\phi^2 \, ,
\end{align}
where
\begin{align}
T \in (-\infty,\infty) \, , & \quad R \in [0,\infty) \, , \quad \theta \in [0,\pi] \, , \non \\
& \textrm{and} \quad \phi \in [0, 2\pi) \, .
\end{align}
Notice that the outgoing and incoming characteristic speeds here are~$C^R_\pm = \pm 1$. The region~$T \rightarrow \pm \infty$ corresponds to future and past timelike infinities, denoted with~$i^+$ and~$i^-$, respectively. Since the standard Cauchy slices, defined by the level sets of~$T$, reach spacelike (or spatial) infinity, denoted with~$i^0$, as~$R \rightarrow \infty$, whereas all the null signals reach~$\mathscr{I}^+$ in their infinite future, the former are incompatible when it comes to extracting null signals at infinity. We, therefore, introduce a foliation consisting of hyperboloidal slices described below.

\subsection{Hyperboloidal Slices}\label{sec:Hyperboloidal_slices}

As stated above, hyperboloidal slices are the ones that are spacelike everywhere but reach~$\mathscr{I}^+$ as~$R \rightarrow \infty$. These slices are defined as the level sets of a function~$t$, such that~$\nabla_a t \, \nabla^a t < 0$ everywhere but vanishes at~$\mathscr{I}^+$. Since this~$t$ plays the same role as~$T$ in defining Cauchy slices, we call it the hyperboloidal time. Along these slices, we introduce a compactified radial coordinate~$r$, such that~$R = R(r)$, and monotonically transitions from~$0$ to a finite value~$r_\mathscr{I}$, as~$R$ ranges from~$0$ to~$\infty$. Therefore, the hyperboloidal coordinates,~$x^\mu = (t,r,\theta,\phi)$, can be defined as
\begin{align}
t = t(T,R) \, , \quad \textrm{and} \quad r  = r(R) \, . 
\end{align}
The standard~$3+1$ quantities, namely the lapse~$\alpha$, shift~$\beta^i$, unit normal~$n^a$, and spatial metric~$\gamma_{ij}$, associated with this foliation, can be defined in a standard way. The extrinsic and mean curvatures can also be obtained from the standard definitions
\begin{align}
K_{ab} \equiv - \frac{1}{2} \Lie_\mathbf{n} \gamma_{ab} \, , \quad \textrm{or} \quad K_{ab} \equiv - \gamma_a{}^c \gamma_b{}^d \nabla_c n_d \, ,
\end{align}
and
\begin{align}
K \equiv K^a{}_a = \gamma^{ab} K_{ab} \, .
\end{align}
One way to assure that these slices are hyperboloidal meeting~$\mathscr{I}^+$ is to demand that they are constant mean curvature,~$K = constant$, with the constant being negative definite \cite{Yor72, BriCavIse80}. Another relative way, that we shall adopt here, is to introduce the height function~$H(R)$~\cite{BeiMur98, MalMur03, Zen07}, defined by
\begin{equation}\label{eq:Hyperboloidal_Coords}
T = t + H(R) \, , \quad \text{with} \, , \quad R = R(r) \, ,
\end{equation}
which gives the difference between the Cauchy and hyperboloidal times on a slice at any radial coordinate. In other words, the height function~$H(R)$ tilts the Cauchy slices toward the outgoing null direction so that they meet~$\mathscr{I}^+$. In terms of this height function and compactification, the metric takes the following form
\begin{align}
ds^2 = & -dt^2 - 2 H' R' dt \, dr + (R')^2 (1 - (H')^2) \, dr^2 + R^2 \, d \theta^2 \non \\
& + R^2 \, \sin^2 \theta \, d\phi^2 \, ,
\end{align}
which gives the following outgoing and ingoing characteristic speeds
\begin{align}
c^r_{\pm} = \pm \frac{1}{R'\left(1\mp H'\right)} \, .
\end{align}
To ensure that these slices are hyperboloidal, we demand that~$c_+^r = C_+^R$ at least asymptotically. For the present purposes, we demand that~$c_+^r = C_+^R = 1$ everywhere, which gives~\cite{Zen11, Hil15, HilHarBug16}
\begin{align}
H' = 1 - \frac{1}{R'} \, .
\end{align}
The price we pay is that it gives
\begin{equation}
c_-^r = - \frac{1}{2R'-1} \, ,
\end{equation}
which makes a smooth transition from~$-1$ to~$0$ as we move from the origin to~$\mathscr{I}^+$. However, this choice assures that the outgoing waves are well resolved near~$\mathscr{I}^+$. Since no physical signal can enter the domain from~$\mathscr{I}^+$, we hope that a vanishing~$c_-^r$ will not create any problem.

Demanding~$H$ being an even function of~$R$, and~$H(0) = 0$, we get the following functional form for~$H$
\begin{align}\label{eq:Height}
H(R(r)) = \left\{ \begin{array}{c}
 R(r) - r \, , \quad \textrm{for} \quad r \geq 0 \\
 r - R(r) \, , \quad \textrm{for} \quad r < 0 \, . 
\end{array} \right.
\end{align}
Since this function is defined piecewise at the origin, the required smoothness there will depend on the accuracy of the numerical scheme. For example, a second-order~(FD) scheme requires that~$H(R(r))$ vanishes at the origin up to its third-order radial derivative. However, a pseudo-spectral scheme requires an arbitrarily high accuracy. We shall, therefore, make it~$C^\infty$ at the origin for a general numerical scheme. This can be achieved by choosing~$R(r)$ that satisfies
\begin{align}\label{eq:R_r}
R(0) = 0 \, , \, R'(0) = 1 \, , \textrm{ and } R^{(k)}(0) = 0 \textrm{ for all } k \geq 2 \, .
\end{align}

One can introduce  the radial compactification as in~\cite{CalGunHil06}
\begin{align}\label{eq:Compactification}
& R(r) = \frac{r}{\Omega^{\frac{1}{n-1}}(r)} \, , \textrm{ with } r \in [0,r_\mathscr{I}] \, , \textrm{ and } 1 < n \leq 2 \, , \non \\
& \Omega(r) > 0 \textrm{ and } \Omega'(r) \leq 0 \textrm{ for } 0 < r < r_\mathscr{I} \, , \textrm{ and} \non \\
& \Omega(r_\mathscr{I}) = 0 \, .
\end{align}
The height function~\eqref{eq:Height} then couples the choice of compactification with that of slicing. The parameter~$n$ controls the asymptotic behaviour of compactification, as it gives~$R' \sim 2 R^n/(n-1)$ for large~$R$. The requirements~\eqref{eq:R_r} on~$R(r)$ at the origin translate to those on~$\Omega$, as
\begin{align}\label{eq:Omega_Origin}
\Omega(0) = 1 \textrm{ and } \Omega^{(k)}(0) = 0 \, , \textrm{ for all } k > 0 \, .
\end{align}
All these requirements on~$\Omega$ can be fulfilled if we take
\begin{equation}\label{eq:Omega}
\Omega(r) = 1 - \frac{1}{2}\frac{r^2}{r_{\mathscr{I}}}\biggl[\tanh \biggl\{\tan \left(\pi\left(\frac{r}{r_{\mathscr{I}}} - \frac{1}{2}\right)\right)\biggl\}+1\biggl] \, ,
\end{equation}
where, the subscript~`$\mathscr{I}$' is used to remind that~$\mathscr{I}^+$ is now mapped to a finite radial coordinate~$r_\mathscr{I}$. For all our purposes, we shall keep~$r_\mathscr{I} = 1$ and~$n=2$, unless stated otherwise. In terms of these coordinates, the metric now becomes
\begin{align}\label{eq:metric_hyp}
g_{\mu \nu} = \begin{pmatrix}
-1 & 1-R' & 0 & 0\\
 1-R' & 2R'-1 & 0 & 0\\
0 & 0 & R^2 & 0\\
0 & 0 & 0 & R^2\sin^2\theta
\end{pmatrix},
\end{align}
with the lapse $\alpha$, and shift $\beta^i$, given by
\begin{align}\label{eq:lapse_shift}
\alpha = \frac{R'}{\left(2R'-1\right)^{1/2}} \, , \quad \beta^\mu = \bigg\{ 0, -\frac{R'-1}{2R'-1}, 0, 0 \bigg\} \, .
\end{align}
The remaining components, like the unit normal~$n^a$ and the spatial metric~$\gamma_{ab}$ on these slices, can be calculated from the standard definitions.

\subsection{Linear Wave Equation}\label{sec:LWE}

We derive the~SBP scheme for a scalar field~$\psi$ satisfying the following class of linear wave equations~(LWEs)
\begin{equation}\label{eq:LWEP}
\left(\Box - F\right)\psi = 0,
\end{equation}
where~$\Box$ denotes the standard d'Alembertian, or the `Wave', operator in Minkowski spacetime, and~$F$ is a potential. We confine ourselves only to those~$F$ that are defined everywhere and are functions of spatial coordinates, and keep time-dependent cases for future work.

When written in spherical polar coordinates,
\begin{align}\label{eq:LWEP_Sph_Coord}
& \bigg[-\p_T^2 + \frac{1}{R^2} \p_R \left( R^2 \p_R \right) + \frac{1}{R^2 \sin \theta} \p_\theta \left( \sin\theta \p_\theta \right) \non \\
& + \frac{1}{R^2 \sin^2 \theta} \p_\phi^2 - F \bigg] \psi = 0 \, ,
\end{align}
this box operator becomes singular on the~$z$-axis, where~$\sin\theta = 0$, and at the origin, where~$R = 0$. However, the regularity of~$\Box \psi$ demands that we need to redefine this operator at these points. From the~$z$-axis, the only possible directions a particle can move are radial, which is along the~$z$-axis, and transverse, which is along the~$\hat{\theta}$ direction. As the~$\hat{\phi}$ direction is not defined on the~$z$-axis, we anticipate that~\cref{eq:LWEP_Sph_Coord} assumes the following form on the~$z$-axis
\begin{align}\label{eq:LWEP_Sph_Coord_z-axis}
\bigg[ - \p_T^2 + \frac{1}{R^2} \p_R \left( R^2 \p_R \right) + \frac{2}{R^2} \p_\theta^2 - F \bigg] \psi = 0 \, .
\end{align}
Similarly, since all directions are radial for a particle sitting at the origin, we expect~\cref{eq:LWEP_Sph_Coord} to take the following form there 
\begin{align}\label{eq:LWEP_Sph_Coord_origin}
\left[ - \p_T^2 + 3 \p_R^2 - F \right] \psi = 0 \, .
\end{align}

We, therefore, define the d'Alembertian operator on the~$z$-axis and at the origin as
\begin{align}
& \Box_z \equiv - \p_T^2 + \frac{1}{R^2} \p_R \left( R^2 \p_R \right) + \frac{2}{R^2} \p_\theta^2 \, , \non \\
& \textrm{and} \quad \Box_O \equiv - \p_T^2 + 3 \p_R^2 \, .
\end{align}
However, we see that for a general solution~$\psi$,~$(\Box \psi)_{z=0} \neq \Box_z \psi$, and~$(\Box \psi)_{R = 0} \neq \Box_O \psi$. For example, rewriting the general solution to the case~$F = F(R)$ as a linear superposition of spherical harmonics,
\begin{align}\label{eq:Psi_Sph_Harmonics}
\psi(T,R,\theta,\phi) = \sum_{l,m} \psi_{l,m}(T,R) Y_{l,m}(\theta,\phi) \, ,
\end{align}
with every~$Y_{l,m}$'s satisfying
\begin{align}
\nabla^2_{S^2} Y_{l,m} = - l(l+1) Y_{l,m} \, ,
\end{align}
where~$\nabla^2_{S^2}$ denotes the Laplacian over the unit~$2$-sphere
\begin{align}\label{eq:Laplacian_S2}
\nabla^2_{S^2} \equiv \frac{1}{\sin\theta} \p_\theta \left(\sin\theta \p_\theta \right) + \frac{1}{\sin^2\theta} \p_\phi^2 \, ,
\end{align}
and each partial wave~$\psi_{l,m}$ satisfying the following spherically symmetric wave equation
\begin{align}\label{eq:Partial_Wave_Equation}
\bigg[ -\p_T^2 + \frac{1}{R^2} \p_R \left( R^2 \p_R \right) - \frac{l(l+1)}{R^2} - F \bigg] \psi_{l,m} = 0 \, ,
\end{align}
we see that~$(\Box \psi)_{z=0} \neq \Box_z \psi$ for~$m = 1,2$ modes, and, similarly~$(\Box \psi)_{R=0} \neq \Box_O \psi$ for~$l = 1,2$ modes. However, it turns out that the standard Laplacian
\begin{align}\label{eq:Laplacian}
\nabla^2 \psi \equiv & \bigg[ \frac{1}{R^2} \p_R \left( R^2 \p_R \right) + \frac{1}{R^2 \sin \theta} \p_\theta \left( \sin\theta \p_\theta \right) \non \\
& + \frac{1}{R^2 \sin^2 \theta} \p_\phi^2 \bigg] \psi
\end{align}
vanishes identically for all~$m \neq 0$ modes on the~$z$-axis, and, similarly, for all~$l > 0$ modes at the origin. (As every~$Y_{l,m} \sim (\sin \theta)^m$ near the~$z$-axis, and every~$\psi_{l,m} \sim R^l$ near the origin, one can see that in the case of~$m=1,2$ modes, the $2 \cot\theta \, \p_\theta \psi$ term cancels with~$(1/\sin^2\theta) \p^2_\phi \psi$ on the~$z$-axis, and for the case of~$l = 1,2$ modes,~$(2/R) \p_R \psi$ term cancels with~$(1/R^2)\nabla_{S^2}^2 \psi$ at the origin.) We can, therefore, rewrite this equation on the~$z$-axis and at the origin as
\begin{align}\label{eq:LWEP_Sph_Coord_z-axis_m=0}
\bigg[ - \p_T^2 + \frac{1}{R^2} \p_R \left( R^2 \p_R \right) + \frac{2}{R^2} \p_\theta^2 - F \bigg] \psi_{(m=0)} = 0 \, ,
\end{align}
and
\begin{align}\label{eq:LWEP_Sph_Coord_origin_l=0}
\left[ - \p_T^2 + 3 \p_R^2 - F \right] \psi_{(l=0)} = 0 \, ,
\end{align}
where~$\psi_{(m=0)}$ contains all the~$m=0$ modes in the expansion~\eqref{eq:Psi_Sph_Harmonics}, 
\begin{align}\label{eq:Psi_Sph_Harmonics_m=0}
\psi_{(m=0)} = \psi_{(m=0)}(T,R,\theta) \equiv \sum_l \psi_{l,0}(T,R) Y_{l,0}(\theta) \, ,
\end{align}
and,~$\psi_{(l=0)}$ contains only~$l=0$ mode,
\begin{align}\label{eq:Psi_Sph_Harmonics_l=0}
\psi_{(l=0)} = \psi_{(l=0)}(T,R) \equiv \psi_{0,0}(T,R) Y_{0,0} \, .
\end{align}
They can be easily extracted from a data, provided it is defined at all points in the domain
\begin{align}\label{eq:m=0_mode}
\psi_{(m=0)}(T, R, \theta) = \frac{1}{2\pi} \int_0^{2\pi} \psi(T, R, \theta, \phi) \, d\phi \, ,
\end{align}
and
\begin{align}\label{eq:l=0_mode}
\psi_{(l=0)}(T, R) = \frac{1}{4\pi} \int_{\theta = 0}^{\pi} \int_{\phi = 0}^{2\pi} \psi(T, R, \theta, \phi) \, \sin\theta \, d\theta \, d\phi \, .
\end{align}

As we shall see in the next section, defining the discrete equations on the~$z$-axis and at the origin will require extracting these modes numerically, cf.~\cref{eq:Discrete_m=0_mode} and~\eqref{eq:Discrete_l=0_mode}. Equations~\eqref{eq:LWEP_Sph_Coord},~\eqref{eq:LWEP_Sph_Coord_z-axis_m=0} and~\eqref{eq:LWEP_Sph_Coord_origin_l=0} constitute the final system of equations we will discretize. For the case~$F = F(R,\theta,\phi)$, we shall use the decomposition~\eqref{eq:Psi_Sph_Harmonics} for~$F$ as well, which will replace~$F$ in~\eqref{eq:Partial_Wave_Equation} by~$F_{l,m}$, keeping all these results unchanged.

\subsection{First-Order Reduction}\label{sec:FOR}

It is convenient to evolve a second-order hyperbolic system numerically by introducing an equivalent first-order reduction~(FOR) system. In terms of our coordinates, a natural~FOR is
\begin{align}\label{eq:FOR_TR_comps}
& \psi_T \equiv \p_{T} \psi \, , \; \psi_R \equiv \p_{R} \psi \, , \; \psi_{\theta} \equiv \p_{\theta} \psi \, , \; \psi_{\phi} \equiv \p_{\phi} \psi \, ,
\end{align}
which we shall refer to as the Cauchy~FOR variables. It gives the following~FOR system
\begin{align}\label{eq:LWEP_FOR}
\p_T \psi = &\ \psi_T \, , \non \\
\p_T \psi_T = &\ \frac{1}{R^2} \p_R \left( R^2 \, \psi_R \right)
 + \frac{1}{R^2 \sin \theta} \p_\theta \left( \sin\theta \: \psi_\theta \right) \non \\ 
& + \frac{1}{R^2 \sin^2 \theta} \p_\phi \psi_\phi - F \psi \, , \non \\
\p_T \psi_R = &\ \p_R \psi_T \, , \non \\
\p_T \psi_\theta = &\ \p_\theta \psi_T \, , \non \\
\p_T \psi_\phi = &\ \p_\phi \psi_T \, .
\end{align}
This system is equivalent to the original one~\eqref{eq:LWEP_Sph_Coord} iff all~FOR constraints~$\mathcal{C}_{i'} \equiv \psi_{i'} - \p_{i'} \psi = 0$ are satisfied throughout the evolution. It is, therefore, helpful if we introduce the constraint damping terms in the above system to ensure that any constraint violation damps away with time, making the overall system now look like
\begin{align}\label{eq:LWEP_FOR_Constraint_damping}
\p_T \psi = &\ \psi_T \, , \non \\
\p_T \psi_T = &\ \frac{1}{R^2} \p_R \left( R^2 \, \psi_R \right)
 + \frac{1}{R^2 \sin \theta} \p_\theta \left( \sin\theta \: \psi_\theta \right) \non \\ 
& + \frac{1}{R^2 \sin^2 \theta} \p_\phi \psi_\phi - F \psi \, , \non \\
\p_T \psi_R = &\ \p_R \psi_T + \zeta_R (\p_R \psi - \psi_R) \, , \non \\
\p_T \psi_\theta = &\ \p_\theta \psi_T + \zeta_\theta (\p_\theta \psi -  \psi_\theta) \, , \non \\
\p_T \psi_\phi = &\ \p_\phi \psi_T + \zeta_\phi (\p_\phi \psi -  \psi_\phi) \, .
\end{align}
The free parameters~$\zeta_R, \zeta_\theta$, and~$\zeta_\phi$ determine the rate at which constraint violations are damped, via the following evolution
\begin{align}
\p_T C_R = - \zeta_R \, C_R \quad \Rightarrow C_R (T) = C_{R} (0) \, e^{-\zeta_R T} \, ,
\end{align}
and, similarly,
\begin{align}
\hspace{-0.8em} C_\theta (T) = C_{\theta} (0) \, e^{-\zeta_\theta T} \, , \textrm{ and } C_\phi (T) &\ = C_{\phi} (0) \, e^{-\zeta_\phi T} \, .
\end{align}
Therefore, the necessary and sufficient condition that all the constraints are satisfied for all times is that they are satisfied in the initial data~(ID), even for~$\zeta_R = \zeta_\theta = \zeta_\phi = 0$. However, constraint violations might occur due to discretization, and adding these constraint damping terms would, therefore, be useful, especially in the nonlinear systems. 

On the~$z$-axis, the above~FOR system reduces to
\begin{align}\label{eq:LWEP_FOR_Constraint_damping_z-axis}
\p_T \psi = &\ \psi_T \, , \non \\
\p_T \psi_T = &\ \frac{1}{R^2} \p_R \left( R^2 \, \psi_{R(m=0)} \right)
 + \frac{2}{R^2} \p_\theta \psi_{\theta(m=0)} - F \psi_{(m=0)} \, , \non \\
\p_T \psi_R = &\ \p_R \psi_T + \zeta_R (\p_R \psi - \psi_R) \, , \non \\
\p_T \psi_\theta = &\ \p_\theta \psi_T + \zeta_\theta (\p_\theta \psi -  \psi_\theta) \, , \non \\
\p_T \psi_\phi = &\ 0 \, ,
\end{align}
with~$\psi_{i'(m=0)}$ and~$\psi_{(m=0)}$ being the~$m=0$ modes. Similarly, it reduces to the following set of equations at the origin
\begin{align}\label{eq:LWEP_FOR_Constraint_damping_origin}
\p_T \psi = &\ \psi_T \, , \non \\
\p_T \psi_T = &\ 3 \, \p_R \psi_{R(l=0)} - F \psi_{(l=0)} \, , \non \\
\p_T \psi_R = &\ \p_R \psi_T + \zeta_R (\p_R \psi - \psi_R) \, , \non \\
\p_T \psi_\theta = &\ 0 \, , \non \\
\p_T \psi_\phi = &\ 0 \, ,
\end{align}
with~$\psi_{R(l=0)}$ being the~$l=0$ mode in~$\psi_R$.

In general,~$\psi$ satisfies the following parity conditions on the~$z$-axis and at the origin
\begin{align}\label{eq:Parity}
& \psi(T, R, -\theta, \phi) = \left\{
\begin{array}{cc}
\psi(T, R, \theta, \phi + \pi) \, , & \textrm{for} \, 0 \leq \phi < \pi \\
\psi(T, R, \theta, \phi - \pi) \, , & \textrm{for} \, \pi \leq \phi < 2\pi \, ,
\end{array}
\right. \non \\
& \psi(T, R, \pi + \theta, \phi) = \left\{
\begin{array}{c}
\psi(T, R,\pi - \theta, \phi + \pi) \, , \\
\textrm{for} \, 0 \leq \phi < \pi \\
\psi(T, R,\pi - \theta, \phi - \pi) \, , \\
\textrm{for} \, \pi \leq \phi < 2\pi \, ,
\end{array}
\right. \non \\
& \psi(T,-R,\theta,\phi) = \left\{
\begin{array}{c}
\psi(T, R, \pi - \theta, \phi + \pi) \, , \\
\textrm{for} \, 0 \leq \phi < \pi \\
\psi(T, R, \pi - \theta, \phi - \pi) \, , \\
\textrm{for} \, \pi \leq \phi < 2\pi \, .
\end{array}
\right.
\end{align}
The parity conditions of the~FOR variables can be determined accordingly, with~$\p_R$ introducing an additional minus sign to these conditions at the origin, and~$\p_\theta$ doing the same on the~$z$-axis. Therefore, the parity conditions of the~FOR variables become
\begin{align}\label{eq:Parity_FOR_R_theta}
& \psi_\theta (T, R, -\theta, \phi) = \left\{
\begin{array}{cc}
- \psi_\theta (T, R, \theta, \phi + \pi) \, , & \textrm{for} \, 0 \leq \phi < \pi \\
- \psi_\theta (T, R, \theta, \phi - \pi) \, , & \textrm{for} \, \pi \leq \phi < 2\pi \, ,
\end{array}
\right. \non \\
& \psi_\theta (T, R, \pi + \theta, \phi) = \left\{
\begin{array}{c} 
- \psi_\theta (T, R,\pi - \theta, \phi + \pi) \, , \\
\textrm{for} \, 0 \leq \phi < \pi \\
- \psi_\theta (T, R,\pi - \theta, \phi - \pi) \, , \\
\textrm{for} \, \pi \leq \phi < 2\pi \, ,
\end{array}
\right. \non \\
& \psi_R (T,-R,\theta,\phi) = \left\{
\begin{array}{c}
- \psi_R (T, R, \pi - \theta, \phi + \pi) \, , \\
\textrm{for} \, 0 \leq \phi < \pi \\
- \psi_R (T, R, \pi - \theta, \phi - \pi) \, , \\
\textrm{for} \, \pi \leq \phi < 2\pi \, ,
\end{array}
\right.
\end{align}
keeping them unchanged for the rest of the variables as in~\eqref{eq:Parity}. These parity conditions will later be used to populate the ghost points beyond the domain of~$\theta$ and for negative~$R$ while introducing discretization. We shall use the periodicity conditions to populate the ghost points beyond the domain of~$\phi$
\begin{align}\label{eq:Parity_phi}
& \psi(T, R, \theta, - \phi) = \psi(T, R, \theta, 2\pi - \phi) \, , \non \\
& \psi(T, R, \theta, 2\pi + \phi) = \psi(T, R, \theta, \phi) \, ,
\end{align}
and the~$\phi$-derivatives do not change these conditions.

\subsection{Characteristic Variables}\label{sec:FOR_Characteristic}

It will be beneficial to define the following characteristic variables to obtain a fully regular~FOR system on hyperboloidal slices
\begin{align}\label{eq:FOR_null_comps}
\psi_+ \equiv \psi_T + \psi_R \, , \quad \textrm{and,} \quad \psi_- \equiv \psi_T - \psi_R \, ,
\end{align}
which leads to the following~FOR system
\begin{align}\label{eq:LWEP_Characteristic_FOR_Constraint_damping}
\p_T \psi = &\ \frac{\psi_+ + \psi_-}{2} \, , \non \\
\p_T \psi_+ = &\ \p_R \psi_+ + \frac{\psi_+ - \psi_-}{R} + \frac{1}{R^2 \sin \theta} \p_\theta \left( \sin\theta \: \psi_\theta \right) \non \\ 
& + \frac{1}{R^2 \sin^2 \theta} \p_\phi \psi_\phi - F \psi + \zeta_R \left( \p_R \psi - \frac{\psi_+ - \psi_-}{2} \right) \, , \non \\
\p_T \psi_- = &\ - \p_R \psi_- + \frac{\psi_+ - \psi_-}{R} + \frac{1}{R^2 \sin \theta} \p_\theta \left( \sin\theta \: \psi_\theta \right) \non \\ 
& + \frac{1}{R^2 \sin^2 \theta} \p_\phi \psi_\phi - F \psi - \zeta_R \left( \p_R \psi - \frac{\psi_+ - \psi_-}{2} \right) \, , \non \\
\p_T \psi_\theta = &\ \frac{\p_\theta (\psi_+ + \psi_-)}{2} + \zeta_\theta (\p_\theta \psi -  \psi_\theta) \, , \non \\
\p_T \psi_\phi = &\ \frac{\p_\phi (\psi_+ + \psi_-)}{2} + \zeta_\phi (\p_\phi \psi -  \psi_\phi) \, .
\end{align}
These equations now reduce to the following form on the~$z$-axis
\begin{align}\label{eq:LWEP_Characteristic_FOR_Constraint_damping_z-axis}
\p_T \psi = &\ \frac{\psi_+ + \psi_-}{2} \, , \non \\
\p_T \psi_+ = &\ \p_R \psi_+ + \frac{\psi_+ - \psi_-}{R} + \frac{2}{R^2} \p_\theta \psi_{\theta(m=0)} - F \psi_{(m=0)} \non \\ 
& + \zeta_R \left( \p_R \psi - \frac{\psi_+ - \psi_-}{2} \right) \, , \non \\
\p_T \psi_- = &\ - \p_R \psi_- + \frac{\psi_+ - \psi_-}{R} + \frac{2}{R^2} \p_\theta \psi_{\theta(m=0)} - F \psi_{(m=0)} \non \\ 
& - \zeta_R \left( \p_R \psi - \frac{\psi_+ - \psi_-}{2} \right) \, , \non \\
\p_T \psi_\theta = &\ \frac{\p_\theta (\psi_+ + \psi_-)}{2} + \zeta_\theta (\p_\theta \psi -  \psi_\theta) \, , \non \\
\p_T \psi_\phi = &\ 0 \, .
\end{align}
Notice that, this time, we did not extract the~$m=0$ modes in all the terms on the right as not all of them come from the Laplace operator in~\eqref{eq:LWEP_Sph_Coord}. The same arguments hold in deriving the equations at the origin
\begin{align}\label{eq:LWEP_Characteristic_FOR_Constraint_damping_origin}
\p_T \psi = &\ \frac{\psi_+ + \psi_-}{2} \, , \non \\
\p_T \psi_+ = &\ \frac{3 \, \p_R (\psi_+ - \psi_-)_{(l=0)}}{2} + \frac{\p_R (\psi_+ + \psi_-)}{2} - F \psi_{(l=0)} \non \\ 
& + \zeta_R \left( \p_R \psi - \frac{\psi_+ - \psi_-}{2} \right) \, , \non \\
\p_T \psi_- = &\ \frac{3 \, \p_R (\psi_+ - \psi_-)_{(l=0)}}{2} - \frac{\p_R (\psi_+ + \psi_-)}{2} - F \psi_{(l=0)} \non \\ 
& - \zeta_R \left( \p_R \psi - \frac{\psi_+ - \psi_-}{2} \right) \, , \non \\
\p_T \psi_\theta = &\ 0 \, , \non \\
\p_T \psi_\phi = &\ 0 \, .
\end{align}
The parity conditions for these~FOR variables become
\begin{align}\label{eq:Parity_FOR_Null_R}
\psi_\pm (T,-R,\theta,\phi) = \left\{
\begin{array}{c}
\psi_\mp (T, R, \pi - \theta, \phi + \pi) \, , \\
\textrm{for} \, 0 \leq \phi < \pi \\
\psi_\mp (T, R, \pi - \theta, \phi - \pi) \, , \\
\textrm{for} \, \pi \leq \phi < 2\pi \, ,
\end{array}
\right.
\end{align}
and the rest of the conditions remain the same as before. Due to these parity conditions, $\p_R (\psi_+ - \psi_-)_{(l=0)} = \p_R (\psi_+)_{(l=0)} = - \p_R (\psi_-)_{(l=0)}$ at the origin. So the~$\psi_+$ and~$\psi_-$ equations in~\eqref{eq:LWEP_Characteristic_FOR_Constraint_damping_origin} could be rewritten accordingly.

\subsection{Introducing Hyperboloidal Coordinates: Dual Foliation approach}\label{sec:LWEP_Hyp_Slices}

We now cast these equations on hyperboloidal slices by employing the dual-foliation formulation described in~\cite{Hil15, HilHarBug16} used in our previous works~\cite{GasGauHil20, GauVanHil21, PetGauRai23, PetGauVan24, PetHil25}. This formulation uses the~$3+1$ split of the Jacobian~$J^{\mu'}_\mu = \p X^{\mu'}/ \p x^\mu$ relating the two coordinate systems~$X^{\mu'} = (T,R, \theta, \phi)$ and~$x^\mu = (t, r, \theta, \phi)$, and thereby decouples the choice of gauge from the choice of coordinates. As a result, one can use the same~FOR variables, defined in~$X^{\nu'}$, in the coordinate system~$x^\mu$. In terms of the hyperboloidal coordinates described in Sec.~\ref{sec:Hyperboloidal_slices}, the time and radial derivatives transform as
\begin{align}
\p_T = \p_t \, , \quad \textrm{and} \quad \p_R = \frac{(R'-1) \, \p_t + \p_r}{R'} \, .
\end{align}
giving rise to the following~FOR system
\begin{align}\label{eq:hyp_LWEP_FOR}
\p_t \psi = &\ \psi_T \, , \non  \\
\p_t \psi_T = &\ \frac{R'}{2R'-1} \bigg[ - \frac{(R'-1)}{R'} \, \p_r \psi_T + \left( \p_r + \frac{2R'}{R} \right) \psi_R \non \\
& + \frac{R'}{R^2} \bigg( \p_\theta \psi_\theta + \cot\theta \, \psi_\theta + \frac{1}{\sin^2\theta} \p_\phi \psi_\phi \bigg) - R' F \, \psi \non \\
& + \zeta_R \frac{R'-1}{R'} \Big( (R'-1) \psi_T + R' \psi_R - \p_r \psi \Big) \bigg] \, , \non \\
\p_t \psi_R = &\ \frac{R'-1}{2R'-1} \bigg[ \frac{R'}{R'-1} \p_r \psi_T - \left( \p_r + \frac{2R'}{R} \right) \psi_R  \non \\
& - \frac{R'}{R^2} \bigg( \p_\theta \psi_\theta + \cot\theta \, \psi_\theta + \frac{1}{\sin^2\theta} \p_\phi \psi_\phi \bigg) + R' F \psi  \non \\
& - \zeta_R \frac{R'}{R'-1} \Big( (R'-1) \psi_T + R' \psi_R - \p_r \psi \Big) \bigg] \, , \non \\
\p_t \psi_\theta = &\ \p_\theta \psi_T + \zeta_\theta \left( \p_\theta \psi -\psi_\theta \right) \, , \non \\
\p_t \psi_\phi = &\ \p_\phi \psi_T + \zeta_\phi \left( \p_\phi \psi -\psi_\phi \right) \, ,
\end{align}
or, equivalently,
\begin{align}\label{eq:hyp_LWEP_FOR_null}
\p_t \psi = &\ \frac{1}{2} \left( \psi_+ + \psi_- \right) \, , \non \\
\p_t \psi_+ = &\ \frac{1}{2R'-1} \bigg[ \p_r \psi_+ + \frac{R'}{R} (\psi_+ - \psi_-) - R' F \psi \non \\
& + \frac{R'}{R^2} \bigg( \p_\theta \psi_\theta + \cot\theta \, \psi_\theta + \frac{1}{\sin^2\theta} \p_\phi \psi_\phi \bigg) \non \\
& + \zeta_R \left( \p_r \psi - \frac{(2R'-1) \psi_+}{2} + \frac{\psi_-}{2} \right) \bigg] \, , \non \\
\p_t \psi_- = &\ - \p_r \psi_- + \frac{R'}{R} (\psi_+ - \psi_-) + \frac{R'}{R^2} \bigg( \p_\theta \psi_\theta \non \\
& + \cot\theta \, \psi_\theta + \frac{1}{\sin^2\theta} \p_\phi \psi_\phi \bigg) - R' F \psi \non \\
& - \zeta_R \left( \p_r \psi - \frac{(2R'-1) \psi_+}{2} + \frac{\psi_-}{2} \right)  \, , \non \\
\p_t \psi_\theta = &\ \frac{1}{2} \p_\theta (\psi_+ + \psi_-) + \zeta_\theta \left( \p_\theta \psi -\psi_\theta \right) \, , \non \\
\p_t \psi_\phi = &\ \frac{1}{2} \p_\phi (\psi_+ + \psi_-) + \zeta_\phi \left( \p_\phi \psi -\psi_\phi \right) \, .
\end{align}
The equations on the poles and at the origin can be written in the same way as before. In fact, equations at the origin remain the same whenever conditions~\eqref{eq:R_r} are satisfied.

\subsection{Regularisation at~$\mathscr{I}^+$}
\label{sec:Reg_LWE_Hyp_Slices}

As pointed out in Sec.~\ref{sec:Hyperboloidal_slices},~$R' \sim 2 R^n/(n-1)$ for large~$R$, where~$1 < n \leq 2$. This makes many coefficients in eq.~\eqref{eq:hyp_LWEP_FOR} and~\eqref{eq:hyp_LWEP_FOR_null} singular at~$\mathscr{I}^+$. However, these terms are only formally singular as the solution to the linear wave equation falls off like~$1/R$ towards~$\mathscr{I}^+$, and any positive~$F$ makes this fall-off even faster. Therefore, one hopes that these equations can be regularized at~$\mathscr{I}^+$ by rescaling the variables as
\begin{align}\label{eq:Rescaled_FOR_TR_comps}
& \tilde{\psi} \equiv \chi \, \psi \, , \quad \tilde{\psi}_T \equiv \chi \, \psi_T \, , \quad \tilde{\psi}_R \equiv \chi \, \psi_R \, , \non \\
&  \tilde{\psi}_\theta \equiv \chi \, \psi_\theta \, , \quad \tilde{\psi}_\phi \equiv \chi \, \psi_\phi \, ,
\end{align}
where~$\chi = \chi(r)$ is an even function of~$r$, monotonically increasing, with~$\chi(0) = 1$ and~$\sim R$ for large~$R$. The evenness property is required to keep the parity condition of the rescaled variables the same at the origin as that of the original ones. The resulting equations become
\begin{align}\label{eq:hyp_LWEP_rescaled_FOR}
\p_t \tilde{\psi} = &\ \tilde{\psi}_T \, , \non  \\
\p_t \tilde{\psi}_T = &\ \frac{R'}{2R'-1} \bigg[ \left( \p_r + \frac{2R'}{R} - \frac{\chi'}{\chi} \right) \tilde{\psi}_R - \frac{(R'-1)}{R'} \bigg( \p_r \non \\
& - \frac{\chi'}{\chi}  \bigg) \tilde{\psi}_T + \frac{R'}{R^2} \bigg( \p_\theta \tilde{\psi}_\theta + \cot\theta \, \tilde{\psi}_\theta + \frac{1}{\sin^2\theta} \p_\phi \tilde{\psi}_\phi \bigg) \non \\
& - R' F \, \tilde{\psi} + \zeta_R \frac{R'-1}{R'} \bigg( (R'-1) \tilde{\psi}_T + R' \tilde{\psi}_R + \frac{\chi'}{\chi} \tilde{\psi} \non \\
& - \p_r \tilde{\psi} \bigg) \bigg] \, , \non \\
\p_t \tilde{\psi}_R = &\ \frac{R'-1}{2R'-1} \bigg[ \frac{R'}{R'-1} \left( \p_r - \frac{\chi'}{\chi} \right) \tilde{\psi}_T - \bigg( \p_r + \frac{2R'}{R} \non \\
& - \frac{\chi'}{\chi} \bigg) \tilde{\psi}_R - \frac{R'}{R^2} \bigg( \p_\theta \tilde{\psi}_\theta + \cot\theta \, \tilde{\psi}_\theta + \frac{1}{\sin^2\theta} \p_\phi \tilde{\psi}_\phi \bigg) \non \\
& + R' F \tilde{\psi} - \zeta_R \frac{R'}{R'-1} \bigg( (R'-1) \tilde{\psi}_T + R' \tilde{\psi}_R + \frac{\chi'}{\chi} \tilde{\psi} \non \\
& - \p_r \tilde{\psi} \bigg) \bigg] \, , \non \\
\p_t \tilde{\psi}_\theta = &\ \p_\theta \tilde{\psi}_T + \zeta_\theta \left( \p_\theta \tilde{\psi} -\tilde{\psi}_\theta \right) \, , \non \\
\p_t \tilde{\psi}_\phi = &\ \p_\phi \tilde{\psi}_T + \zeta_\phi \left( \p_\phi \tilde{\psi} -\tilde{\psi}_\phi \right) \, .
\end{align}

As~$\chi \sim R$ and~$\chi' \sim R'$ for large~$R$, these equations take the following limit at~$\mathscr{I}^+$
\begin{align}\label{eq:hyp_LWEP_rescaled_FOR_scri}
\p_t \tilde{\psi} = &\ \tilde{\psi}_T \, , \non  \\
\p_t \tilde{\psi}_T = &\ \frac{1}{2} \bigg[ \left( \p_r + \frac{R'}{R} \right) \tilde{\psi}_R - \bigg( \p_r - \frac{R'}{R} \bigg) \tilde{\psi}_T + \frac{R'}{R^2} \bigg( \p_\theta \tilde{\psi}_\theta \non \\
& + \cot\theta \, \tilde{\psi}_\theta + \frac{1}{\sin^2\theta} \p_\phi \tilde{\psi}_\phi \bigg) - R' F \, \tilde{\psi} \non \\
& + \zeta_R \bigg( R' \tilde{\psi}_T + R' \tilde{\psi}_R + \frac{R'}{R} \tilde{\psi} - \p_r \tilde{\psi} \bigg) \bigg] \, , \non \\
\p_t \tilde{\psi}_R = &\ \frac{1}{2} \bigg[ \left( \p_r - \frac{R'}{R} \right) \tilde{\psi}_T - \bigg( \p_r + \frac{R'}{R} \bigg) \tilde{\psi}_R - \frac{R'}{R^2} \bigg( \p_\theta \tilde{\psi}_\theta \non \\
& + \cot\theta \, \tilde{\psi}_\theta + \frac{1}{\sin^2\theta} \p_\phi \tilde{\psi}_\phi \bigg) + R' F \tilde{\psi} \non \\
& - \zeta_R \bigg( R' \tilde{\psi}_T + R' \tilde{\psi}_R + \frac{R'}{R} \tilde{\psi} - \p_r \tilde{\psi} \bigg) \bigg] \, , \non \\
\p_t \tilde{\psi}_\theta = &\ \p_\theta \tilde{\psi}_T + \zeta_\theta \left( \p_\theta \tilde{\psi} -\tilde{\psi}_\theta \right) \, , \non \\
\p_t \tilde{\psi}_\phi = &\ \p_\phi \tilde{\psi}_T + \zeta_\phi \left( \p_\phi \tilde{\psi} -\tilde{\psi}_\phi \right) \, .
\end{align}
which is formally singular, as the combination~$\tilde{\psi}_T + \tilde{\psi}_R$ falls-off like~$1/R$ towards~$\mathscr{I}^+$~\cite{GasHil19}. This finite limit could be computed by using l'H{\^o}pital's rule at~$\mathscr{I}^+$, which will break the~SBP scheme that we will derive in the next section. For this reason, we will work with the rescaled characteristic~FOR system in our numerical setup over this one.

As the characteristic variables defined in~\eqref{eq:FOR_null_comps} capture even a more aggressive fall-off~\cite{GasHil19} towards~$\mathscr{I}^+$, one can introduce the rescaling
\begin{align}\label{eq:Rescaled_FOR_null_comps}
& \tilde{\psi}_+ \equiv \chi_+ \psi_+ \, , \quad \tilde{\psi}_- \equiv \chi_- \psi_- \, , \non \\
\textrm{with} \quad & \chi_+ = \chi^2 \quad \textrm{and} \quad \chi_- = \chi \quad \textrm{for} \quad r > 0 \, .
\end{align}
To keep the parity conditions on these variables the same as those of the original ones~\eqref{eq:Parity_FOR_Null_R}, we impose the following parity conditions on~$\chi$ and~$\chi_\pm$
\begin{align}\label{eq:Parity_Chi}
\chi(-r) = \chi(r) \, , \quad \chi_\pm(-r) = \chi_\mp (r) \, .
\end{align}
The variables~$\psi$,~$\psi_\theta$ and~$\psi_\phi$ are rescaled the same as in~\cref{eq:Rescaled_FOR_TR_comps}.

To keep the rescaled equations smooth at the origin, the rescaling factors~$\chi_\pm$ should make a smooth transition from~$r > 0$ to~$r < 0$. For the same reason as before, we make this transition~$C^\infty$, by choosing
\begin{align}\label{eq:chi}
\chi = \sqrt{ 1 + \frac{R^2}{2} \left( \tanh \left( \tan \left( \left( r - \frac{1}{2} \right) \pi \right) \right) + 1 \right)}
\end{align}
One can see that this choice gives~$\chi(0) = 1$ and~$\chi^{(k)} (0) = 0$ for all~$k > 0$. This choice also assures that~$\chi(r) \sim R$ for large~$R$, or as~$r \rightarrow r_\mathscr{I} = 1$ in our choice of compactification\footnote{Another such choice, which we realised in the end is the most natural one, is
\begin{align}\label{eq:Chi_natural}
\chi = \Omega^{\frac{1}{1-n}} \, .
\end{align}
This choice corresponds to conformal rescaling in the case of conformal compactification~\cite{Zen07, Zen08, ZenKid10, Zen11, ZenGal12, YanZimZen13, Rin25, MonRin08, Rin09, Rin10, RinMon13, Rin14, BaaRin16, MalRin18, VanHusHil14, VanHus14, Van15, VanHus16, VanHus17, Van23, Van23a, VanVal24, AlvVan25, AlvVan25a}, where~$n=2$, and satisfies all the required properties described above for~$\Omega$ satisfying~\eqref{eq:Omega_Origin}. It also gives~$R/\chi = r$, which makes the final equations and the volume elements, given in~\eqref{eq:Energy_Density_Rescaled_Variables}-\eqref{eq:Stokes_diff_form_rescaled}, look the simplest. We shall use this specific choice of~$\chi$ in all our future works.}.

This gives the following~FOR system
\begin{align}\label{eq:hyp_LWEP_rescaled_FOR_null}
\p_t \tilde{\psi} = &\ \frac{1}{2} \left( \frac{\tilde{\psi}_+}{\chi} + \tilde{\psi}_- \right) \, , \non  \\
\p_t \tilde{\psi}_+ = &\ \frac{1}{2R'-1} \bigg[ \bigg( \p_r + \frac{R'}{R} - \frac{2 \chi'}{\chi} \bigg) \tilde{\psi}_+ - \frac{\chi R'}{R} \tilde{\psi}_- \non \\
& + \frac{\chi R'}{R^2} \bigg( \p_\theta \tilde{\psi}_\theta + \cot\theta \, \tilde{\psi}_\theta + \frac{1}{\sin^2\theta} \p_\phi \tilde{\psi}_\phi \bigg) - \chi R' F \tilde{\psi} \non \\
& + \zeta_R \bigg( \chi \p_r \tilde{\psi} - \chi' \tilde{\psi} - \frac{(2R'-1) \tilde{\psi}_+}{2} + \frac{\chi \tilde{\psi}_-}{2} \bigg) \bigg] \, , \non \\
\p_t \tilde{\psi}_- = &\ \bigg[ - \bigg( \p_r + \frac{R'}{R} - \frac{\chi'}{\chi} \bigg) \tilde{\psi}_- + \frac{R'}{R \chi} \tilde{\psi}_+ \non \\
& + \frac{R'}{R^2} \bigg( \p_\theta \tilde{\psi}_\theta + \cot\theta \, \tilde{\psi}_\theta + \frac{1}{\sin^2\theta} \p_\phi \tilde{\psi}_\phi \bigg) - R' F \tilde{\psi} \non \\
& - \zeta_R \bigg( \p_r \tilde{\psi} - \frac{\chi'}{\chi} \tilde{\psi} - \frac{(2R'-1) \tilde{\psi}_+}{2 \chi} + \frac{\tilde{\psi}_-}{2} \bigg) \bigg] \, , \non \\
\p_t \tilde{\psi}_\theta = &\ \frac{1}{2} \p_\theta \left( \frac{\tilde{\psi}_+}{\chi} + \tilde{\psi}_- \right) + \zeta_\theta \left( \p_\theta \tilde{\psi} -\tilde{\psi}_\theta \right) \, , \non \\
\p_t \tilde{\psi}_\phi = &\ \frac{1}{2} \p_\phi \left( \frac{\tilde{\psi}_+}{\chi} + \tilde{\psi}_- \right) + \zeta_\phi \left( \p_\phi \tilde{\psi} -\tilde{\psi}_\phi \right) \, .
\end{align}
which, for~$n=2$, takes the following limit at~$\mathscr{I}^+$
\begin{align}\label{eq:hyp_LWEP_rescaled_FOR_null_scri}
\p_t \tilde{\psi} = &\ \frac{1}{2} \tilde{\psi}_- \, , \non  \\
\p_t \tilde{\psi}_+ = &\ - \frac{1}{2} \tilde{\psi}_- - \frac{R F}{2} \tilde{\psi} \, , \non \\
\p_t \tilde{\psi}_- = &\ \bigg[ - \p_r \tilde{\psi}_- + 2 \tilde{\psi}_+ + 2 \bigg( \p_\theta \tilde{\psi}_\theta + \cot\theta \, \tilde{\psi}_\theta \non \\
& + \frac{1}{\sin^2\theta} \p_\phi \tilde{\psi}_\phi \bigg) - R' F \tilde{\psi} + 2 \, \zeta_r \, ( \tilde{\psi} + \tilde{\psi}_+ ) \bigg] \, , \non \\
\p_t \tilde{\psi}_\theta = &\ \frac{1}{2} \p_\theta \tilde{\psi}_- + \zeta_\theta \left( \p_\theta \tilde{\psi} -\tilde{\psi}_\theta \right) \, , \non \\
\p_t \tilde{\psi}_\phi = &\ \frac{1}{2} \p_\phi \tilde{\psi}_- + \zeta_\phi \left( \p_\phi \tilde{\psi} -\tilde{\psi}_\phi \right) \, ,
\end{align}
where
\begin{align}\label{eq:gamma_r}
\zeta_r \equiv \zeta_R \, \chi \, .
\end{align}
This system is completely regular at~$\mathscr{I}^+$ provided~$\zeta_r, \zeta_\theta, \zeta_\phi = O(1)$ everywhere, and the potential~$F$ falls-off like~$1/R^2$ for large~$R$.

For~$F$ falling-off like~$1/R^{1+\epsilon}$, with~$0 < \epsilon < 1$ towards~$\mathscr{I}^+$, the above system can be made regular by taking~$n \in (1, 1 + \epsilon]$, cf~\cite{GauVanHil21}. However, the potential terms become singular for an even slower fall-off. One such case which is physically relevant is~$F = m^2$. In such cases, we need to restrict the~ID to the class that gives a sufficient decay of the solution towards~$\mathscr{I}^+$ for all times~\cite{Kla93, GauVanHil21}.

\subsection{Introducing Regularized Covariant Divergence Operators}
\label{sec:Reg_Cov_Div}

Now, the next step is to introduce operators that will be discretized to obtain the~$3$D~SBP scheme. The key is to discretize the components of covariant divergence:
\begin{align}
\bigg( \p_r + \frac{2R'}{R} \bigg) f = \frac{1}{R^2} \p_r (R^2 \, f) \, , \quad \frac{1}{\sin\theta} \p_\theta (\sin\theta \, f) \, , \quad \p_\phi f \, .
\end{align}
However, as~$r \rightarrow r_\mathscr{I}$,~$R \rightarrow \infty$, and numerically it is infeasible to define the first operator at~$\mathscr{I}^+$. Therefore, we define the regularized covariant divergence operators\footnote{For the choice for~$\chi$ given in~\eqref{eq:Chi_natural}, the above definition of~$\tilde{\p}_r$ would give
\begin{align}\label{eq:Tilded_Operator_simplified}
\tilde{\p}_r f = r^{-2} \p_r (r^2 f) \, ,
\end{align}
which looks much simpler and a natural covariant divergence operator in terms of the hyperboloidal radial coordinate.}
\begin{align}\label{eq:Tilded_Operators}
& \tilde{\p}_r f \equiv \frac{\chi^2}{R^2} \p_r \left( \frac{R^2}{\chi^2} f \right) = \p_r f + 2 \left( \frac{R'}{R} - \frac{\chi'}{\chi} \right) f \, , \non \\
& \tilde{\p}_\theta f \equiv \frac{1}{\sin\theta} \p_\theta \left( (\sin\theta) \, f \right) = \p_\theta f + (\cot\theta) \, f \, , \quad \textrm{and} \non \\
& \tilde{\p}_\phi \equiv \p_\phi \, .
\end{align}
Additionally, we write the resulting equations of motion~(EOMs) in the following form
\begin{align}\label{eq:hyp_LWEP_rescaled_FOR_null_Tilded}
\p_t \tilde{\psi} = &\ \frac{1}{2} \left( \frac{\tilde{\psi}_+}{\chi} + \tilde{\psi}_- \right) \, , \non  \\
\p_t \tilde{\psi}_+ = &\ \frac{\chi}{2R'-1} \bigg[ \bigg( \frac{\p_r + \tilde{\p}_r}{2} \bigg) \left( \frac{\tilde{\psi}_+}{\chi} \right) + \bigg( \frac{\p_r - \tilde{\p}_r}{2} \bigg) \tilde{\psi}_- \non \\
& - \frac{\chi'}{\chi} \tilde{\psi}_- + \frac{R'}{R^2} \bigg( \tilde{\p}_\theta \tilde{\psi}_\theta + \frac{1}{\sin^2\theta} \p_\phi \tilde{\psi}_\phi \bigg) - R' F \tilde{\psi} \non \\
& + \zeta_R \bigg( \p_r \tilde{\psi} - \frac{\chi'}{\chi} \tilde{\psi} - \frac{(2R'-1) \tilde{\psi}_+}{2 \chi} + \frac{\tilde{\psi}_-}{2} \bigg) \bigg] \, , \non \\
\p_t \tilde{\psi}_- = &\ \bigg[ - \bigg( \frac{\p_r + \tilde{\p}_r}{2} \bigg) \tilde{\psi}_- - \bigg( \frac{\p_r - \tilde{\p}_r}{2} \bigg) \left( \frac{\tilde{\psi}_+}{\chi} \right) \non \\
& + \frac{\chi'}{\chi^2} \tilde{\psi}_+ + \frac{R'}{R^2} \bigg( \tilde{\p}_\theta \tilde{\psi}_\theta + \frac{1}{\sin^2\theta} \p_\phi \tilde{\psi}_\phi \bigg) - R' F \tilde{\psi} \non \\
& - \zeta_R \bigg( \p_r \tilde{\psi} - \frac{\chi'}{\chi} \tilde{\psi} - \frac{(2R'-1) \tilde{\psi}_+}{2 \chi} + \frac{\tilde{\psi}_-}{2} \bigg) \bigg] \, , \non \\
\p_t \tilde{\psi}_\theta = &\ \frac{1}{2} \p_\theta \left( \frac{\tilde{\psi}_+}{\chi} + \tilde{\psi}_- \right) + \zeta_\theta \left( \p_\theta \tilde{\psi} -\tilde{\psi}_\theta \right) \, , \non \\
\p_t \tilde{\psi}_\phi = &\ \frac{1}{2} \p_\phi \left( \frac{\tilde{\psi}_+}{\chi} + \tilde{\psi}_- \right) + \zeta_\phi \left( \p_\phi \tilde{\psi} -\tilde{\psi}_\phi \right) \, ,
\end{align}
everywhere,
\begin{align}\label{eq:hyp_LWEP_rescaled_FOR_null_Tilded_z-axis}
\p_t \tilde{\psi} = &\ \frac{1}{2} \left( \frac{\tilde{\psi}_+}{\chi} + \tilde{\psi}_- \right) \, , \non  \\
\p_t \tilde{\psi}_+ = &\ \frac{\chi}{2R'-1} \bigg[ \bigg( \frac{\p_r + \tilde{\p}_r}{2} \bigg) \left( \frac{\tilde{\psi}_+}{\chi} \right) + \bigg( \frac{\p_r - \tilde{\p}_r}{2} \bigg) \tilde{\psi}_- \non \\
& - \frac{\chi'}{\chi} \tilde{\psi}_- + \frac{2 R'}{R^2} \p_\theta \tilde{\psi}_{\theta (m=0)} - R' F \tilde{\psi} \non \\
& + \zeta_R \bigg( \p_r \tilde{\psi} - \frac{\chi'}{\chi} \tilde{\psi} - \frac{(2R'-1) \tilde{\psi}_+}{2 \chi} + \frac{\tilde{\psi}_-}{2} \bigg) \bigg] \, , \non \\
\p_t \tilde{\psi}_- = &\ \bigg[ - \bigg( \frac{\p_r + \tilde{\p}_r}{2} \bigg) \tilde{\psi}_- - \bigg( \frac{\p_r - \tilde{\p}_r}{2} \bigg) \left( \frac{\tilde{\psi}_+}{\chi} \right) \non \\
& + \frac{\chi'}{\chi^2} \tilde{\psi}_+ + \frac{2 R'}{R^2} \p_\theta \tilde{\psi}_{\theta (m=0)} - R' F \tilde{\psi} \non \\
& - \zeta_R \bigg( \p_r \tilde{\psi} - \frac{\chi'}{\chi} \tilde{\psi} - \frac{(2R'-1) \tilde{\psi}_+}{2 \chi} + \frac{\tilde{\psi}_-}{2} \bigg) \bigg] \, , \non \\
\p_t \tilde{\psi}_\theta = &\ \frac{1}{2} \p_\theta \left( \frac{\tilde{\psi}_+}{\chi} + \tilde{\psi}_- \right) + \zeta_\theta \left( \p_\theta \tilde{\psi} -\tilde{\psi}_\theta \right) \, , \non \\
\p_t \tilde{\psi}_\phi = &\ 0 \, ,
\end{align}
on the~$z$-axis, and
\begin{align}\label{eq:hyp_LWEP_rescaled_FOR_null_Tilded_origin}
\p_t \tilde{\psi} = &\ \frac{\tilde{\psi}_+ + \tilde{\psi}_-}{2} \, , \non  \\
\p_t \tilde{\psi}_+ = &\ \frac{3 \, \p_r (\tilde{\psi}_+ - \tilde{\psi}_-)_{(l=0)}}{2} + \frac{\p_r (\tilde{\psi}_+ + \tilde{\psi}_-)}{2} - F \tilde{\psi}_{(l=0)} \non \\
& + \zeta_R \bigg( \p_r \tilde{\psi} - \frac{\tilde{\psi}_+ - \tilde{\psi}_-}{2} \bigg) \, , \non \\
\p_t \tilde{\psi}_- = &\ \frac{3 \, \p_r (\tilde{\psi}_+ - \tilde{\psi}_-)_{(l=0)}}{2} - \frac{\p_r (\tilde{\psi}_+ + \tilde{\psi}_-)}{2} - F \tilde{\psi}_{(l=0)} \non \\
& - \zeta_R \bigg( \p_r \tilde{\psi} - \frac{\tilde{\psi}_+ - \tilde{\psi}_-}{2} \bigg) \, , \non \\
\p_t \tilde{\psi}_\theta = &\ 0 \, , \non \\
\p_t \tilde{\psi}_\phi = &\ 0 \, ,
\end{align}
at the origin. For~$n=2$, these equations reduce to the following forms at~$\mathscr{I}^+$,
\begin{align}\label{eq:hyp_LWEP_rescaled_FOR_null_Tilded_scri}
\p_t \tilde{\psi} = &\ \frac{\tilde{\psi}_-}{2} \, , \non  \\
\p_t \tilde{\psi}_+ = &\ - \frac{\tilde{\psi}_-}{2} - \frac{R F}{2} \tilde{\psi} \, , \non \\
\p_t \tilde{\psi}_- = &\ - \bigg( \frac{\p_r + \tilde{\p}_r}{2} \bigg) \tilde{\psi}_- - \bigg( \frac{\p_r - \tilde{\p}_r}{2} \bigg) \left( \frac{\tilde{\psi}_+}{\chi} \right) + 2 \, \tilde{\psi}_+ \non \\
& + 2 \bigg( \tilde{\p}_\theta \tilde{\psi}_\theta + \frac{1}{\sin^2\theta} \p_\phi \tilde{\psi}_\phi \bigg) - R' F \tilde{\psi} + 2 \, \zeta_r \, ( \tilde{\psi} + \tilde{\psi}_+ ) \, , \non \\
\p_t \tilde{\psi}_\theta = &\ \frac{1}{2} \p_\theta \tilde{\psi}_- + \zeta_\theta \left( \p_\theta \tilde{\psi} -\tilde{\psi}_\theta \right) \, , \non \\
\p_t \tilde{\psi}_\phi = &\ \frac{1}{2} \p_\phi \tilde{\psi}_- + \zeta_\phi \left( \p_\phi \tilde{\psi} -\tilde{\psi}_\phi \right) \, ,
\end{align}
everywhere, and
\begin{align}\label{eq:hyp_LWEP_rescaled_FOR_null_Tilded_z-axis_scri}
\p_t \tilde{\psi} = &\ \frac{\tilde{\psi}_-}{2} \, , \non  \\
\p_t \tilde{\psi}_+ = &\ - \frac{\tilde{\psi}_-}{2} - \frac{R F}{2} \tilde{\psi} \, , \non \\
\p_t \tilde{\psi}_- = &\ - \bigg( \frac{\p_r + \tilde{\p}_r}{2} \bigg) \tilde{\psi}_- - \bigg( \frac{\p_r - \tilde{\p}_r}{2} \bigg) \left( \frac{\tilde{\psi}_+}{\chi} \right) + 2 \, \tilde{\psi}_+ \non \\
& + 4 \, \p_\theta \tilde{\psi}_{\theta (m=0)} - R' F \tilde{\psi} + 2 \, \zeta_r \, ( \tilde{\psi} + \tilde{\psi}_+ ) \, , \non \\
\p_t \tilde{\psi}_\theta = &\ \frac{1}{2} \p_\theta \tilde{\psi}_- + \zeta_\theta \left( \p_\theta \tilde{\psi} -\tilde{\psi}_\theta \right) \, , \non \\
\p_t \tilde{\psi}_\phi = &\ 0 \, ,
\end{align}
on the~$z$-axis, with~$\zeta_r$ defined in~\eqref{eq:gamma_r}. These are the final equation we will discretize to derive the~SBP scheme.

\subsection{Conserved energy on hyperboloidal slices}
\label{sec:Cons_Energy_hyp_slices}

As the principal idea of the~SBP scheme is to preserve a conserved energy at the discrete level, the next and final step is to find a conserved energy norm on hyperboloidal slices, before proceeding to discretization. Demanding this conserved energy to remain conserved even at the discrete level will facilitate obtaining a stable time evolution even for the systems where the equations are singular at~$\mathscr{I}^+$. One such example of interest is~$F = m^2$~\cite{Win88}. Following the same steps as in~\cite{GunGarGar10, GauVanHil21}, we derive this conserved energy expression from the standard stress-energy tensor for~$\psi$ satisfying the~LWE~\eqref{eq:LWEP}
\begin{align}\label{eq:Stress_Energy}
T_{\mu\nu} = \p_\mu \psi \, \p_\nu \psi - \frac{1}{2} g_{\mu\nu} \, ( \p^\alpha \psi \, \p_\alpha \psi + F \psi^2) \, .
\end{align}
As before, we shall use the vector field method, described in~\cite{Are13}. We first contract this stress-energy tensor with a vector field~$K^\mu$ and then take the total divergence of the resulting quantity. Using the~EOM~\eqref{eq:LWEP} then gives
\begin{align}\label{eq:Total_Div}
\nabla^\mu (T_{\mu\nu} \, K^\nu) = &\ \left( \nabla^\mu T_{\mu\nu} \right) K^\nu + T_{\mu\nu} (\nabla^\mu K^\nu) \non \\
= &\ - \frac{1}{2} K^\nu \left( \nabla_\nu F \right) \psi^2 + T_{\mu\nu} \nabla^\mu K^\nu \, .
\end{align}
As~$T_{\mu\nu}$ is symmetric, the second term on the right vanishes whenever~$K^\mu$ satisfies~$\nabla^\mu K^\nu + \nabla^\mu K^\nu = 0$, or, whenever~$K^\mu$ is Killing. The first term on the right vanishes for a time-independent~$F$ whenever~$K^\mu = (K^0,0,0,0)$. All these properties are satisfied by the timelike Killing vector field~$K^\mu = \p_T^\mu = (1,0,0,0)$ of Minkowski spacetime.

\begin{figure}\label{Fig:Spacetime_Region}
\includegraphics[scale=0.5]{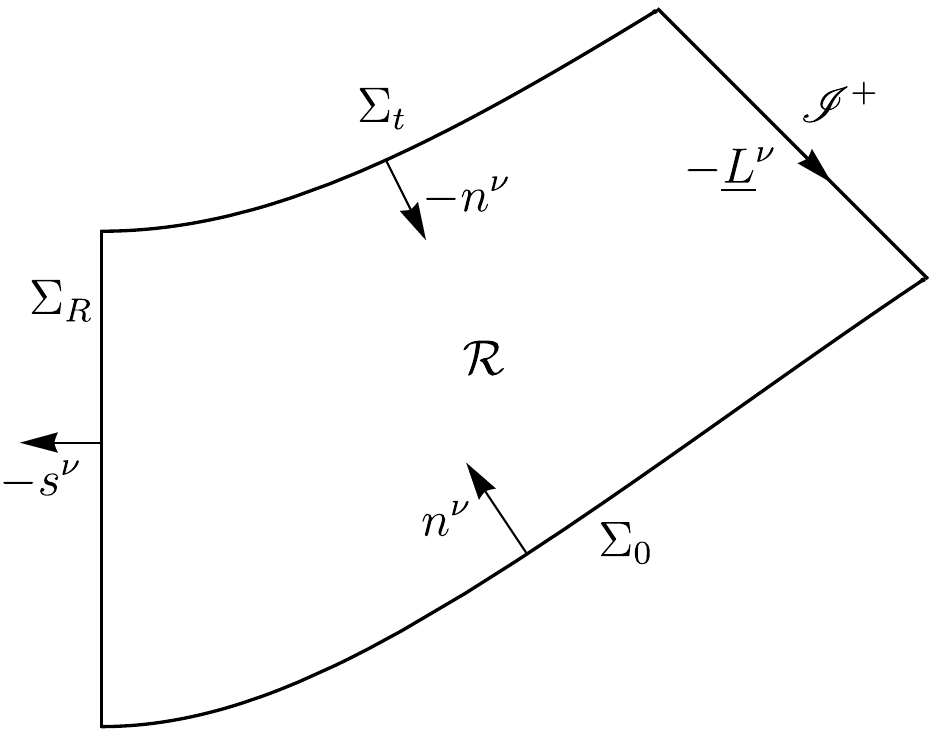}
\caption{Spacetime region where Stoke's theorem is applied, cf.~\cite{GauVanHil21}. The normal vectors here follow the conventions of~\cite{Are13}.}
\end{figure}

We now choose a spacetime region~$\mathcal{R}$ surrounded by the initial hyperboloidal slice~$\Sigma_0$, final hyperboloidal slice~$\Sigma_t$, with~$t>0$, future null-infinity~$\mathscr{I}^+$, and a timelike inner boundary~$\Sigma_R$, corresponding to an~$R = constant$ hypersurface, as shown in Fig.~\ref{Fig:Spacetime_Region}. Taking~$R = 0$ on~$\Sigma_R$, we get rid of~$\Sigma_R$, and the spacetime region is now surrounded by~$\Sigma_0$,~$\Sigma_t$ and~$\mathscr{I}^+$. The subscripts in~$\Sigma_0$ and~$\Sigma_t$ denote the hyperboloidal time~$t$.

We now integrate~\eqref{eq:Total_Div}, with~$K^\mu = \p_T^\mu$, over~$\mathcal{R}$. Using Stokes theorem, and the conventions of~\cite{Are13}, we get
\begin{align}\label{eq:Stokes_Thm}
\int_{\Sigma_0} T_{\mu\nu} \, K^\mu n^\nu - \int_{\Sigma_t} T_{\mu\nu} \, K^\mu n^\nu - \int_{\mathscr{I}^+} T_{\mu\nu} \, K^\mu \underline{L}^\nu = 0 \, ,
\end{align}
where~$n^\mu$ denotes the future-directed unit normals to the hyperboloidal slices~$\Sigma_0$ and~$\Sigma_t$, and~$\underline{L}^\mu$ denotes the incoming-null vector at~$\mathscr{I}^+$, as defined in~\cite{GasGauHil20}. For the spacetime metric given in~\eqref{eq:metric_hyp}, we can compute
\begin{align}\label{eq:normals}
n^\mu = (\alpha^{-1}, \alpha^{-1} \beta^i) \, , \quad \textrm{and,} \quad \underline{L}^\mu = (\p_T - \p_R)^\mu \, ,
\end{align}
where~$\alpha$ and~$\beta^i$ are given in~\eqref{eq:lapse_shift}. Substituting all these quantities, and~\eqref{eq:Stress_Energy} for the stress-energy tensor, and expressing everything in terms of the characteristic variables,~\cref{eq:Stokes_Thm} gives
\begin{align}\label{eq:Stokes_Thm_Expanded}
& \left( \int_{\Sigma_t} - \int_{\Sigma_0} \right) \frac{\alpha}{2} \Bigg( F \, \psi^2 + \left( \frac{2R'-1}{2R'} \right) \psi_+^2 + \left( \frac{1}{2R'} \right) \psi_-^2 \non \\
& + \frac{1}{R^2} \bigg( \psi_\theta^2 + \frac{1}{\sin^2 \theta} \psi_\phi^2  \bigg) \Bigg) = - \int_{\mathscr{I}^+} \frac{1}{2} \Bigg( F \, \psi^2 + \psi_-^2 \non \\
& + \frac{1}{R^2} \bigg( \psi_\theta^2 + \frac{1}{\sin^2 \theta} \psi_\phi^2 \bigg) \Bigg) \, .
\end{align}

As~$R' \geq 1$ throughout the domain,~$F \geq 0$ gives positive-definite integrands on the left, and the right side becomes negative semi-definite. This implies that total energy on a hyperboloidal slice is always upper bounded by the energy on the previous slice, with the difference being the flux at~$\mathscr{I}^+$. Interestingly, this flux contains contributions from~$\psi$ via~$F$, the angular~FOR variables and the outgoing characteristic mode~$\psi_-$. The absence of the incoming mode~$\psi_+$ simply implies that energy can never enter the domain from~$\mathscr{I}^+$. This result is expected because~$\mathscr{I}^+$ itself is an incoming-null hypersurface.

We define the total energy on a hyperboloidal slice as
\begin{align}\label{eq:Energy_norm}
E(t) \equiv \int_{\Sigma_t} \varepsilon(t,r,\theta,\phi) \, dr \, d\theta \, d\phi \, ,
\end{align}
with
\begin{align}
\varepsilon(t,r,\theta,\phi) \equiv &\ \frac{1}{2} \Bigg( F \, \psi^2 + \left( \frac{2R'-1}{2R'} \right) \psi_+^2 + \left( \frac{1}{2R'} \right) \psi_-^2 \non \\
& + \frac{1}{R^2} \bigg( \psi_\theta^2 + \frac{1}{\sin^2 \theta} \psi_\phi^2  \bigg) \Bigg) R^2 \, R' \, \sin\theta \, .
\end{align}
Just like in~\cite{GauVanHil21}, we adopt the unusual convention of absorbing the volume-form into~$\varepsilon$. It will help us to obtain the volume form on~$\mathscr{I}^+$ in the integral on the right of~\eqref{eq:Stokes_Thm_Expanded}, as we shall see shortly.

To begin with, we truncate the outer boundary at~$r_o < r_\mathscr{I}$, and take~$r_o = r_o(t)$. The total energy on these truncated slices can now be defined as
\begin{align}
E(t,r_o (t)) \equiv \int_{r=0}^{r_o(t)} \int_{\theta=0}^\pi \int_{\phi=0}^{2\pi} \varepsilon(t,r,\theta,\phi) \, dr \, d\theta \, d\phi \, ,
\end{align}
giving
\begin{align}\label{eq:Energy_change}
\frac{d}{dt} E(t,r_o (t)) = \p_t E(t,r_o (t)) + \p_{r_o} E(t,r_o (t)) \frac{d r_o}{dt} \, .
\end{align}
Substituting the~EOMs, with~$\zeta_R = \zeta_\theta = \zeta_\phi = 0$, and using Gauss's divergence theorem on the total spatial divergence, the first term on the right gives
\begin{align}
\p_t E(t,r_o (t)) \equiv &\ \int_{r=0}^{r_o(t)} \int_{\theta=0}^\pi \int_{\phi=0}^{2\pi} \p_t \varepsilon \, dr \, d\theta \, d\phi \non \\
= &\ \frac{1}{4} \int_{\theta=0}^\pi \int_{\phi=0}^{2\pi} (\psi_+^2 - \psi_-^2) \, R^2 \, \sin\theta \, d\theta \, d\phi \, \bigg|_{r = r_o} \, ,
\end{align}
while the definition of~$E(t,r_o (t))$ gives
\begin{align}\label{eq:Moving_Boundary}
& \p_{r_o} E(t,r_o (t)) = \non \\
& \lim_{\delta r_o \rightarrow 0} \frac{1}{\delta r_o} \left( \int_{r=0}^{r_o + \delta r_o} - \int_{r=0}^{r_o(t)} \right) \int_{\theta=0}^\pi \int_{\phi=0}^{2\pi} \varepsilon \, dr \, d\theta \, d\phi \non \\
& = \int_{\theta=0}^\pi \int_{\phi=0}^{2\pi} \varepsilon \, d\theta \, d\phi \, \bigg|_{r = r_o} \, .
\end{align}

If~$dr_o/dt = 0$, then the second term in~\eqref{eq:Energy_change} vanishes. If, instead, the outer boundary is an incoming-null hypersurface, then~$dr_o/dt = c_-^r|_{r = r_o} = -1/(2R'(r_o) - 1)$, giving the time rate of change of total energy to be
\begin{align}\label{eq:Energy_change_expanded}
\frac{d}{dt} E(t,r_o (t)) = &\ - \int_{\theta=0}^\pi \int_{\phi=0}^{2\pi} \frac{1}{2} \Bigg( F \, \psi^2 + \psi_-^2 + \frac{1}{R^2} \bigg( \psi_\theta^2 \non \\
& + \frac{1}{\sin^2 \theta} \psi_\phi^2  \bigg) \Bigg) \, \frac{R^2 \, R'}{2R' - 1} \, \sin\theta \, d\theta \, d\phi \, \bigg|_{r = r_o} \, .
\end{align}

The total change in energy on an entire hyperboloidal slice can be obtained by taking the limit~$r_o \rightarrow r_\mathscr{I}$, which gives~$R'/(2R'-1) \rightarrow 1/2$, reducing the entire expression to
\begin{align}\label{eq:Stokes_diff_form}
\frac{d}{dt} E(t) = &\ - \int_{\theta=0}^\pi \int_{\phi=0}^{2\pi} \frac{1}{4} \Bigg( F \, \psi^2 + \psi_-^2 + \frac{1}{R^2} \bigg( \psi_\theta^2 \non \\
& + \frac{1}{\sin^2 \theta} \psi_\phi^2  \bigg) \Bigg) \, R^2 \, \sin\theta \, d\theta \, d\phi \, \bigg|_{r = r_\mathscr{I}} \, .
\end{align}
which is consistent with the integral form~\eqref{eq:Stokes_Thm_Expanded}. As expected, it suggests the volume element on~$\mathscr{I}^+$ to be~$dV_{\mathscr{I}^+} = (1/2) \, R^2 \, \sin\theta \, dt \, d\theta \, d\phi$.

In terms of the rescaled variables, these expressions become
\begin{align}\label{eq:Energy_Density_Rescaled_Variables}
\varepsilon(t,r,\theta,\phi) = &\ \frac{1}{2} \Bigg( F R' \, \tilde{\psi}^2 + \left( \frac{2R'-1}{2 \chi^2} \right) \tilde{\psi}_+^2 + \frac{\tilde{\psi}_-^2}{2} \non \\
& + \frac{R'}{R^2} \bigg( \tilde{\psi}_\theta^2 + \frac{1}{\sin^2 \theta} \tilde{\psi}_\phi^2  \bigg) \Bigg) \frac{R^2}{\chi^2} \, \sin\theta \, ,
\end{align}
for the energy density,
\begin{align}\label{eq:E-dot_rescaled}
\p_t E(t,r_o (t)) = &\ \frac{1}{4} \int_{\theta=0}^\pi \int_{\phi=0}^{2\pi} \left( \frac{\tilde{\psi}_+^2}{\chi^2} - \tilde{\psi}_-^2 \right) \, \frac{R^2}{\chi^2} \, \sin\theta \, d\theta \, d\phi \, \bigg|_{r = r_o} \, ,
\end{align}
for the energy evolution on a slice with outer boundary at~$r_o < r_\mathscr{I}$, and
\begin{align}\label{eq:Energy_change_expanded_rescaled}
\frac{d}{dt} E(t,r_o (t)) = &\ - \int_{\theta=0}^\pi \int_{\phi=0}^{2\pi} \frac{1}{2} \Bigg( F \, \tilde{\psi}^2 + \tilde{\psi}_-^2 + \frac{1}{R^2} \bigg( \tilde{\psi}_\theta^2 \non \\
+ &\ \frac{1}{\sin^2 \theta} \tilde{\psi}_\phi^2 \bigg) \Bigg) \, \frac{R^2}{\chi^2} \, \frac{R'}{2R' - 1} \, \sin\theta \, d\theta \, d\phi \, \bigg|_{r = r_o} \, ,
\end{align}
for the total energy change, when the outer boundary is an incoming-null hypersurface, which gives
\begin{align}\label{eq:Stokes_diff_form_rescaled}
\frac{d}{dt} E(t) = &\ - \int_{\theta=0}^\pi \int_{\phi=0}^{2\pi} \frac{1}{4} \left( F \, \tilde{\psi}^2 + \tilde{\psi}_-^2 \right) \, \sin\theta \, d\theta \, d\phi \, \bigg|_{r = r_\mathscr{I}} \, ,
\end{align}
for the total energy flux at~$\mathscr{I}^+$. Here, we used~$R/\chi \rightarrow 1$ as~$r \rightarrow r_\mathscr{I}$. Interestingly, this energy flux at~$\mathscr{I}^+$ has a much simplified form, and does not have contributions from the angular variables~$\tilde{\psi}_\theta$ and~$\tilde{\psi}_\phi$. As we can see, this happens because the contribution from the angular variables falls off like~$1/R^2$ towards~$\mathscr{I}^+$.

The only variables contributing to the flux at~$\mathscr{I}^+$ are the outgoing mode~$\tilde{\psi}_-$, and~$\tilde{\psi}$, through the potential~$F$. As we shall see in Sec.~\ref{sec:numerics}, this term will contribute to the late-time tails when~$F$ falls-off like~$O(1/R^2)$, and both of these terms vanish at~$\mathscr{I}^+$ for a sufficiently decaying~ID whenever~$F = m^2$, cf~\cite{Win88, Kla93}.

\subsection{Testing Hyperbolicity}\label{sec:Hyperbolicity}

As the~LWE is a classic example of a hyperbolic second-order partial differential equation~(PDE), we expect the corresponding~FOR system, \cref{eq:hyp_LWEP_rescaled_FOR_null}, with~$\zeta_R = \zeta_\theta = \zeta_\phi = 0$, to constitute a hyperbolic system of~PDEs. This section, therefore, explores the conditions under which the resulting system, obtained by adding the constraint damping terms, remains symmetric hyperbolic.

In terms of the state vector~$\mathbf{u} \equiv (\tilde{\psi}, \tilde{\psi}_+, \tilde{\psi}_-, \tilde{\psi}_\theta, \tilde{\psi}_\phi)^T$, where the superscript~$T$ here denotes the transpose, the system~\eqref{eq:hyp_LWEP_rescaled_FOR_null} can be written in a concise form as
\begin{align}\label{eq:LWEP_vector_form}
\p_t \mathbf{u} = \mathbf{A}^i \, \p_i \mathbf{u} + \mathbf{S \, u} \, .
\end{align}
Here,~$\mathbf{A}^i = \mathbf{A}^i(x^\mu) \equiv \left( \mathbf{A}^r (x^\mu), \mathbf{A}^\theta (x^\mu), \mathbf{A}^\phi (x^\mu) \right)$, denote the velocity matrices along the~$\hat{r}, \hat{\theta}$ and~$\hat{\phi}$ directions, respectively, and~$\mathbf{S} \equiv \mathbf{S} (x^\mu)$ denotes the source matrix. We use boldface letters here to denote the vectors in the state space defined by the state vector~$\mathbf{u}$, and the operators on that space. The velocity matrices associated with~\eqref{eq:hyp_LWEP_rescaled_FOR_null} are
\begin{align}
& \mathbf{A}^r \equiv \left( \begin{array}{ccccc}
0 & 0 & 0 & 0 & 0 \\
\frac{\zeta_R \, \chi}{2R'-1} & \frac{1}{2R'-1} & 0 & 0 & 0 \\
-\zeta_R & 0 & -1 & 0 & 0 \\
0 & 0 & 0 & 0 & 0 \\
0 & 0 & 0 & 0 & 0 \\
\end{array} \right) \non \\
& \mathbf{A}^\theta \equiv \left( \begin{array}{ccccc}
0 & 0 & 0 & 0 & 0 \\
0 & 0 & 0 & \frac{\chi \, R'}{R^2 \, (2R'-1)} & 0 \\
0 & 0 & 0 & \frac{R'}{R^2} & 0 \\
\zeta_\theta & \frac{1}{2 \, \chi} & \frac{1}{2} & 0 & 0 \\
0 & 0 & 0 & 0 & 0 \\
\end{array} \right) \quad \textrm{and,} \non \\
& \mathbf{A}^\phi \equiv \left( \begin{array}{ccccc}
0 & 0 & 0 & 0 & 0 \\
0 & 0 & 0 & 0 & \frac{\chi \, R'}{R^2 \, \sin^2\theta \, (2R'-1)} \\
0 & 0 & 0 & 0 & \frac{R'}{R^2 \, \sin^2\theta} \\
0 & 0 & 0 & 0 & 0 \\
\zeta_\phi & \frac{1}{2 \, \chi} & \frac{1}{2} & 0 & 0 \\
\end{array} \right) \, .
\end{align}
while the source matrix is given in~\cref{eq:Source_Matrix}. Eq.~\eqref{eq:LWEP_vector_form} then constitutes a symmetric-hyperbolic system of~PDEs whenever there exists a symmetrizer, a symmetric positive-definite matrix with real entries,~$\mathbf{H} = \mathbf{H}(x^\mu)$ such that all the products~$\mathbf{H} \mathbf{A}^r, \, \mathbf{H} \mathbf{A}^\theta$ and~$\mathbf{H} \mathbf{A}^\phi$ are symmetric, and~$\mathbf{H}$ defines a conserved energy norm, given by
\begin{align}\label{eq:Energy_norm_vector_form}
E(t) = \frac{1}{2} (\mathbf{u},\mathbf{u})_\mathbf{H} \equiv \frac{1}{2} \int_{\Sigma_t} \mathbf{u}^T \, \mathbf{H} \, \mathbf{u} \, dr \, d\theta \, d\phi \, .
\end{align}
Eq.~\eqref{eq:Energy_Density_Rescaled_Variables} then suggests the symmetrizer associated with~\eqref{eq:LWEP_vector_form}, and gives a conserved energy for~$\zeta_R = \zeta_\theta = \zeta_\phi = 0$
\begin{align}\label{eq:Symmetrizer}
\mathbf{H} = \frac{R^2 \, \sin\theta}{2 \, \chi^2} \left( \begin{array}{ccccc}
F \, R' & 0 & 0 & 0 & 0 \\
0 & \frac{2R'-1}{2 \, \chi^2} & 0 & 0 & 0 \\
0 & 0 & \frac{1}{2} & 0 & 0 \\
0 & 0 & 0 & \frac{R'}{R^2} & 0 \\
0 & 0 & 0 & 0 & \frac{R'}{R^2 \, \sin^2 \theta} \\
\end{array} \right) \, .
\end{align}

This symmetrizer is positive-semi-definite for~$F = 0$, and positive definite only for~$F > 0$, and its product with the velocity matrices gives
\begin{align}
& \mathbf{H} \mathbf{A}^r = \frac{R^2 \, \sin\theta}{4 \, \chi^2} \left( \begin{array}{ccccc}
0 & 0 & 0 & 0 & 0 \\
\frac{\zeta_R}{\chi} & \frac{1}{\chi^2} & 0 & 0 & 0 \\
-\zeta_R & 0 & -1 & 0 & 0 \\
0 & 0 & 0 & 0 & 0 \\
0 & 0 & 0 & 0 & 0 \\
\end{array} \right) \non \\
& \mathbf{H} \mathbf{A}^\theta = \frac{R' \, \sin\theta}{4 \, \chi^2} \left( \begin{array}{ccccc}
0 & 0 & 0 & 0 & 0 \\
0 & 0 & 0 & \frac{1}{\chi} & 0 \\
0 & 0 & 0 & 1 & 0 \\
2 \zeta_\theta & \frac{1}{\chi} & 1 & 0 & 0 \\
0 & 0 & 0 & 0 & 0 \\
\end{array} \right) \non \\
& \mathbf{H} \mathbf{A}^\phi = \frac{R'}{4 \, \chi^2 \, \sin\theta} \left( \begin{array}{ccccc}
0 & 0 & 0 & 0 & 0 \\
0 & 0 & 0 & 0 & \frac{1}{\chi} \\
0 & 0 & 0 & 0 & 1 \\
0 & 0 & 0 & 0 & 0 \\
2 \, \zeta_\phi & \frac{1}{\chi} & 1 & 0 & 0 \\
\end{array} \right) \, ,
\end{align}
which are symmetric for~$\zeta_R = \zeta_\theta = \zeta_\phi = 0$. For nonzero~$\zeta_R$,~$\zeta_\theta$ and~$\zeta_\phi$, these matrices remain non-symmetric, and the system seems to be weakly hyperbolic. The constraint damping terms also change the energy evolution to
\begin{align}
\frac{d}{dt} E(t) = &\ \ldots - \zeta_r \int_{\Sigma_t} \frac{R^2 \sin\theta}{2 \chi^3} \left( \tilde{\psi}_- - \frac{\tilde{\psi}_+}{\chi} \right) \bigg( \p_r \tilde{\psi} \non \\
& - \frac{\chi'}{\chi} \tilde{\psi} - \frac{2R'-1}{2 \chi} \tilde{\psi}_+ + \frac{\tilde{\psi}_-}{2} \bigg) \non \\
& - \zeta_\theta \int_{\Sigma_t} \frac{R' \sin\theta}{\chi^2} \tilde{\psi}_\theta (\tilde{\psi}_\theta - \p_\theta \tilde{\psi}) \non \\
& - \zeta_\phi \int_{\Sigma_t} \frac{R'}{\chi^2 \sin\theta} \tilde{\psi}_\phi (\tilde{\psi}_\phi - \p_\phi \tilde{\psi}) \, ,
\end{align}
with~$\zeta_r$ is defined in~\eqref{eq:gamma_r}, and the ellipsis~`$\ldots$' denotes the flux through~$\mathscr{I}^+$ given in~\eqref{eq:Stokes_diff_form_rescaled}.

In the presence of the constraint damping terms, the only possible symmetrizer is
\begin{align}\label{eq:Symmetrizer_modified}
\mathbf{H\zeta} = \frac{R^2 \, \sin\theta}{2 \, \chi^2} \left( \begin{array}{ccccc}
F_\zeta \, R' & \frac{\zeta \, (2R'-1)}{2\chi} & \frac{\zeta}{2} & 0 & 0 \\
\frac{\zeta \, (2R'-1)}{2\chi} & \frac{2R'-1}{2 \, \chi^2} & 0 & 0 & 0 \\
\frac{\zeta}{2} & 0 & \frac{1}{2} & 0 & 0 \\
0 & 0 & 0 & \frac{R'}{R^2} & 0 \\
0 & 0 & 0 & 0 & \frac{R'}{R^2 \, \sin^2 \theta} \\
\end{array} \right) \, ,
\end{align}
which also requires the additional condition~$\zeta_R = \zeta_\theta = \zeta_\phi = \zeta$. This result implies that the constraint damping is introduced in the original system~\eqref{eq:LWEP_FOR} geometrically via vector addition~$\zeta \, C_{i'}$ to the spatial part of the system~$\psi_{i'}$, with~$\zeta$ being the constraint damping parameter. For this reason, we shall now take~$\zeta_R = \zeta_\theta = \zeta_\phi = \zeta \equiv \tilde{\zeta}/\chi$ everywhere, with~$\tilde{\zeta}$ being either~$0$ or~$1$, for all our purposes.

The condition under which this symmetrizer is positive semi-definite is~$F_\zeta \geq \zeta^2 = \frac{\tilde{\zeta}^2}{\chi^2}$. Therefore, we take the following functional form for~$F_\zeta$
\begin{align}\label{eq:F_gamma}
F_\zeta = \left \lbrace \begin{array}{cc}
\frac{\tilde{\zeta}^2}{\chi^2} \, , & \textrm{whenever } F < \zeta^2 = \frac{\tilde{\zeta}^2}{\chi^2} \\
F \, , & \textrm{otherwise}
\end{array}  \right. .
\end{align}
With this new definition
\begin{align}
E_\zeta \equiv \int_{\Sigma_t} \mathbf{u}^T \, \mathbf{H}\zeta \, \mathbf{u} \, dr \, d\theta \, d\phi \, ,
\end{align}
and,
\begin{align}
&\ \p_t E_\zeta(t,r_o (t)) = \ldots \non \\
& + \frac{\tilde{\zeta}}{2} \int_{\theta=0}^\pi \int_{\phi=0}^{2\pi} \frac{\tilde{\psi}}{\chi} \left( \frac{\tilde{\psi}_+}{\chi} - \tilde{\psi}_- \right) \, \frac{R^2}{\chi^2} \, \sin\theta \, d\theta \, d\phi \, \bigg|_{r = r_o} \non \\
& + \frac{1}{2} \int_{\Sigma_t} (F_\zeta - F) \, R' \, \tilde{\psi} \, \Bigg( \tilde{\psi}_- + \frac{\tilde{\psi}_+}{\chi} \Bigg) \left( \frac{R^2}{\chi^2} \right) \sin\theta \, dr \, d\theta \, d\phi \non \\
& - \tilde{\zeta} \int_{\Sigma_t} \Bigg( \frac{R' \, F}{\chi} \, \tilde{\psi}^2 + \frac{\chi' \, \tilde{\psi}}{\chi^2} \left( \tilde{\psi}_- - \frac{\tilde{\psi}_+}{\chi} \right) - \frac{R'}{\chi^2} \, \tilde{\psi}_- \, \tilde{\psi}_+ \non \\
& \quad \quad \quad \, + \frac{R'}{\chi \, R^2} \bigg( \tilde{\psi}_\theta^2 + \frac{\tilde{\psi}_\phi^2}{\sin^2\theta} \bigg) \Bigg) \frac{R^2}{\chi^2} \, \sin\theta \, dr \, d\theta \, d\phi \, .
\end{align}
The limit~$r_o \rightarrow r_\mathscr{I}$ then gives
\begin{align}\label{eq:Energy_flux_modified}
&\ \frac{d}{dt} E_\zeta(t) = \ldots - \frac{1}{2} \int_{\theta = 0}^\pi \int_{\phi=0}^{2\pi} (F_\zeta - F) \, \tilde{\psi}^2 \, \sin\theta \, d\theta \, d\phi \, \bigg|_{r = r_\mathscr{I}} \non \\
& + \frac{1}{2} \int_{\Sigma_t} (F_\zeta - F) \, R' \, \tilde{\psi} \, \Bigg( \tilde{\psi}_- + \frac{\tilde{\psi}_+}{\chi} \Bigg) \left( \frac{R^2}{\chi^2} \right) \sin\theta \, dr \, d\theta \, d\phi \non \\
& - \tilde{\zeta} \int_{\Sigma_t} \Bigg( \frac{R' \, F}{\chi} \, \tilde{\psi}^2 + \frac{\chi' \, \tilde{\psi}}{\chi^2} \left( \tilde{\psi}_- - \frac{\tilde{\psi}_+}{\chi} \right) - \frac{R'}{\chi^2} \, \tilde{\psi}_- \, \tilde{\psi}_+ \non \\
& \quad \quad \quad \, + \frac{R'}{\chi \, R^2} \bigg( \tilde{\psi}_\theta^2 + \frac{\tilde{\psi}_\phi^2}{\sin^2\theta} \bigg) \Bigg) \frac{R^2}{\chi^2} \, \sin\theta \, dr \, d\theta \, d\phi \, .
\end{align}
Here, we used the same methods as in Sec.~\ref{sec:Cons_Energy_hyp_slices}. The ellipses~`$\ldots$' again denote the part corresponding to~$\zeta = 0$. Interestingly, the modified flux at~$\mathscr{I}^+$ is still negative semidefinite, as~$F_\zeta \geq F$, but taking~$F_\zeta = F$, and, thereby, relaxing the condition for~$E_\zeta$ to be positive semidefinite for all~$F$, we get rid of the first two integrals on the right. The last integral can be upper bounded by using Cauchy-Schwarz and Gr\"onwall's inequalities.

In summary, this subsection shows that the resulting system is symmetric hyperbolic and there is a natural energy norm for the system with constraint damping terms. It also unveils that the constraint damping should be added to the spatial part of the original system geometrically via vector addition to ensure that the resulting system remains symmetric hyperbolic.

\section{Discretization}\label{sec:Discretization}

The above section derived the continuum set-up ready for discretization with all the desired properties, like regularity, at least for a certain class of~$F$, and hyperbolicity. The domain is set to be a certain class of hyperboloidal slices with a compactified radial coordinate~$r$ and characterised by the hyperboloidal time~$t$. All operators defined in the equations are also regular everywhere, and there is a conserved energy associated with this system of equations for the special case~$\zeta = 0$. The energy flux at~$\mathscr{I}^+$ for this special case takes an astonishingly simple form even in full~$3$D, and the energy norm on these slices decreases monotonically with time. For a nonzero~$\zeta$, this setup was also proved to be symmetric hyperbolic and bounds were given on the modified energy norm. This section now discretizes the above setup to get the relevant~SBP scheme that preserves this energy conservation even at the discrete level. A thorough discussion of the~SBP scheme in~NR can be found in~\cite{SarTig12}.

This section also generalizes the concept of dissipation to include the boundary points, to more general coordinates, such that it satisfies the dissipative property~(DP) in our energy norm. It also explains which equations should be augmented with dissipative terms and which variables the dissipation operators should act on. Interestingly, the form of these operators depends on the variables they act on, and not every equation should be added with the dissipative terms. A thorough discussion is given in Sec.~\ref{sec:Dissipation}.

\subsection{Summation-by-Parts scheme overview}\label{sec:SBP_Overview}

We derive here a semi-discrete approximation to the continuum equations that discretizes only spatial coordinates, keeping the time continuity intact. As mentioned in Sec.~\ref{sec:intro}, this can be accomplished by adopting the discrete time integrator that satisfies certain continuum properties required by the system. In such a setup, the spatial coordinates take discrete values~$r_I$,~$\theta_J$ and~$\phi_K$ in the intervals~$[0,r_\mathscr{I}]$ with~$r_\mathscr{I} = 1$,~$[0,\pi]$ and~$[0,2\pi)$, respectively. Here,~$I, J$ and~$K$ are integer values in~$[0, N_r]$,~$[0, N_\theta]$ and~$[0, N_\phi)$, where~$N_r, N_\theta$ and~$N_\phi$ are again positive integers, satisfying
\begin{align}
& 0 \leq r_0 < r_1 < r_2 < \ldots < r_{N_r} = 1 \non \\
& 0 \leq \theta < \theta_1 < \theta_2 < \ldots < \theta_{N_\theta} = \pi \non \\
& 0 \leq \phi_0 < \phi_1 < \phi_2 < \ldots < \phi_{N_\phi} = 2\pi \, .
\end{align}

All the variables that were originally functions of these coordinates are now replaced with arrays of integers and are denoted in uppercase. For example, we define~$(\Psi_\alpha)_{IJK} \equiv \psi_\alpha(r_I, \theta_J, \phi_K)$, where~$\psi_\alpha$ here stands for any of the variables~$\psi$, $\psi_+$, $\psi_-$, $\psi_\theta$, or~$\psi_\phi$. In general, the state vector~$\mathbf{u}$ is now replaces with~$\mathbf{U}_{I J K} \equiv \mathbf{u}(r_I, \theta_J, \phi_K)$, with~$\mathbf{U} = \{ U_\alpha \} \equiv (\tilde{\Psi}, \tilde{\Psi}_+, \tilde{\Psi}_-, \tilde{\Psi}_\theta, \tilde{\Psi}_\phi)^T$.

Defining the discrete analogue of the symmetrizer~$\mathbf{H}$ in~\eqref{eq:Symmetrizer} in a similar way, and denoting it with~$\mathbf{\hat{H}}$, the discrete energy norm in these variables is now given by
\begin{align}\label{eq:Energy_norm_vector_form_Discrete}
\hat{E}(t) \equiv &\ \frac{1}{2} (\mathbf{U}, \mathbf{U})_\mathbf{H} \equiv \frac{1}{2} (\mathbf{U}, \mathbf{\hat{H}} \Upsilon \mathbf{U}) \non \\
\equiv &\ \frac{1}{2} \sum_{\substack{I', J', K' \\ I'', J'', K'' \\ I, \, J, \, K}} \mathbf{U}_{I'J'K'}^T \, \Upsilon_{\substack{I' J' K' \\ I'' J'' K''}} \, \mathbf{\hat{H}}_{\substack{I'' J'' K'' \\ I \, J \, K}} \, \mathbf{U}_{IJK} \, ,
\end{align}
where~$(\mathbf{U}^T, \mathbf{U})$ denotes the standard~$L^2$-norm on the state space spanned by~$U_\alpha$. Additionally,~$\Upsilon$ denotes the quadrature matrix on the grid space, a discrete version of the differential~$dr \, d\theta \, d\phi$, and encodes the information of the grid spacing at the point~$(I,J,K)$. On the state space, it is just a scalar.

Without losing generality, we can demand that the quadrature decomposes on the grid space as
\begin{align}\label{eq:Quadrature}
\Upsilon_{I'J'K'IJK} \equiv \Upsilon^r_{I'I} \, \Upsilon^\theta_{J'J} \, \Upsilon^\phi_{K'K} \, ,
\end{align}
and is symmetric in the~$(I',I)$,~$(J',J)$, and~$(K',K)$ indices,
\begin{align}\label{eq:Quadrature_symmetric}
(\Upsilon^r)^T = \Upsilon^r \, , \quad (\Upsilon^\theta)^T = \Upsilon^\theta \, , \quad (\Upsilon^\phi)^T = \Upsilon^\phi \, .
\end{align}
We also assume the same symmetry conditions on the symmetrizer~$\mathbf{\hat{H}}$
\begin{align}\label{eq:Symmetrizer_symmetric}
\mathbf{\hat{H}}_{\substack{I' J' K' \\ I \, J \, K}} = \mathbf{\hat{H}}_{\substack{I \, J' K' \\ I' J \, K}} = \mathbf{\hat{H}}_{\substack{I' J \, K' \\ I \, J' K}} = \mathbf{\hat{H}}_{\substack{I' J' K \\ I \, J \, K'}} \, ,
\end{align}
and demand that the quadrature commutes with the components of the symmetrizer on the grid space
\begin{align}\label{eq:Quadraure_Commutations}
\Upsilon \, \mathbf{\hat{H}}^{\alpha\beta} = \mathbf{\hat{H}}^{\alpha\beta} \, \Upsilon \, .
\end{align}
We denote the discrete analogues of the partial derivatives~$\p_r$,~$\p_\theta$ and~$\p_\phi$ with~$(\Upsilon^r)^{-1} D_r$,~$(\Upsilon^\theta)^{-1} D_\theta$ and~$(\Upsilon^\phi)^{-1} D_\phi$, respectively, and the regularized covariant derivatives~$\tilde{\p}_r$,~$\tilde{\p}_\theta$ and~$\tilde{\p}_\phi$ with~$(\Upsilon^r)^{-1} \tilde{D}_r$,~$(\Upsilon^\theta)^{-1} \tilde{D}_\theta$ and~$(\Upsilon^\phi)^{-1} \tilde{D}_\phi$, respectively, to keep the discrete operators~$D_r$,~$D_\theta$ and~$D_\phi$, and, similarly,~$\tilde{D}_r$,~$\tilde{D}_\theta$ and~$\tilde{D}_\phi$, dimensionless.

The discrete analogue of~\cref{eq:LWEP_vector_form} can now be written as
\begin{align}\label{eq:LWEP_vector_form_Discrete}
\p_t \mathbf{U} = \mathbf{\hat{A}}^i \, (\Upsilon^i)^{-1} \, D_i \mathbf{U} + \mathbf{\hat{S} \, U} \, ,
\end{align}
where~$\mathbf{\hat{A}}^i$ and~$\mathbf{\hat{S}}$ have obvious meanings. Taking the time derivative of the discrete energy~\eqref{eq:Energy_norm_vector_form_Discrete}, and substituting the discrete~EOMs~\eqref{eq:LWEP_vector_form_Discrete}, one obtains
\begin{align}\label{eq:Energy_change_Discrete}
\frac{d}{dt} \hat{E}(t) = &\ (\mathbf{U}, \mathbf{\hat{A}}^i \, (\Upsilon^i)^{-1} \, D_i \mathbf{U} + \mathbf{\hat{S} \, U})_\mathbf{H} + \frac{1}{2} (\mathbf{U}, \mathbf{\dot{\hat{H}}} \Upsilon \mathbf{U}) \non \\
& + \frac{1}{2} (\mathbf{U}, \mathbf{\hat{H}} \, \dot{\Upsilon} \mathbf{U}) \, ,
\end{align}
where~$\dot{\Upsilon} \equiv \p_t \Upsilon$ and~$\mathbf{\dot{\hat{H}}} \equiv \p_t \mathbf{\hat{H}}$. The first two terms on the right give a discrete analogue of the first term in~\eqref{eq:Energy_change}, and the third term here corresponds to the second term in~\eqref{eq:Energy_change}. Since the matrix~$\mathbf{H}$ in~\eqref{eq:Symmetrizer} is time independent, we can choose the same for~$\mathbf{\hat{H}}$, giving~$\mathbf{\dot{\hat{H}}} = 0$ for all times.

We can also fix the grid points in the bulk to constant spatial coordinate values while letting the ones on the outer boundary move with time. In this case, the effect of a moving outer boundary gets captured in~$\dot{\Upsilon}$, so that the third term in~\eqref{eq:Energy_change_Discrete} can be made a discrete analogue to~\eqref{eq:Moving_Boundary}.

Finally, we can now choose a discretization that gives
\begin{align}
(\mathbf{U}, \mathbf{\hat{A}}^i \, (\Upsilon^i)^{-1} \, D_i \mathbf{U} + \mathbf{\hat{S} \, U})_\mathbf{H} + \frac{1}{2} (\mathbf{U}, \mathbf{\dot{\hat{H}}} \Upsilon \mathbf{U}) = (\mathbf{U}, \mathbf{U})_\mathbf{\p H} \, ,
\end{align}
where~$(\mathbf{U}, \mathbf{U})_\mathbf{\p H}$ gives the discrete analogue to a surface integral at the outer boundary, defined by
\begin{align}
(\mathbf{U}, \mathbf{U})_\mathbf{\p H} \equiv \frac{1}{2} \sum_{\substack{I', J', K' \\ I'', J'', K'' \\ I, \, J, \, K}} \mathbf{U}_{I'J'K'}^T \, \Upsilon_{\substack{I' J' K' \\ I'' J'' K''}} \, \mathbf{\hat{B}}_{\substack{I'' J'' K'' \\ I \, J \, K}} \, \mathbf{U}_{IJK} \, ,
\end{align}
for some boundary matrix~$\mathbf{\hat{B}}$ which is nontrivial only at the outer boundary, corresponding to all the grid points with~$I, I'' = N_r$, and having the same symmetry conditions as~$\mathbf{\hat{H}}$. If these conditions are satisfied, then all the bulk summations are reduced to the summations over the boundary points, and we call it the~SBP scheme.

\subsection{Grid Structure}\label{sec:Grid}

As stated previously, we will focus on finite-difference~(FD) methods and introduce a grid structure suitable for them. A suitable grid could also be specified for pseudo-spectral methods, which will be addressed in future work.

Our computational domain spans the full three-dimensional space in compactified spherical polar coordinates,~$r \in [0, r_{\mathscr{I}}]$,~$\theta \in [0, \pi]$, and~$\phi \in [0, 2\pi)$, with future null infinity,~$\mathscr{I}^+$, located at $r_{\mathscr{I}}$. As mentioned before, we take~$r_\mathscr{I} = 1$. We introduce a non-staggered grid, the one with grid points at the boundaries, in this domain, with~$N_r + 1$,~$N_{\theta} + 1$, and~$N_{\phi}$ grid points along the~$\hat{r}$,~$\hat{\theta}$, and~$\hat{\phi}$ directions, respectively,~$N_r$,~$N_\theta$ and~$N_\phi$ being positive integers. For simplicity, we construct a uniformly spaced grid along all these directions, which fixes grid points on the following coordinate locations
\begin{align}\label{eq:grid}
& r_I \in \{0,\Delta r,2\Delta r,\ldots,N_r\Delta r\} \, , \quad \textrm{with,} \quad \Delta r = 1/N_r \, , \nonumber \\
& \theta_J \in \{0,\Delta \theta,2\Delta \theta,\ldots,N_\theta \Delta \theta\} \, , \quad \textrm{with,} \quad \Delta \theta = \pi/N_\theta \, , \nonumber \\
& \phi_K \in \{0,\Delta \phi,2\Delta \phi,\ldots,(N_\phi - 1)\Delta \phi\} \, , \textrm{with } \Delta \phi = 2\pi/N_\phi \,.
\end{align}
This choice of indexing ensures that grid points are placed explicitly at the origin $r=0$, at the outer boundary $r=r_{\mathscr{I}}$, and on the polar axis $\theta=0,\pi$, while preserving the periodicity~\eqref{eq:Parity_phi} along the azimuthal direction. Imposing the parity conditions~\eqref{eq:Parity},~\eqref{eq:Parity_FOR_R_theta} and~\eqref{eq:Parity_FOR_Null_R} numerically requires~$N_\phi$ being even and for every~$J$, there must be a grid point at~$N_\theta - J$, such that~$J \, \Delta \theta + (N_\theta - J) \Delta \theta = \pi$, which is true in our case for all~$N_\theta \geq 2$.

All spatial differential operators and coefficients appearing in the continuum equations~\eqref{eq:hyp_LWEP_rescaled_FOR_null_Tilded}-\eqref{eq:hyp_LWEP_rescaled_FOR_null_Tilded_z-axis_scri}, which in general are functions of~$r$,~$\theta$ and~$\phi$, are now approximated with discrete operators, which are the outer products of~$(N_r + 1) \times (N_r +1)$,~$(N_\theta + 1) \times (N_\theta +1)$ and~$N_\phi \times N_\phi$ matrices along the~$\hat{r}$,~$\hat{\theta}$ and~$\hat{\phi}$ directions, respectively. All differential operators~$\p_i$ and~$\tilde{\p}_i$ in the continuum equations are now denoted with~$(\Upsilon^i)^{-1} D_i$ and~$(\Upsilon^i)^{-1} \tilde{D}_i$, respectively, and all coefficients~$f(r,\theta,\phi)$ in the continuum equations are now denoted with the operators~$[f]$ in the discrete equations.

For simplicity, we take all the operators~$[f(r,\theta,\phi)]$ to be diagonal matrices with components
\begin{align}\label{eq:Discrete_Coefficients}
[f(r,\theta,\phi)]_{I J K I' J' K'} = \delta^r_{I I'} \, \delta^\theta_{J J'} \, \delta^\phi_{K K'} \, f(r_I, \theta_J, \phi_K) \, ,
\end{align}
where~$\delta^r_{II'}$,~$\delta^\theta_{JJ'}$ and~$\delta^\phi_{KK'}$ are Kronecker deltas of obvious dimensionalities. The same construction applies to the elements of the symmetrizer~$\mathbf{\hat{H}}$
\begin{align}
\mathbf{\hat{H}}^{\alpha\beta}_{IJKI'J'K'} = \delta^r_{II'} \, \delta^\theta_{JJ'} \, \delta^\phi_{KK'} \, \mathbf{H}^{\alpha\beta}(r_I, \theta_J, \phi_K) \, ,
\end{align}
with~$\alpha\beta$ being indices over the state space. We define all our norms on a centred grid, where all the intervals of size~$\Delta r$,~$\Delta\theta$, and~$\Delta\phi$ are symmetrically distributed about the grid points. So the boundary points along~$\hat{\theta}$, which correspond to~$\theta = 0$ and~$\pi$, are left with one-sided intervals of size~$\Delta\theta/2$ towards the bulk. Similarly, the boundary points along~$\hat{r}$, corresponding to~$r = 0$ and~$1$, are left with one-sided intervals of size~$\Delta r/2$ towards the bulk. From the periodicity property, all grid points along~$\hat{\phi}$ are internal, and so have the same quadrature~$\Delta\phi$ at all grid points. With this construction, the quadrature matrix reduces to
\begin{align}\label{eq:Quadrature_diagonal}
& \Upsilon^r = {\rm diag}(\Delta r/2, \Delta r, \ldots, \Delta r, \Delta r/2) \, , \non \\
& \Upsilon^\theta = {\rm diag}(\Delta\theta/2, \Delta\theta, \ldots, \Delta\theta, \Delta\theta/2) \, , \non \\
& \Upsilon^\phi = {\rm diag}(\Delta\phi, \Delta\phi, \ldots, \Delta\phi, \Delta\phi) \, ,
\end{align}
with obvious dimensionalities, and all the properties on~$\mathbf{\hat{H}}$ and~$\Upsilon$ demanded above are automatically satisfied.

We apply the same construction as in Sec.~IV-B of~\cite{GauVanHil21} to incorporate the cases where the outer boundary of the domain lies much inside~$\mathscr{I}^+$, corresponding to the radial coordinate~$r_o < r_\mathscr{I}$, and moves with time. In this case, too, we keep all our grid points in the bulk fixed to their coordinate values, while letting the radial coordinate at the boundary points change with time,~$r = r_o(t)$. The radial quadrature in this case modifies to
\begin{align}\label{eq:Quadrature_Moving_boundary}
\Upsilon^r = {\rm diag}(\Delta r_0, \Delta r_1, \ldots, \Delta r_{N-1}, \Delta r_N) \, ,
\end{align}
with~$\Delta r_0 = \Delta r/2$,~$\Delta r_I = \Delta r$ for~$I = 1,\ldots, N-1$, and~$\Delta r_N = \Delta r_N(t)$ can take values only in~$0 < \Delta r_N \leq \Delta r$. Whenever~$r_N - r_{N-1} > \Delta r/2$,~$\Delta r_N$ is given by
\begin{align}
\Delta r_N = r_N - r_{N-1} - \Delta r/2 \, ,
\end{align}
and whenever~$r_N - r_{N-1} < \Delta r/2$, the last grid point does not contribute to the norm, and~$\Delta r_{N-1}$ takes the value
\begin{align}
\Delta r_{N-1} = r_N - r_{N-1} + \Delta r/2 \, .
\end{align}

Whenever~$r_N$ merges with~$r_{N-1}$, the dimensionality along the radial direction of all the operators and vectors on the grid space reduces by~$1$, and~$r_{N-1}$ becomes the new boundary point. The radial quadrature at~$r_{N-1}$ is now calculated in the same way as for~$r_N$ above. Whereas, whenever~$r_N$ moves further away from~$r_{N-1}$ with~$r_N - r_{N-1} > \Delta r$, we fix~$r_N$ at~$r_{N-1} + \Delta r$ and create a new grid point~$r_{N+1}$ at the outer boundary, with radial coordinate~$r_{N+1} = r_o - r_N$. In this case, the dimensionality of all the operators and vectors on the grid space increases by~$1$ along the radial direction, and the radial quadrature at the grid points~$r_{N+1}$ is again calculated in the same way as for~$r_N$ above. We take~$\Delta r_N = \Delta r/2$ on the initial slice.

All these cases give
\begin{align}
\dot{\Upsilon} = \frac{\p \Upsilon}{\p r_N} \dot{r}_N = \Upsilon^\theta \, \Upsilon^\phi \, \delta^r_{I N} \, \delta^r_{N I'} \, \dot{r}_N \, ,
\end{align}
where~$I,I' = 0, \ldots, N$, and~$N$ is the index of the last radial grid point. We take~$N = N_r$ at~$t=0$, and then let it change depending on the trajectory of~$r_o$, as described above. The third term on the right of~\eqref{eq:Energy_change_Discrete} now becomes
\begin{align}
& \frac{1}{2} (\mathbf{U}, \mathbf{\hat{H}} \, \dot{\Upsilon} \mathbf{U}) \non \\
&= \frac{\dot{r}_N}{2} \sum_{JK} (\mathbf{U}_\alpha)_{NJK} \mathbf{\hat{H}}^{\alpha\beta}(r_N, \theta_J, \phi_K) \, \Upsilon^\theta_{JJ} \, \Upsilon^\phi_{KK} (\mathbf{U}_\beta)_{NJK} \, .
\end{align}
The same arguments apply for a more general case, where~$r_o$ also depends on the angles,~$r_o = r_o(t,\theta,\phi)$. In this case, the above expression will still hold with the additional modification that the dimensionality along the radial direction will now depend on the angles as well,~$N = N(t,\theta,\phi)$.

Whenever the outer boundary traverses along the incoming null geodesics, we get
\begin{align}
\dot{r}_N = -\frac{1}{2 R_N'-1} \, , \quad \textrm{where} \quad R_N' = R'(r_N) \, . 
\end{align}
This value vanishes whenever the outer boundary starts at~$\mathscr{I}^+$, giving~$N = N_r$ for all times, and the third term on the right of~\eqref{eq:Energy_change_Discrete} adds to the other terms in the same way both terms on the right of~\eqref{eq:Energy_change} combine to give the energy flux~\eqref{eq:Stokes_diff_form_rescaled} at~$\mathscr{I}^+$.

\subsection{Discrete equations: The SBP scheme}\label{sec:SBP_Scheme}

We now discretize the~EOMs,~\eqref{eq:hyp_LWEP_rescaled_FOR_null_Tilded}-\eqref{eq:hyp_LWEP_rescaled_FOR_null_Tilded_z-axis_scri}, and the energy norm~\eqref{eq:Energy_norm}, with the energy density~$\varepsilon$ given by~\eqref{eq:Energy_Density_Rescaled_Variables}, to obtain the~SBP scheme. Most of the derivation remains independent of the choice of discretization, whether~FD or pseudo-spectral, and, hence, of the grid structure. We begin with the case of zero constraint damping,~$\tilde{\zeta} = 0$, and then propose how the scheme is modified when the constraint damping is switched on, by taking~$\tilde{\zeta} = 1$.

As stated before, we denote the discrete analogues of~$\p_i$ and~$\tilde{\p}_i$ with $\Upsilon^i D_i$ and~$\Upsilon^i \tilde{D}_i$, respectively, and the time derivative with a dot overhead. Also, all the coefficients in the continuum equations are now replaced by operators, with the representation given in~\cref{eq:Discrete_Coefficients}. This way, all the multiplicative operators commute with each other, and have the same multiplicative properties as the continuum functions,~$[f][g] = [g][f] = [fg]$. The discrete~EOMs in the bulk then become
\begin{align}\label{eq:Discrete_EOM_bulk}
\dot{\tilde{\Psi}} = &\ \frac{1}{2} \left( \left[ \frac{1}{\chi} \right] \tilde{\Psi}_+ + \tilde{\Psi}_- \right) \, , \non  \\
\dot{\tilde{\Psi}}_+ = &\ \left[ \frac{\chi}{2R'-1} \right] \Bigg( (\Upsilon^r)^{-1} \left[ \frac{D_r + \tilde{D}_r}{2} \right] \left[ \frac{1}{\chi} \right] \tilde{\Psi}_+ \non \\
& + (\Upsilon^r)^{-1} \left[ \frac{D_r - \tilde{D}_r}{2} \right] \tilde{\Psi}_- - \left[ \frac{\chi'}{\chi} \right] \tilde{\Psi}_- - \left[ R' F \right] \tilde{\Psi} \non \\
& + \left[ \frac{R'}{R^2} \right] \bigg( (\Upsilon^\theta)^{-1} \, \tilde{D}_\theta \tilde{\Psi}_\theta + \left[ \frac{1}{\sin^2\theta} \right] (\Upsilon^\phi)^{-1} \, \tilde{D}_\phi \tilde{\Psi}_\phi \bigg) \non \\
& + \tilde{\zeta} \left[ \frac{1}{\chi} \right] \bigg( (\Upsilon^r)^{-1} \, D_r \tilde{\Psi} - \left[ \frac{\chi'}{\chi} \right] \tilde{\Psi} - \left[ \frac{2R'-1}{2 \chi} \right] \tilde{\Psi}_+ \non \\
& + \frac{\tilde{\Psi}_-}{2} \bigg) \Bigg) \, , \non \\
\dot{\tilde{\Psi}}_- = &\ - (\Upsilon^r)^{-1} \left[ \frac{D_r + \tilde{D}_r}{2} \right] \tilde{\Psi}_- + \left[ \frac{R'}{R^2} \right] \bigg(  (\Upsilon^\theta)^{-1} \, \tilde{D}_\theta \tilde{\Psi}_\theta \non \\
& + \left[ \frac{1}{\sin^2\theta} \right]  (\Upsilon^\phi)^{-1} \, \tilde{D}_\phi \tilde{\Psi}_\phi \bigg) + \left[ \frac{\chi'}{\chi^2} \right] \tilde{\Psi}_+ - \left[ R' F \right] \tilde{\Psi} \non \\
& - (\Upsilon^r)^{-1} \left[ \frac{D_r - \tilde{D}_r}{2} \right] \left[ \frac{1}{\chi} \right] \tilde{\Psi}_+ - \tilde{\zeta} \left[ \frac{1}{\chi} \right] \bigg( - \left[ \frac{\chi'}{\chi} \right] \tilde{\Psi} \non \\
& + (\Upsilon^r)^{-1} \, D_r \tilde{\Psi} - \left[ \frac{2R'-1}{2 \chi} \right] \tilde{\Psi}_+ + \frac{\tilde{\Psi}_-}{2} \bigg) \, , \non \\
\dot{\tilde{\Psi}}_\theta = &\ \frac{1}{2} \, (\Upsilon^\theta)^{-1} \, D_\theta \left( \left[ \frac{1}{\chi} \right] \tilde{\Psi}_+ + \tilde{\Psi}_- \right) \non \\
& + \tilde{\zeta} \left[ \frac{1}{\chi} \right] \bigg( (\Upsilon^\theta)^{-1} \, D_\theta \tilde{\Psi} - \tilde{\Psi}_\theta \bigg) \, , \non \\
\dot{\tilde{\Psi}}_\phi = &\ \frac{1}{2} \, (\Upsilon^\phi)^{-1} \, D_\phi \left( \left[ \frac{1}{\chi} \right] \tilde{\Psi}_+ + \tilde{\Psi}_- \right) \non \\
& + \tilde{\zeta} \left[ \frac{1}{\chi} \right] \bigg( (\Upsilon^\phi)^{-1} \, D_\phi \tilde{\Psi} - \tilde{\Psi}_\phi \bigg) \, ,
\end{align}
everywhere,
\begin{align}\label{eq:Discrete_EOM_bulk_z-axis}
\dot{\tilde{\Psi}} = &\ \frac{1}{2} \left( \left[ \frac{1}{\chi} \right] \tilde{\Psi}_+ + \tilde{\Psi}_- \right) \, , \non  \\
\dot{\tilde{\Psi}}_+ = &\ \left[ \frac{\chi}{2R'-1} \right] \Bigg( (\Upsilon^r)^{-1} \left[ \frac{D_r + \tilde{D}_r}{2} \right] \left[ \frac{1}{\chi} \right] \tilde{\Psi}_+ \non \\
& + (\Upsilon^r)^{-1} \left[ \frac{D_r - \tilde{D}_r}{2} \right] \tilde{\Psi}_- - \left[ \frac{\chi'}{\chi} \right] \tilde{\Psi}_- - \left[ R' F \right] \tilde{\Psi} \non \\
& + \left[ \frac{2 R'}{R^2} \right] (\Upsilon^\theta)^{-1} \, D_\theta \tilde{\Psi}_{\theta (m=0)} + \tilde{\zeta} \left[ \frac{1}{\chi} \right] \bigg( \frac{\tilde{\Psi}_-}{2} \non \\
& + (\Upsilon^r)^{-1} \, D_r \tilde{\Psi} - \left[ \frac{\chi'}{\chi} \right] \tilde{\Psi} - \left[ \frac{2R'-1}{2 \chi} \right] \tilde{\Psi}_+ \bigg) \Bigg)  \, , \non \\
\dot{\tilde{\Psi}}_- = &\ - (\Upsilon^r)^{-1} \left[ \frac{D_r + \tilde{D}_r}{2} \right] \tilde{\Psi}_- + \left[ \frac{\chi'}{\chi^2} \right] \tilde{\Psi}_+ - \left[ R' F \right] \tilde{\Psi} \non \\
& - (\Upsilon^r)^{-1} \left[ \frac{D_r - \tilde{D}_r}{2} \right] \left[ \frac{1}{\chi} \right] \tilde{\Psi}_+ \non \\
& + \left[ \frac{2 R'}{R^2} \right] (\Upsilon^\theta)^{-1} \, D_\theta \tilde{\Psi}_{\theta (m=0)} - \tilde{\zeta} \left[ \frac{1}{\chi} \right] \bigg( \frac{\tilde{\Psi}_-}{2} \non \\
& + (\Upsilon^r)^{-1} \, D_r \tilde{\Psi} - \left[ \frac{\chi'}{\chi} \right] \tilde{\Psi} - \left[ \frac{2R'-1}{2 \chi} \right] \tilde{\Psi}_+ \bigg) \, , \non \\
\dot{\tilde{\Psi}}_\theta = &\ \frac{1}{2} \, (\Upsilon^\theta)^{-1} \, D_\theta \left( \left[ \frac{1}{\chi} \right] \tilde{\Psi}_+ + \tilde{\Psi}_- \right) \non \\
& + \tilde{\zeta} \left[ \frac{1}{\chi} \right] \left( (\Upsilon^\theta)^{-1} \, D_\theta \tilde{\Psi} -\tilde{\Psi}_\theta \right) \, , \non \\
\dot{\tilde{\Psi}}_\phi = &\ 0 \, ,
\end{align}
on the~$z$-axis, and
\begin{align}\label{eq:Discrete_EOM_origin}
\dot{\tilde{\Psi}} = &\ \frac{\tilde{\Psi}_+ + \tilde{\Psi}_-}{2} \, , \non  \\
\dot{\tilde{\Psi}}_+ = &\ \frac{1}{2} \, (\Upsilon^r)^{-1} \, D_r \left( 3 \, (\tilde{\Psi}_+ - \tilde{\Psi}_-)_{(l=0)} + (\tilde{\Psi}_+ + \tilde{\Psi}_-) \right) \non \\
& - [F] \tilde{\Psi}_{(l=0)} + \tilde{\zeta} \, \bigg( (\Upsilon^r)^{-1} \, D_r \tilde{\Psi} - \frac{\tilde{\Psi}_+ - \tilde{\Psi}_-}{2} \bigg) \, , \non \\
\dot{\tilde{\Psi}}_- = &\ \frac{1}{2} \, (\Upsilon^r)^{-1} \, D_r \left( 3 \, (\tilde{\Psi}_+ - \tilde{\Psi}_-)_{(l=0)} - (\tilde{\Psi}_+ + \tilde{\Psi}_-) \right) \non \\
& - [F] \tilde{\Psi}_{(l=0)} - \tilde{\zeta} \, \bigg( (\Upsilon^r)^{-1} \, D_r \tilde{\Psi} - \frac{\tilde{\Psi}_+ - \tilde{\Psi}_-}{2} \bigg) \, , \non \\
\dot{\tilde{\Psi}}_\theta = &\ 0 \, , \non \\
\dot{\tilde{\Psi}}_\phi = &\ 0 \, ,
\end{align}
at the origin. The ones at~$\mathscr{I}^+$, for~$n=2$, become
\begin{align}\label{eq:Discrete_EOM_scri}
\dot{\tilde{\Psi}} = &\ \frac{\tilde{\Psi}_-}{2} \, , \non  \\
\dot{\tilde{\Psi}}_+ = &\ - \frac{\tilde{\Psi}_-}{2} - \left[ \frac{R F}{2} \right] \tilde{\Psi} \, , \non \\
\dot{\tilde{\Psi}}_- = &\ - (\Upsilon^r)^{-1} \left[ \frac{D_r + \tilde{D}_r}{2} \right] \tilde{\Psi}_- + 2 \, \tilde{\Psi}_+ - \left[ R' F \right] \tilde{\Psi} \non \\
& - (\Upsilon^r)^{-1} \left[ \frac{D_r - \tilde{D}_r}{2} \right] \left[ \frac{1}{\chi} \right] \tilde{\Psi}_+ + 2 \, \bigg( (\Upsilon^\theta)^{-1} \, \tilde{D}_\theta \tilde{\Psi}_\theta \non \\
& + \left[ \frac{1}{\sin^2\theta} \right] (\Upsilon^\phi)^{-1} \, \tilde{D}_\phi \tilde{\Psi}_\phi \bigg) + 2 \, \tilde{\zeta} \, ( \tilde{\Psi} + \tilde{\Psi}_+ ) \, , \non \\
\dot{\tilde{\Psi}}_\theta = &\ \frac{1}{2} \, (\Upsilon^\theta)^{-1} \, D_\theta \tilde{\Psi}_- \, , \non \\
\dot{\tilde{\Psi}}_\phi = &\ \frac{1}{2} \, (\Upsilon^\phi)^{-1} \, D_\phi \tilde{\Psi}_- \, ,
\end{align}
everywhere, and
\begin{align}\label{eq:Discrete_EOM_z-axis_scri}
\dot{\tilde{\Psi}} = &\ \frac{\tilde{\Psi}_-}{2} \, , \non  \\
\dot{\tilde{\Psi}}_+ = &\ - \frac{\tilde{\Psi}_-}{2} - \left[ \frac{R F}{2} \right] \tilde{\Psi} \, , \non \\
\dot{\tilde{\Psi}}_- = &\ - (\Upsilon^r)^{-1} \left[ \frac{D_r + \tilde{D}_r}{2} \right] \tilde{\Psi}_- + 2 \, \tilde{\Psi}_+ - \left[ R' F \right] \tilde{\Psi} \non \\
& - (\Upsilon^r)^{-1} \left[ \frac{D_r - \tilde{D}_r}{2} \right] \left[ \frac{1}{\chi} \right] \tilde{\Psi}_+ \non \\
& + 4 \, (\Upsilon^\theta)^{-1} \, D_\theta \tilde{\Psi}_{\theta (m=0)} + 2 \, \tilde{\zeta} \, ( \tilde{\Psi} + \tilde{\Psi}_+ ) \, , \non \\
\dot{\tilde{\Psi}}_\theta = &\ \frac{1}{2} \, (\Upsilon^\theta)^{-1} \, D_\theta \tilde{\Psi}_- \, , \non \\
\dot{\tilde{\Psi}}_\phi = &\ 0 \, ,
\end{align}
on the~$z$-axis.

We compute here the~$m=0$ and~$l=0$ modes the same way we did in the continuum case, eqs.~\eqref{eq:m=0_mode} and~\eqref{eq:l=0_mode}. As stated above, extracting the~$m=0$ mode correctly from the relation
\begin{align}\label{eq:Discrete_m=0_mode}
(\Psi_{m=0})_{IJK} = \frac{1}{2 \pi} \sum_{K'=0}^{N_\phi - 1} (\Upsilon^\phi)_{KK'} \Psi_{IJK'} \, ,
\end{align}
requires that, for every grid point~$K$, located at~$\phi_K$, there exists a grid point~$K + N_\phi/2$ located at~$\phi_K + \pi$, for~$\phi_K < \pi$, or~$K - N_\phi/2$, located at~$\phi_K - \pi$, for~$K\Delta\phi \geq \pi$, so that all~$m>0$ modes are cancelled perfectly. This condition requires~$N_\phi$ to be even.

Similarly, the~$l=0$ mode, extracted correctly from the relation
\begin{align}\label{eq:Discrete_l=0_mode}
(\Psi_{l=0})_{IJK} = \frac{1}{4\pi} \sum_{\substack{J',J'' \\ =0}}^{N_\theta} \sum_{K'=0}^{N_\phi - 1} (\Upsilon^\theta)_{JJ'} \, (\Upsilon^\phi)_{KK'} \, [\sin\theta]_{J'J''} \, \Psi_{IJ''K'} \, ,
\end{align}
requires, additionally, that for every grid point~$J$, located at~$\theta_J$, there must be a grid point~$N_\theta - J$ located at~$\pi - \theta_J$, so that all the~$l>0$ modes are cancelled perfectly. The modes~$\Psi_{m=0}$ are, thereby, independent of~$K$, and~$\Psi_{l=0}$ are independent of both~$J$ and~$K$.

Inspired from~\eqref{eq:Energy_norm} and~\eqref{eq:Energy_Density_Rescaled_Variables}, and using the prescription~\eqref{eq:Energy_norm_vector_form_Discrete}, we define our discrete energy norm as
\begin{align}\label{eq:Discrete_Energy_Norm}
\hspace{-1.0em}\hat{E}(t) = &\ \frac{1}{2} \bigg( \tilde{\Psi}^T \Upsilon [F] \tilde{W} \tilde{\Psi} + \tilde{\Psi}_+^T \Upsilon \tilde{W}_+ \tilde{\Psi}_+ + \tilde{\Psi}_-^T \Upsilon \tilde{W}_- \tilde{\Psi}_- \non \\
& + \tilde{\Psi}_\theta^T \Upsilon \tilde{W}_\theta \tilde{\Psi}_\theta + \tilde{\Psi}_\phi^T \Upsilon \tilde{W}_\phi \tilde{\Psi}_\phi \bigg) \, ,
\end{align}
where
\begin{align}\label{eq:Discrete_Weights}
& \tilde{W} = \left[ \frac{R' \, R^2 \, \sin\theta}{\chi^2} \right] \, , \, \tilde{W}_+ = \left[ \frac{(2R'-1) \, R^2 \, \sin\theta}{2\, \chi^4} \right] , \non \\
& \tilde{W}_- = \left[ \frac{R^2 \, \sin\theta}{2 \, \chi^2} \right] , \, \tilde{W}_\theta = \left[ \frac{R' \, \sin\theta}{\chi^2} \right] , \, \tilde{W}_\phi = \left[ \frac{R'}{\chi^2 \, \sin\theta} \right] .
\end{align}
The quadratic form~\eqref{eq:Discrete_Energy_Norm} suggests that all these weight matrices can be taken to be symmetric on the grid space, without loss of generality. In other words, we can take
\begin{align}
W_{II'JJ'KK'} = W_{I'IJJ'KK'} = W_{II'J'JKK'} = W_{II'JJ'K'K} \, .
\end{align}
As we take all of these matrices to be diagonal, this property is automatically satisfied.

Taking time derivative and substituting the EOMs \eqref{eq:Discrete_EOM_bulk} with~$\tilde{\zeta} = 0$, and using the properties~\eqref{eq:Quadrature}-\eqref{eq:Quadraure_Commutations}, we get
\begin{align}
\p_t \hat{E}(t) = &\ \tilde{\Psi}_+^T \left[ \frac{1}{\chi} \right] \Upsilon^\theta \, \Upsilon^\phi \, \tilde{W}_- \left[ \frac{D_r + \tilde{D}_r}{2} \right] \left[ \frac{1}{\chi} \right] \tilde{\Psi}_+ \non \\
& - \tilde{\Psi}_-^T \, \Upsilon^\theta \, \Upsilon^\phi \, \tilde{W}_- \left[ \frac{D_r + \tilde{D}_r}{2} \right] \tilde{\Psi}_- \non \\
& + \tilde{\Psi}_+^T \left[ \frac{1}{\chi} \right] \Upsilon^\theta \, \Upsilon^\phi \, \tilde{W}_- \left[ \frac{D_r - \tilde{D}_r}{2} \right] \tilde{\Psi}_- \non \\
& - \tilde{\Psi}_-^T \, \Upsilon^\theta \, \Upsilon^\phi \, \tilde{W}_- \left[ \frac{D_r - \tilde{D}_r}{2} \right] \left[ \frac{1}{\chi} \right] \tilde{\Psi}_+ \non \\
& + \tilde{\Psi}_+^T \left[ \frac{1}{\chi} \right] \Upsilon^r \, \Upsilon^\phi \, \frac{\tilde{W}_\theta}{2} \, \tilde{D}_\theta \tilde{\Psi}_\theta \non \\
& + \tilde{\Psi}_\theta^T \, \Upsilon^r \, \Upsilon^\phi \, \frac{\tilde{W}_\theta}{2} D_\theta \left[ \frac{1}{\chi} \right] \tilde{\Psi}_+ \non \\
& + \tilde{\Psi}_-^T \, \Upsilon^r \, \Upsilon^\phi \, \frac{\tilde{W}_\theta}{2} \, \tilde{D}_\theta \tilde{\Psi}_\theta + \tilde{\Psi}_\theta^T \, \Upsilon^r \, \Upsilon^\phi \, \frac{\tilde{W}_\theta}{2} D_\theta \tilde{\Psi}_- \non \\
& + \tilde{\Psi}_+^T \left[ \frac{1}{\chi} \right] \Upsilon^r \, \Upsilon^\theta \, \frac{\tilde{W}_\phi}{2} \, \tilde{D}_\phi \tilde{\Psi}_\phi \non \\
& + \tilde{\Psi}_\phi^T \, \Upsilon^r \, \Upsilon^\phi \, \frac{\tilde{W}_\phi}{2} D_\phi \left[ \frac{1}{\chi} \right] \tilde{\Psi}_+ \non \\
& + \tilde{\Psi}_-^T \, \Upsilon^r \, \Upsilon^\theta \, \frac{\tilde{W}_\phi}{2} \, \tilde{D}_\phi \tilde{\Psi}_\phi + \tilde{\Psi}_\phi^T \, \Upsilon^r \, \Upsilon^\phi \, \frac{\tilde{W}_\phi}{2} D_\phi \tilde{\Psi}_- \, .
\end{align}
The~SBP scheme is now derived by demanding that this energy change equals the flux at the outer boundary that approaches~\eqref{eq:E-dot_rescaled} in the continuum limit. The discrete analogue can be written as
\begin{align}\label{eq:Boundary_Matrix_Definition}
\p_t \hat{E}(t) = \tilde{\Psi}_+^T \left[ \frac{1}{\chi} \right] \Upsilon^\theta \, \Upsilon^\phi \, B \left[ \frac{1}{\chi} \right] \tilde{\Psi}_+ - \tilde{\Psi}_-^T \, \Upsilon^\theta \, \Upsilon^\phi \, B \, \tilde{\Psi}_- \, ,
\end{align}
where~$B$ is called the `boundary matrix', which is nontrivial only at the boundary points, i.e.~$B_{II'} = 0$ for both~$I,I' < N_r$. The quadratic from~\eqref{eq:Boundary_Matrix_Definition} also suggests that one can take~$\tilde{B}$ to be symmetric
\begin{align}
B_{IJKI'J'K'} = B_{I'JKIJ'K'} = B_{IJ'KI'JKK'} = B_{IJK'I'J'K} \, ,
\end{align}
without losing generality. This condition gives us the following relations
\begin{align}\label{eq:SBP_Scheme}
& \tilde{W}_- \left[ \frac{D_r + \tilde{D}_r}{2} \right] + \left( \tilde{W}_- \left[ \frac{D_r + \tilde{D}_r}{2} \right] \right)^T = 2\, B \, , \non \\
& \tilde{W}_- \left[ \frac{D_r - \tilde{D}_r}{2} \right] - \left( \tilde{W}_- \left[ \frac{D_r - \tilde{D}_r}{2} \right] \right)^T = 0 \, , \non \\
& \tilde{W}_\theta \, \tilde{D}_\theta + (\tilde{W}_\theta \, D_\theta)^T = 0 \, , \quad \tilde{W}_\phi \, \tilde{D}_\phi + (\tilde{W}_\phi \, D_\phi)^T \, .
\end{align}
Here, we used~\eqref{eq:Quadrature_symmetric},~\eqref{eq:Quadraure_Commutations}, and the fact that~$\Upsilon^\theta$ and~$\Upsilon^\phi$ are scalars along the~$\hat{r}$-direction, and so they commute with all radial operators. The same property holds for~$\Upsilon^r$ and~$\Upsilon^\phi$ along the~$\hat{\theta}$-direction, and for~$\Upsilon^r$ and~$\Upsilon^\theta$ along the~$\hat{\phi}$-direction. For the same reason, the quadratures~$\Upsilon^r$,~$\Upsilon^\theta$, and~$\Upsilon^\phi$ also commute with each other. The system of equations~\eqref{eq:SBP_Scheme} constitutes what we call our final \textit{\textbf{summation-by-parts~(SBP) scheme}}.

The next task is to solve these equations for various operators. Since the metric components, and hence the~$\tilde{W}$'s do not have any~$\phi$-dependence,~$\tilde{W}_\phi$ commutes with~$D_\phi$ and~$\tilde{D}_\phi$, and the last equation in~\eqref{eq:SBP_Scheme} simply gives
\begin{align}
\tilde{D}_\phi = - D_\phi^T \, .
\end{align}
This equation gives~$\tilde{D}_\phi = D_\phi$ whenever~$(D_\phi)^T = - D_\phi$. This result is consistent with the continuum limit~\eqref{eq:Tilded_Operators}. The centred-grid~FD operators generally satisfy this property in the bulk. As, due to periodicity, all the grid points along~$\hat{\phi}$ are internal, we can, in general, define~$D_\phi$ to be anti-symmetric everywhere, and get~$\tilde{D}_\phi = D_\phi$. We shall demonstrate this construction in the next subsection.

The third equation gives
\begin{align}\label{eq:Dtilde_theta_abstract}
\tilde{D}_\theta = - \tilde{W}_\theta^{-1} \, D_\theta^T \, \tilde{W}_\theta \, .
\end{align}
Since all radial operators~$[f(r)]$ are scalars along~$\hat{\theta}$, they commute with~$D_\theta$ and~$\tilde{D}_\theta$, giving the simple relation
\begin{align}\label{eq:Dtilde_theta}
\tilde{D}_\theta = - \left[ \frac{1}{\sin\theta} \right] D_\theta^T \bigg[ \sin\theta \bigg] \, .
\end{align}
As we observe, eqs~\eqref{eq:hyp_LWEP_rescaled_FOR_null_Tilded_z-axis},~\eqref{eq:hyp_LWEP_rescaled_FOR_null_Tilded_z-axis_scri},~\eqref{eq:Discrete_EOM_bulk_z-axis} and~\eqref{eq:Discrete_EOM_z-axis_scri}, both~$\tilde{\p}_\theta$ and~$\tilde{D}_\theta$ do not appear in the equations on the~$z$-axis, we do not need to worry about their singular behavior there. Again, if we choose~$D_\theta$ such that~$(D_\theta)^T = - D_\theta$, we obtain the relation
\begin{align}\label{eq:Dtilde_theta_applied}
\tilde{D}_\theta = \left[ \frac{1}{\sin\theta} \right] D_\theta \bigg[ \sin\theta \bigg] \, ,
\end{align}
which is consistent with~\eqref{eq:Tilded_Operators}. Again, since all the grid-points along the~$\hat{\theta}$-direction are internal, we can define a centred-grid~FD operator~$D_\theta$ satisfying this anti-symmetric property everywhere, as we shall show in the next subsection.

The first two equations are interestingly not independent, but give a single equation
\begin{align}\label{eq:SBP_Scheme_1}
\tilde{D}_r = - \tilde{W}_-^{-1} \, D_r^T \, \tilde{W}_- + 2 \, \tilde{W}_-^{-1} \, B \, ,
\end{align}
with two unknowns,~$\tilde{D}_r$ and~$B$. Since all these operators are radial, any angular dependence in~$\tilde{W}_-$ will commute with them and cancel out, leaving
\begin{align}
\tilde{D}_r = - \left[ \frac{\chi^2}{R^2} \right] D_r^T \left[ \frac{R^2}{\chi^2} \right] + 2\, \tilde{W}_-^{-1} \, B \, .
\end{align}
Any symmetric~$B$, coming from the first equation in \eqref{eq:SBP_Scheme}, gives a~$\tilde{D}_r$ that does not necessarily converge to~$\tilde{\p}_r$, as defined in~\eqref{eq:Tilded_Operators}, in the continuum limit, unless we decrease the accuracy of~$D_r$ at the outer boundary. To overcome this situation, we define~$\tilde{D}_r$ analogous to the continuum definition
\begin{align}\label{eq:Dtilde_r}
\tilde{D}_r = \left[ \frac{\chi^2}{R^2} \right] D_r \left[ \frac{R^2}{\chi^2} \right] \, ,
\end{align}
and then solve~\eqref{eq:SBP_Scheme_1} to get
\begin{align}\label{eq:Boundary_Matrix}
B = \frac{1}{2} \, \tilde{W}_- \left[ \frac{\chi^2}{R^2} \right] (D_r + D_r^T) \left[ \frac{R^2}{\chi^2} \right] \, .
\end{align}
The price we pay is that this new~$B$ is not symmetric for a general~$D_r$. As a result, the right sides of the first two equations in~\eqref{eq:SBP_Scheme} are replaced by~$B^T+B$ and~$B^T-B$, respectively. However, with a consistent discretization, both~$B^T+B$ and~$B^T-B$ will converge to the continuum limit at the desired accuracy. This will become clearer in the next subsection.

Equation~\eqref{eq:Boundary_Matrix} gives~$B_{II'} = 0$ for~$I,I' < N_r$ whenever~$D_r$ is antisymmetric in the bulk, and gives nontrivial~$B_{IN_r}$ and $B_{N_r I}$, at least for some indices~$I$ in~$[0, N_r]$, depending on the definition of~$D_r$ at the outer boundary. Such a construction is possible for an~FD scheme, as demonstrated in the next subsection.

The final task is to study how this scheme gets modified in the presence of constraint damping terms, and with the symmetrizer~$\mathbf{\hat{H}}\zeta$, a discrete analogue of~$\mathbf{H}\zeta$ defined in~\eqref{eq:Symmetrizer_modified}. Interestingly, all the above results remain unaltered even in the presence of constraint damping. Also, using the~SBP relations above, we obtain a discrete version of~\eqref{eq:Energy_flux_modified} with the same boundary matrix~$B$ defined above.

\subsection{Discrete Operators and Stencils}\label{sec:Disc_Ops}

For numerical implementation, we choose finite-difference~(FD) methods, and confine ourselves to the second-order accuracy. The~FD methods with higher accuracy can be constructed similarly. Constructing a similar pseudo-spectral scheme is kept for the future.

The~FD operators~$D_r$,~$D_\theta$ and~$D_\phi$, when applied to a function~$f_{IJK} \equiv f(r_I, \theta_J, \phi_K)$, can be defined in the bulk on a centred grid as
\begin{align}\label{eq:Dr_Bulk}
\left( D_r f \right)_{I,J,K} = \frac{f_{I+1,J,K} - f_{I-1,J,K}}{2} \, ,
\end{align}
for~$I = 1, \ldots, N_r-1$, and~$\forall$ $J$ and~$K$,
\begin{align}\label{eq:Dtheta_Bulk}
\left( D_\theta f \right)_{I,J,K} = \frac{f_{I,J+1,K} - f_{I,J-1,K}}{2} \, ,
\end{align}
for~$J = 1, \ldots, N_\theta-1$, and~$\forall$ $I$ and~$K$, and,
\begin{align}\label{eq:Dphi}
\left( D_\phi f \right)_{I,J,K} = \frac{f_{I,J,K+1} - f_{I,J,K-1}}{2} \, ,
\end{align}
for~$K = 0, \ldots, N_\phi-1$, and~$\forall$ $I$ and~$J$.

To define these operators at the endpoints (the origin, $r=0$, polar axis,~$\theta =0$ and $\pi$, and branch points~$\phi = 0$ and~$2\pi$), we extend the computational grid virtually across the domain by introducing ghost points in the regions corresponding to~$r<0$,~$\theta < 0$ and~$\theta > \pi$, and~$\phi < 0$ and~$\phi \geq 2\pi$, with the same grid width~$\Delta r$, $\Delta \theta$ and $\Delta \phi$, and populate them with grid functions, with the parity conditions~\eqref{eq:Parity}, \eqref{eq:Parity_FOR_R_theta}, \eqref{eq:Parity_phi}, \eqref{eq:Parity_FOR_Null_R} and \eqref{eq:Parity_Chi}. To ensure that~$(\Upsilon^r)^{-1} D_r$ remains a discrete approximation to~$\p_r$ everywhere, and so do~$(\Upsilon^\theta)^{-1} D_\theta$ and~$(\Upsilon^\phi)^{-1} D_\phi$ for~$\p_\theta$ and~$\p_\phi$, respectively, we define these operators at the endpoints as
\begin{align}\label{eq:FD_Endpoints}
& \left( D_r f \right)_{0,J,K} = \frac{f_{1,J,K} - f_{-1,J,K}}{4} \, , \quad \forall \, J,K \, , \non \\
& \left( D_\theta f \right)_{I,0,K} = \frac{f_{I,1,K} - f_{I,-1,K}}{4} \, , \quad \forall \, I,K \, , \quad \textrm{and} \non \\
& \left( D_\theta f \right)_{I,N_\theta,K} = \frac{f_{I,N_\theta + 1,K} - f_{I,N_\theta - 1,K}}{4} \, , \quad \forall \, I,K \, .
\end{align}

We see that~$D_\phi$, as defined in~\eqref{eq:Dphi}, is antisymmetric everywhere in the domain, giving~$\tilde{D}_\phi = D_\phi$, whereas~$D_\theta$, as defined in~\eqref{eq:Dtheta_Bulk} and~\eqref{eq:FD_Endpoints}, is antisymmetric in the domain of definition of~$\tilde{D}_\theta$, i.e. for all~$0 < J < N_\theta$, giving \eqref{eq:Dtilde_theta_applied}. Interestingly, the operator~$D_r$, as defined by~\eqref{eq:Dr_Bulk} and~\eqref{eq:FD_Endpoints}, is also antisymmetric in the bulk, for~$0 < I < N_r$, giving~$B_{II'} = 0$ for~$0 < I,I' < N_r$, consistent with our requirements for~$B$ being the boundary matrix, cf.~\eqref{eq:Boundary_Matrix_Definition}. Since~$\tilde{D}_r$ does not appear in the equations at the origin, we do not worry about its definition and properties at~$I=0$. The only remaining task is to define~$D_r$ at~$I=N_r$. Since this point corresponds to the actual boundary of the domain, we need to define it here lopsidedly, by introducing upwinding. We prescribe the methods discussed in~\cite{GauVanHil21}, which lead to two kinds of~SBP schemes: \textit{SBP-TEM} (for \textit{truncation-error-matching}) and \textit{SBP-stable}, depending on the stencil at~$I=N_r$.

\subsubsection{SBP-TEM}\label{sec:SBP_TEM}

In this method, we define~$D_r$ at the last grid point by imposing the~TEM property, suggested in~\cite{Pre02}, so that the order of accuracy of~$D_r$ remains the same throughout the domain. The truncation error associated with~$D_r$, defined by~\eqref{eq:Dr_Bulk}, in the bulk is computed to be
\begin{align}\label{eq:Truncation_Error_Bulk}
\left( (\Upsilon^r)^{-1} D_r f \right)_I = f'_I + \frac{(\Delta r)^2}{6} f'''_I + \frac{(\Delta r)^4}{5!} f_I^{(5)} + \ldots \, ,
\end{align}
for~$I = 0, \ldots, N_r -1$. Here we suppressed the indices~$J$ and~$K$ for succinctness. We need at least~$4$ grid points to define~$D_r$ at~$I = N_r$ so that its series expansion there agrees with~\eqref{eq:Truncation_Error_Bulk} up to order~$(\Delta r)^2$. Exactly~$4$ grid points give
\begin{align}
\left( D_r f \right)_{N_r} = \frac{- f_{N_r-3} + 4 \, f_{N_r-2} - 7 \, f_{N_r-1} + 4 \, f_{N_r}}{4} \, ,
\end{align}
which has the following expansion
\begin{align}
\left( (\Upsilon^r)^{-1} D_r f \right)_{N_r} = f'_{N_r} + \frac{(\Delta r)^2}{6} f'''_{N_r} - \frac{(\Delta r)^3}{2} f_{N-r}^{(4)} + \ldots \, .
\end{align}
But, the~$(\Delta r)^3$ term here does not appear in~\eqref{eq:Truncation_Error_Bulk}, and, as we shall see later, our dissipation operator also acts at the order~$(\Delta r)^3$ at all grid points. This~$(\Delta r)^3$ term will, therefore, interfere with the dissipation, potentially leading the code to blow up at the outer boundary. To overcome this problem, we include another grid point in the stencil to make this~$(\Delta r)^3$ term also vanish, giving the following definition of~$D_r$ at~$I=N_r$,
\begin{align}\label{eq:Dr_TEM}
\left (D_r f \right)_{N_r, J, K} = &\ \frac{1}{4} \, \bigl( f_{N_r-4, J, K} - 5 \, f_{N_r-3, J, K} + 10 \, f_{N_r-2, J, K} \non \\
& - 11 \, f_{N_r-1, J, K} + 5 \, f_{N_r, J, K} \bigl) \, ,
\end{align}
for all~$J$ and~$K$.

The resulting boundary matrix becomes
\begin{widetext}
\begin{align}
B = \left( \begin{array}{ccccccc}
\ldots & . & . & . & . & . & . \\
\ldots & 0 & 0 & 0 & 0 & 0 & 0 \\
\ldots & 0 & 0 & 0 & 0 & 0 & \frac{W_{N_r N_r}}{8} \\
\ldots & 0 & 0 & 0 & 0 & 0 & -\frac{5 W_{N_r N_r}}{8} \\
\ldots & 0 & 0 & 0 & 0 & 0 & \frac{5 W_{N_r N_r}}{4} \\
\ldots & 0 & 0 & 0 & 0 & 0 & -\frac{9 W_{N_r N_r}}{8} \\
\ldots & 0 & \frac{W_{(N_r-4) (N_r-4)}}{8} & -\frac{5 W_{(N_r-3) (N_r-3)}}{8} & \frac{5 W_{(N_r-2) (N_r-2)}}{4} & -\frac{9 W_{(N_r-1) (N_r-1)}}{8} & \frac{5 W_{N_r N_r}}{4}
\end{array} \right) \, ,
\end{align}
\end{widetext}
which is nontrivial in the last row and the last column in their last~$5$ entries.

As we see, the price we pay here is that the resulting boundary matrix is neither diagonal nor positive definite in its symmetric part. As a result,~$\dot{\hat{E}}(t)$ can become positive whenever the state corresponding to the negative eigenvalue dominates over the remaining ones. However, this state corresponds to highly noisy data near~$\mathscr{I}^+$, which is unphysical. One way to eliminate this eigenstate is by introducing dissipation, which we shall introduce shortly. In any case, the resulting flux at~$\mathscr{I}^+$ approaches the continuum limit~\eqref{eq:Stokes_diff_form_rescaled} at the second-order accuracy. We direct the reader to~\cite{GauVanHil21} for more detailed discussions.

\subsubsection{SBP-Stable}\label{sec:SBP_Stable}

This scheme prefers to make~$B$ diagonal over preserving the accuracy of the~$D_r$ at~$I = N_r$. The only way is to define
\begin{align}\label{eq:Dr_Stable}
(D_r f)_{N_r,J,K} = \frac{f_{N_r,J,K} - f_{N_r-1,J,K}}{2} \, , 
\end{align}
for all~$J$ and~$K$, giving the following boundary matrix
\begin{align}
B = {\rm diag}(0, \ldots, 0, (\tilde{W}_-)_{N_r N_r}/2) \, ,
\end{align}
and, thereby, all equations in~\eqref{eq:SBP_Scheme} are satisfied identically. This way, we assure the discrete flux at~$\mathscr{I}^+$ to assume the same form as the continuum expression~\eqref{eq:E-dot_rescaled}, at the cost of the accuracy of the~FD scheme at the outer boundary, which is now first-order accurate. However, this scheme now ensures that~$\dot{\hat{E}}(t)$ is always negative semi-definite for all times~$t$. Ref.~\cite{GauVanHil21} for a more detailed discussion.

\subsection{Constraint Damping}\label{sec:Constraints}

Although we observed that the~FOR constraints are satisfied throughout the evolution in the continuum system whenever they are satisfied in the~ID, and the constraint damping terms damp them exponentially whenever they are violated. We now study this behavior in the discrete system.

The discrete~FOR constraints are now defined as
\begin{align}
\hat{C}_r \equiv &\ \bigg( \left[ \frac{1}{\chi} \right] (\Upsilon^r)^{-1} \, D_r \tilde{\Psi} - \left[ \frac{\chi'}{\chi^2} \right] \tilde{\Psi} - \left[ \frac{2R'-1}{2 \chi^2} \right] \tilde{\Psi}_+ \non \\
& + \left[ \frac{1}{2 \chi} \right] \tilde{\Psi}_- \bigg) \, , \non \\
\hat{C}_{\theta} \equiv &\ \Upsilon_{\theta}^{-1}D_{\theta}\tilde{\Psi} - \tilde{\Psi}_{\theta} \, , \quad
\hat{C}_{\phi} \equiv \Upsilon_{\phi}^{-1}D_{\phi}\tilde{\Psi} - \tilde{\Psi}_{\phi} \, ,
\end{align}
and the discrete~EOMs lead to the solution~$\dot{\hat{C}}_i = - \zeta \, \hat{C}_i$, where the subscript~`$i$' again stands for the~$r$,~$\theta$, and~$\phi$ components, giving the overall solution~$\hat{C}_i (t) = \hat{C}_i (0) \, e^{-\zeta \, t}$, as in the continuum case. Therefore, taking~$\tilde{\zeta} = 0$ again leads to the solution~$\hat{C}(t) = \hat{C}(0)$, for all~$t$. So the next task is to compute the~$\hat{C}_i(0)$ for an~ID satisfying~$C_i(0) = 0$.

The discrete constraints differ from the continuum ones at the order of accuracy of the discretization scheme. Therefore, for a second-order~FD scheme, one obtains
\begin{align}
& \hat{C}_r (t) = C_r (t) + O((\Delta r)^2) \, , \non \\
& \hat{C}_\theta (t) = C_\theta (t) + O((\Delta \theta)^2) \, , \non \\
& \hat{C}_\phi (t) = C_\phi (t) + O((\Delta \phi)^2) \, ,
\end{align}
and a similar expression with the first-order truncation error in~$\hat{C}_r$ at~$I = N_r$ whenever~$D_r$ is defined there by~\eqref{eq:Dr_Stable}. This shows that the discrete constraints are always violated for a continuum constraint satisfying~ID, but the difference vanishes with increasing resolution. The constraint damping terms then enforce the solutions to satisfy the discrete constraints over the continuum ones. However, this is the best we can do for a system whose analytical solutions are unknown.

\subsection{Dissipation}\label{sec:Dissipation}

Another artefact of discretization is the generation of spurious high-frequency modes unresolved by the grid. This numerical noise is generally produced in the vicinity of the regions where certain coefficients in the equations become singular, at the boundary points due to the boundary conditions, or at the grid points where a sudden change in accuracy is introduced. These unwanted modes are generally damped by introducing dissipation operators that vanish at a particular order in the continuum limit.

We start by determining which equations in our system~\eqref{eq:hyp_LWEP_rescaled_FOR_null_Tilded}, or, equivalently~\eqref{eq:Discrete_EOM_bulk}, require dissipation terms. One sees in the original system~\eqref{eq:LWEP_FOR}, that the only equation which contains all singular coefficients, has an outer boundary, and is affected by the decrease in accuracy at the outer boundary in the~SBP-stable scheme, is the~$\psi_T$ equation. Interestingly, this is the only equation in our system that comes from the original~EOM~\eqref{eq:LWEP}, as the remaining ones arise from the definition of the~FOR variables~\eqref{eq:FOR_TR_comps}. This suggests that this is the only equation in the system that needs to be introduced with the dissipation.

The next question is which variables in this equation the dissipation operator should act on. To answer this question, we observe that the only variable affected by all sources of numerical noise is~$\psi_T$, and the remaining variables get injected with this noise only via~$\psi_T$, as their evolution equations,~\eqref{eq:LWEP_FOR}, contain~$\psi_T$ or its derivatives on their right sides. This suggests that the dissipation operator, denoted by~$Q$, should act only on~$\psi_T$, making the overall system look like
\begin{align}\label{eq:LWEP_FOR_Constraint_damping_dissipation}
\p_T \psi = &\ \psi_T \, , \non \\
\p_T \psi_T = &\ - F \psi + \frac{1}{R^2} \p_R \left( R^2 \, \psi_R \right)
 + \frac{1}{R^2 \sin \theta} \p_\theta \left( \sin\theta \: \psi_\theta \right) \non \\ 
& + \frac{1}{R^2 \sin^2 \theta} \p_\phi \psi_\phi + Q  \, \psi_T \, , \non \\
\p_T \psi_R = &\ \p_R \psi_T + \zeta \, (\p_R \psi - \psi_R) \, , \non \\
\p_T \psi_\theta = &\ \p_\theta \psi_T + \zeta \, (\p_\theta \psi -  \psi_\theta) \, , \non \\
\p_T \psi_\phi = &\ \p_\phi \psi_T + \zeta \, (\p_\phi \psi -  \psi_\phi) \, .
\end{align}
While adding dissipative terms to the remaining equations is acceptable, the previous arguments indicate that this is the minimum requirement to eliminate high-frequency numerical noise.

We observed in~\cref{sec:Hyperbolicity} that, to ensure that the resulting system remains symmetric hyperbolic, the constraint damping terms should be added to the spatial part of the system via vector addition. Here, we observe that the temporal part of the system should be added with a dissipative term, which consists of a dissipation operator~$Q$ acting on the temporal~FOR variable~$\psi_T$. Therefore, the~$4$-vector, the~$(\psi_T, \psi_R, \psi_\theta, \psi_\phi)$-equations, should be added with the~$4$-vector~$C_{\mu'} \equiv (Q \, \psi_T, C_{i'})$, giving it a geometric structure.

The discrete system~\eqref{eq:Discrete_EOM_bulk}-~\eqref{eq:Discrete_EOM_z-axis_scri} is now modified as
\begin{align}\label{eq:Discrete_EOM_with_Diss_bulk}
\dot{\tilde{\Psi}}_+ = &\ \ldots + \left[ \frac{\chi}{2R'-1} \right] \tilde{Q} \left( \left[ \frac{1}{\chi} \right] \tilde{\Psi}_+ + \tilde{\Psi}_-  \right) \, , \non \\
\dot{\tilde{\Psi}}_- = &\ \ldots + \tilde{Q} \left( \left[ \frac{1}{\chi} \right] \tilde{\Psi}_+ + \tilde{\Psi}_- \right) \, ,
\end{align}
and the remaining equations will stay unaltered. Here, the ellipses~`$\ldots$' here denote the right sides of the systems \eqref{eq:Discrete_EOM_bulk}-~\eqref{eq:Discrete_EOM_z-axis_scri}, and~$\tilde{Q}$ is a regularised form of~$Q$ on hyperboloidal slices, and the above equations suggest that it should remain~$O(1)$ everywhere.

To explore how to define this operator in our system, we first understand how it works in a continuum system. Consider the following~$1+1$-dimensional system
\begin{align}
\p_t \psi(t,r) = a \, (-1)^{q+1} \, \p_r^{2q} \psi(t,r) \, ,
\end{align}
for any positive integer~$q$. Introducing Fourier decomposition
\begin{align}
\psi(t,r) = \int \psi_k(t) \, e^{ikr} \, dk \, ,
\end{align}
we turn this parabolic system into a system of eigenvalue equations, with each mode~$\psi_k(t)$ satisfying
\begin{align}
\p_t \psi_k(t) = - a \, k^{2q} \, \psi_k(t) \, ,
\end{align}
which has the solution~$\psi_k(t) \propto e^{-a k^q t}$. This shows that each mode~$\psi_k$, which is damped like~$e^{-a k^q t}$, needs to be an eigenvector of the operator~$\p_r^q$, with a real positive, or negative (depending on the sign appearing before~$\p_r^q$), definite eigenvalue. This is how the standard Kreiss-Oliger dissipation operator~(KODO)~\cite{GusKreOli95, KreOli73, KreLor89} works.

With a second-order accurate system, we usually add a fourth-order~KODO, which is defined in the bulk as
\begin{align}\label{eq:KODO_4}
Q^{(4)}_{\rm KODO} = - a \, (\Delta r)^3 \, (\Delta r)^{-4} \, (D_+ D_-)^2 \, ,
\end{align}
and it is still unclear how to define it at the boundary points. Here,~$D_\pm$ denote the forward and backward~FD operators, defined as
\begin{align}\label{eq:Dplus_minus}
(D_\pm f)_I \equiv \pm (f_{I \pm 1} - f_I) \, ,
\end{align}
that makes the fourth-order~KODO an~FD approximation of a fourth-order derivative that vanishes at the third-order in the continuum limit
\begin{align}
\big( (\Delta r)^{-4} \, (D_+ D_-)^2 & f \big)_I = \frac{f_{I-2} - f_{I-1} + f_I - f_{I+1} + f_{I+2}}{(\Delta r)^4} \non \\
= & \left( f_I^{(4)} + \frac{(\Delta r)^2}{6} f_I^{(6)} + \ldots \right) \, ,
\end{align}
and~$a$ is the dissipation parameter whose value can be tuned freely during numerical evolutions. As inferred from above, it determines the rate of damping of all the modes. Its value should, therefore, be small enough to keep the physical modes, resolved by the grid, unaltered up to numerical accuracy, and big enough to damp the unresolved modes effectively. Therefore, the smaller the required value of this parameter is, the better the numerical scheme.

Ignoring the boundary points, this operator satisfies the dissipative property~(DP)~\cite{CalLehReu03, GauVanHil21} in the standard~$L^2$ norm as
\begin{align}\label{eq:DP}
(f,Q^{(4)}_{\rm KODO} f) & \equiv \sum_{I} f_I \, (\Delta r) \, (Q^{(4)}_{\rm KODO} f)_I \non \\
& = - a \sum_{I} (D_-^2 f)_I^2 \leq 0 \, .
\end{align}
It is, therefore, essential to (i). define it at the boundary points, and (ii). extend this definition so that it satisfies the~DP in our energy norm.

\subsubsection{In $1D$ with the standard $L^2$ norm}\label{sec:Diss_L2-norm}

Ideally, we wish to define a fourth-order dissipation operator, denoted by~$Q^{(4)}$, such that it
\begin{enumerate}
\item[(i).] agrees with the fourth-order~KODO in the bulk, defined by~\eqref{eq:KODO_4}.
\item[(ii).] should be~$(\Delta r)^3$ times a discrete approximation of the fourth-order derivative at the boundary points.
\item[(iii).] satisfies the~DP,~\eqref{eq:DP}, when boundary points are also included in the norm.
\end{enumerate}
As pointed out in~\cite{GauVanHil21}, no extension of the dissipation operator at the boundary points can satisfy the properties~(ii). and~(iii). simultaneously. Even in the continuum case,
\begin{align}
\hspace{-1.0em} \int_0^{r_\mathscr{I}} f f'''' \, dr = \int_0^{r_\mathscr{I}} (f'')^2 \, dr + (f f''' - f' f'') \bigg|_0^{r_\mathscr{I}} \, ,
\end{align}
where boundary terms are not sign definite. However, it is desirable to demand the property~(ii). along with the~SBP-TEM scheme, so that it acts at~$O(\Delta r)^3$ even at the boundary points, preserving the overall convergence order of the scheme. With this construction,~$Q^{(4)}$ is defined as
\begin{align}\label{eq:Diss_TEM_L2_Norm}
\hspace{-1.0em} Q^{(4)}_1 \equiv \left\{ \begin{array}{l}
- a \, (\Upsilon^r)^3 \, \left( (\Upsilon^r)^{-1} D_+ \right)^2 \, \left( (\Upsilon^r)^{-1} D_- \right)^2 \, , \\
\vspace{0.5em} \textrm{for } I = 0, \ldots, N_r - 2 \, , \\
- a \, (\Upsilon^r)^3 \, (\Upsilon^r)^{-1} D_+ \, \left( (\Upsilon^r)^{-1} D_- \right)^3 \, , \\
\vspace{0.5em} \textrm{for } I = N_r - 1 \, , \textrm{and, } \\
- a \, (\Upsilon^r)^3 \, \left( (\Upsilon^r)^{-1} D_- \right)^4 \, , \textrm{ for } I = N_r \, ,
\end{array} \right.
\end{align}
giving the following matrix representation near the outer boundary
\begin{align}
\hspace{-1.0em} Q^{(4)}_1 = \frac{- a}{(\Delta r)}
\left( \begin{array}{cccccccc}
. & . & . & . & . & . & . & . \\
. & 6 & -4 & 1 & 0 & 0 & 0 & 0 \\
. & -4 & 6 & -4 & 1 & 0 & 0 & 0 \\
. & 1 & -4 & 6 & -4 & 1 & 0 & 0 \\
. & 0 & 1 & -4 & 6 & -4 & 1 & 0 \\
. & 0 & 0 & 1 & -4 & 6 & -4 & 1 \\
. & 0 & 0 & 1 & -4 & 6 & -4 & 1 \\
. & 0 & 0 & 1 & -4 & 6 & -4 & 1 \\
\end{array} \right) .
\end{align}

On the other hand, property~(iii). is desirable with the SBP-stable scheme, leading to the norms decreasing in values with time. To obtain an expression similar to~\eqref{eq:DP} when the boundary points are also included, we define~$Q^{(4)}$ such that it satisfies
\begin{align}
& (\Psi ,Q^{(4)} \, \Psi) \equiv - a \, \Psi^T \, \Upsilon^r \, Q^{(4)} \, \Psi \non \\
& = - a \left( (\Upsilon^r)^2 \left( (\Upsilon^r)^{-1} D_-)^2 \Psi \right) \right)^T \left( (\Upsilon^r)^2 \left( (\Upsilon^r)^{-1} D_-)^2 \Psi \right) \right) ,
\end{align}
which gives
\begin{align}\label{eq:Diss_stable_L2_Norm}
Q^{(4)}_2 = - a \, (\Upsilon^r)^{-1} \, D_-^T \, (\Upsilon^r)^{-1} \, D_-^T \, (\Upsilon^r)^2 \, D_- \, (\Upsilon^r)^{-1} \, D_- \, .
\end{align}
This definition gives the following matrix representation near the outer boundary
\begin{align}
Q^{(4)}_2 = \frac{- a}{(\Delta r)}
\left( \begin{array}{cccccccc}
. & . & . & . & . & . & . & . \\
. & 6 & -4 & 1 & 0 & 0 & 0 & 0 \\
. & -4 & 6 & -4 & 1 & 0 & 0 & 0 \\
. & 1 & -4 & 6 & -4 & 1 & 0 & 0 \\
. & 0 & 1 & -4 & 6 & -4 & 1 & 0 \\
. & 0 & 0 & 1 & -4 & \frac{81}{16} & -\frac{17}{8} & \frac{1}{16} \\
. & 0 & 0 & 0 & 1 & -\frac{17}{8} & \frac{5}{4} & -\frac{1}{8} \\
. & 0 & 0 & 0 & 0 & \frac{1}{8} & -\frac{1}{4} & \frac{1}{8} \\
\end{array} \right) ,
\end{align}
which approximates~$\Delta r$ times the second-order derivative at the last three grid points.

\subsubsection{In $3D$, spherical polar coordinates, and the energy norm}\label{sec:Diss_energy-norm}

As pointed out earlier in this section, for a dissipation operator to work, each mode in an expansion of the general solution should be an eigenvector of the dissipation operator, with a real positive- (or negative-, depending on the sign appearing before the operator) definite eigenvalue. With the decomposition similar to~\eqref{eq:Psi_Sph_Harmonics}, we see that every~$Y_{l,m}$ is simultaneously an eigenvector of~$\p_\phi$ and $\tilde{\p}_\theta \, \p_\theta$, satisfying
\begin{align}
& \p_\phi \, Y_{l,m}(\theta, \phi) = i \, m \, Y_{l,m}(\theta, \phi) \, , \quad \textrm{and} \non \\
& \tilde{\p}_\theta \, \p_\theta \, Y_{l,m}(\theta, \phi) = \left( - l(l+1) + \frac{m^2}{\sin^2\theta} \right)  \, Y_{l,m}(\theta, \phi) \, .
\end{align}
To be defined everywhere on the~$2$-sphere, any fourth-order dissipation operator,~$Q^{(4)}$, should, therefore, contain the $\phi$- and~$\theta$- derivatives in the combination~$\tilde{\p}_\theta \, \p_\theta + (1/\sin^2\theta) \, \p_\phi^2$, which is simply the Laplacian~$\nabla_{S^2}^2$ on the~$2$-sphere, given in~\eqref{eq:Laplacian_S2}.  As the unresolved numerical noise along the angular directions acts like higher-order~$Y_{l,m}$'s, the operator~$\left( \tilde{\p}_\theta \, \p_\theta + (1/\sin^2\theta) \, \p_\phi^2 \right)^2$ will suppress each~$Y_{l,m}$ as~$e^{-a \, l^2(l+1)^2 t}$, which is independent of~$m$. This makes sense, as there are no singular terms in~$\phi$.

To make this operator compatible with the energy norm~\eqref{eq:Discrete_Energy_Norm} with~\eqref{eq:Discrete_Weights}, the radial part should consist of the derivative~$\tilde{\p}_r \, \p_r$, which again shares the same eigenmodes as the previous two operators. Since each eigenvector~$\tilde{\psi}_{l,m}(t,r) \, Y_{l,m}(\theta,\phi)$ has the eigenvalue that behaves near the origin as
\begin{align}
\tilde{\p}_r \, \p_r \, \tilde{\psi}_{l,m}(t,r) = \left( \frac{l(l+1)}{r^2}  + O(1) \right) \tilde{\psi}_{l,m}(t,r) \, ,
\end{align}
one can define the dissipation operator, which is a discrete approximation of
\begin{align}
\tilde{Q}^{(4)} \equiv \left( \tilde{\p}_r \, \p_r + \frac{\chi^2}{R^2} \left( \tilde{\p}_\theta \, \p_\theta + \frac{1}{\sin^2\theta} \, \p_\phi^2 \right) \right)^2 \, .
\end{align}
Along with the~SBP-TEM scheme, one can define this operator as
\begin{align}
\tilde{Q}^{(4)}_{\rm TEM} \equiv &\ - a \, L^2 \, ,
\end{align}
and, for the~SBP-stable scheme, one can choose
\begin{align}
\tilde{Q}^{(4)}_{\rm stable} \equiv - a \, W_-^{-1} \, \Upsilon^{-1} \, L^T \, \Upsilon \, W_- \, L \, ,
\end{align}
where
\begin{align}
L \equiv &\ \bigg( (\Upsilon^r)^{-1} \, \tilde{D}_r \, (\Upsilon^r)^{-1} \, D_r + \left[ \frac{\chi^2}{R^2} \right] \bigg( (\Upsilon^\theta)^{-1} \, \tilde{D}_\theta \, (\Upsilon^\theta)^{-1} \, D_\theta \non \\
& + \left[ \frac{1}{\sin^2\theta} \right] ( (\Upsilon^\phi)^{-1} \, D_\phi)^2 \bigg) \bigg) \non \\
= &\ W_-^{-1} \bigg( (\Upsilon^r)^{-1} \, D_r \, W_- \, (\Upsilon^r)^{-1} \, D_r \non \\
& + \left[ \frac{\chi^2}{R^2} \right] \bigg( (\Upsilon^\theta)^{-1} \, D_\theta \, W_- \, (\Upsilon^\theta)^{-1} \, D_\theta \non \\
& + \left[ \frac{1}{\sin^2\theta} \right] (\Upsilon^\phi)^{-1} \, D_\phi \, W_- \, (\Upsilon^\phi)^{-1} \, D_\phi \bigg) \bigg) \, .
\end{align}

Both~$\tilde{Q}^{(4)}_{\rm TEM}$ and~$\tilde{Q}^{(4)}_{\rm stable}$ agree with each other in the bulk, except for near the boundary points~$I = N_r$. The operator~$D_r$ here can be defined in terms of~$D_\pm$, as in~\eqref{eq:Diss_TEM_L2_Norm}, for the~SBP-TEM scheme, or completely in terms of~$D_-$, as in~\eqref{eq:Diss_stable_L2_Norm}, for the~SBP-stable scheme, with~$D_\pm$ defined in~\eqref{eq:Dplus_minus}.

With these definitions, the modified change in energy becomes
\begin{align}
\dot{\hat{E}}(t) = \ldots + \tilde{\Psi}_T^T \, \Upsilon \, W_- \, \tilde{Q} \, \tilde{\Psi}_T \, ,
\end{align}
which becomes negative semi-definite in the~SBP-stable scheme, given by
\begin{align}
\dot{\hat{E}}(t) = \ldots - a \left( L \, \tilde{\Psi}_T \right)^T \Upsilon \, W_- \left( L \, \tilde{\Psi}_T \right) \, ,
\end{align}
where~$\tilde{\Psi}_T \equiv \left[ \frac{1}{\chi} \right] \tilde{\Psi}_+ + \tilde{\Psi}_-$. Here again, the ellipsis denotes the expression for~$\dot{\hat{E}}$ without dissipation. These definitions are consistent with the special case of spherical symmetry, defined in~\cite{GauVanHil21}, and also justify the fact that the modified dissipation there was defined only for the even function~$\tilde{\psi}_T$. We, therefore, did not need to introduce the dissipation for~$\tilde{\Psi}_R$ in that work. The definition of~$\tilde{Q}$ on the~$z$-axis and at the origin can be extended as in~\eqref{eq:LWEP_Sph_Coord_z-axis_m=0} and~\eqref{eq:LWEP_Sph_Coord_origin_l=0} for the~$\Box$ operator, respectively.

For the same reasons described above, the dissipation operators for the spatial~FOR variables~$\tilde{\psi}_R$,~$\tilde{\psi}_\theta$, and~$\tilde{\psi}_\phi$ can be defined by flipping the order of~$\tilde{\p}_i$ and~$\p_i$. For example,~$\tilde{Q}^{(4)}_\phi$,~$\tilde{Q}_\theta$ and~$\tilde{Q}_R$ now approximate
\begin{align}\label{eq:Dissipation_FOR}
& \tilde{Q}^{(4)}_\phi \approx \p_\phi^4 \, , \quad \tilde{Q}^{(4)}_\theta \approx \p_\theta \, \tilde{\p}_\theta + \frac{1}{\sin^2\theta} \, \p_\phi^2 \, , \non \\
& \tilde{Q}^{(4)}_R \approx \left( \p_r \, \tilde{\p}_r + \frac{\chi^2}{R^2} \left( \p_\theta \, \tilde{\p}_\theta + \frac{1}{\sin^2\theta} \, \p_\phi^2 \right) \right)^2 \, ,
\end{align}
at the discrete level, which can be defined in terms of the operators
\begin{align}
L_\phi \equiv &\ - a_\phi \left( (\Upsilon^\phi)^{-1} D_\phi \right)^2 \, , \non \\
L_\theta \equiv &\ - a_\theta \bigg( (\Upsilon^\theta)^{-1} \, D_\theta \, (\Upsilon^\theta)^{-1} \, \tilde{D}_\theta \non \\
& + \left[ \frac{1}{\sin^2\theta} \right] ( (\Upsilon^\phi)^{-1} \, D_\phi)^2 \bigg) \, , \quad {\rm and} \, , \non \\
L_R \equiv &\ - a_R \bigg( (\Upsilon^r)^{-1} \, D_r \, (\Upsilon^r)^{-1} \, \tilde{D}_r \non \\
& + \left[ \frac{\chi^2}{R^2} \right] \bigg( (\Upsilon^\theta)^{-1} \, D_\theta \, (\Upsilon^\theta)^{-1} \, \tilde{D}_\theta \non \\
& + \left[ \frac{1}{\sin^2\theta} \right] ( (\Upsilon^\phi)^{-1} \, D_\phi)^2 \bigg) \bigg) \, .
\end{align}
The operators~$D_i$ and~$\tilde{D}_i$ here are now defined in terms of~$D_\pm$ given in~\eqref{eq:Dplus_minus}, to make the radial part of dissipation analogous to~\eqref{eq:Diss_TEM_L2_Norm} and~\eqref{eq:Diss_stable_L2_Norm} for the~SBP-TEM and ~SBP stable schemes, respectively. The operators~$\tilde{Q}^{(4)}_\phi$ and~$\tilde{Q}^{(4)}_\theta$ act on the vectors~$\tilde{\Psi}_\phi$ and $\tilde{\Psi}_\theta$, respectively, whereas, the operator~$\tilde{Q}^{(4)}_R$ acts on the linear combination~$\tilde{\Psi}_R \equiv \left[ \frac{1}{\chi} \right] \tilde{\Psi}_+ - \tilde{\Psi}_-$, and appears in the~$\tilde{\Psi}_\pm$ equations as
\begin{align}\label{eq:Diss_QR}
\dot{\tilde{\Psi}}_+ = &\ \ldots + \left[ \frac{\chi}{2R'-1} \right] \tilde{Q}^{(4)}_R \left( \left[ \frac{1}{\chi} \right] \tilde{\Psi}_+ + \tilde{\Psi}_-  \right) \, , \non \\
\dot{\tilde{\Psi}}_- = &\ \ldots - \tilde{Q}^{(4)}_R \left( \left[ \frac{1}{\chi} \right] \tilde{\Psi}_+ + \tilde{\Psi}_- \right) \, .
\end{align}
The stable and~TEM versions of~$\tilde{Q}^{(4)}_R$ can be defined as of the~$Q^{(4)}$ above.

\subsection{Time Integration}\label{sec:Time_Int}

The system of Ordinary Differential Equations~(ODEs) resulting from the spatial discretization is evolved in time using the classical fourth-order Runge-Kutta (RK4) method. The time step $\Delta t$ is strictly limited by the Courant-Friedrichs-Lewy (CFL) condition. In spherical polar coordinates, the the grid spacing along the~$\hat{r}$-, $\hat{\theta}$- and $\hat{\phi}$-directions become~$\Delta r$,~$r \Delta\theta$ and~$r \, \sin\theta \, \Delta\phi$, getting vanishingly small near the poles,~$\theta \rightarrow 0$ or $\pi$, and near the origin,~$r \rightarrow 0$, imposing a severe restriction on the time step. To ensure stability, we choose $\Delta t$ accordingly, which, for our choice of grid~\eqref{eq:grid}, equals
\begin{align}
\Delta t = &\ \text{CFL} \times \min \left( \Delta r_I, \Delta r_I \, \Delta\theta_J, \Delta r_I \, \sin(\Delta \theta_J) \, \Delta\phi_K \right) \non \\
= &\ \text{CFL} \times \min \left( \frac{1}{N_r}, \frac{1}{N_r \, N_\theta}, \frac{1}{N_r \, N_\phi} \sin \left( \frac{1}{N_\theta} \right) \right) \, .
\end{align}

\section{Numerical Implementation and Results}\label{sec:numerics}

In this section, we present a series of numerical experiments to test this scheme empirically. These tests assess stability, energy conservation, convergence, and physical behavior of the solutions propagating towards $\mathscr{I}^+$ for a set of potentials. Specifically, we evolved the system for~$F = 0$,~$F = 1/\chi^2$, and~$F = m^2$, using both SBP-TEM and SBP-Stable discretization schemes, and tried different values of the compactification parameter~$n$, defined in~\eqref{eq:Compactification}. For simplicity, we kept~$r_\mathscr{I} = 1$ in all our cases. In each case, we obtained qualitatively similar results.

Impressively, for each simulation presented in this work, the Courant factor as high as~CFL $= 2.6785$ yields a stable evolution across all resolutions considered. This is the largest value we found empirically.

The second impressive result we found empirically is that all our numerical evolutions stay stable even without dissipation,~$a = 0$. It is expected, as this is the whole idea of the~SBP schemes. However, we observe numerical noise throughout evolutions, which we expect for the reasons described in~\cref{sec:Dissipation}.

The third interesting result is that, for each system, our scheme requires very small amounts of artificial dissipation to eliminate the high-frequency numerical noise, with~$a = 0.002$ in the~SBP-TEM discretization, and~$a = 0.008$ for the SBP-Stable one. All other remaining aspects of numerical implementation are kept identical between the two schemes.

We test our hypothesis on the minimal set of equations required to introduce artificial dissipation and the minimal set of variables it should act on, and find that adding~$\tilde{Q}^{(4)}_\theta \, \tilde{\Psi}_\theta$ and~$\tilde{Q}^{(4)}_\phi \, \tilde{\Psi}_\phi$ to the~$\tilde{\Psi}_\theta$ and~$\tilde{\Psi}_\phi$ equations, respectively, does not affect the numerical noise along these directions, nor does adding the~$\tilde{Q}^{(4)}_R$ term to the~$\tilde{\Psi}_\pm$ equations affect it along~$\hat{r}$. Furthermore, using~$\tilde{Q}^{(4)}_{\rm stable}$ with the~SBP-TEM scheme affects the pointwise convergence near~$r = r_\mathscr{I}$, whereas, using~$\tilde{Q}^{(4)}_{\rm TEM}$ with the~SBP-stable one leads to a slight increase in total energy for small time intervals. However, this effect is very small, and dies away with increasing resolution.

We now turn to the case-specific results for different choices of~$F$. We begin with the massless case, $F=0$, which serves as the baseline for assessing convergence, numerical stability and benchmarking the numerical results against the analytical ones, as the solutions for this case can be obtained analytically. The case~$F = 1/\chi^2$ is interesting as it is the slowest decay in~$F$ that keeps the potential term regular, both in the equations and the energy norm. Moreover, we expect a special behavior of the solutions in this case at~$\mathscr{I}^+$. Finally, we consider~$F=m^2$, as it represents the massive fields but makes the system singular at~$\mathscr{I}^+$. However, we expect the energy associated with its solutions to remain conserved for all times.

Although we tried several types of~ID in our simulations, we shall take~ID corresponding to the distribution
\begin{align}\label{eq:psi_ID_00_20_22}
& \psi(0, R, \theta, \phi) = a \, e^{-\sigma^2 R^2} (1 + R^2 (\cos^2\theta - \sin^2\theta \, \sin(2\phi))) \, , \non \\
& \psi_T (0, R, \theta, \phi) = 0 \, ,
\end{align}
for all our presentation purposes, with~$a = \sigma = 1$, unless stated otherwise. The~ID for all the variables~$\tilde{\Psi}$,~$\tilde{\Psi}_\pm$,~$\tilde{\Psi}_\theta$ and~$\tilde{\Psi}_\phi$ can be calculated from the definitions~\eqref{eq:FOR_TR_comps},~\eqref{eq:FOR_null_comps}, \eqref{eq:Rescaled_FOR_TR_comps} and~\eqref{eq:Rescaled_FOR_null_comps}.

\begin{figure}[t]
\centering
\includegraphics[scale=1.0]{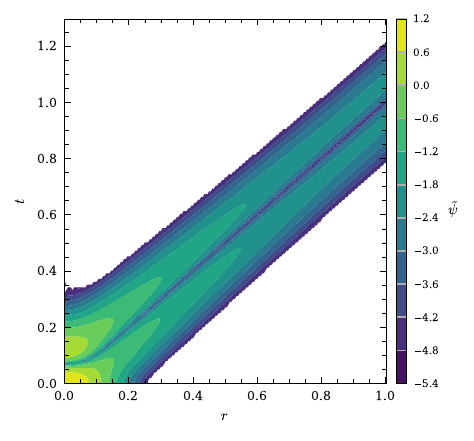}
\caption{Contour plot demonstrating the propagation of a narrow pulse of the scalar field~$\tilde{\psi}$ satisfying the~LWE towards~$\mathscr{I}^+$, located at $r = 1$. This solution is computed using the~SBP-TEM scheme; the corresponding SBP-Stable evolution exhibits, qualitatively, the identical behavior. The plot is truncated at low amplitudes for clarity.}
\label{fig:Contour_TEM}
\end{figure}

\begin{figure}[t]
\centering
\includegraphics[width=1.0\linewidth]{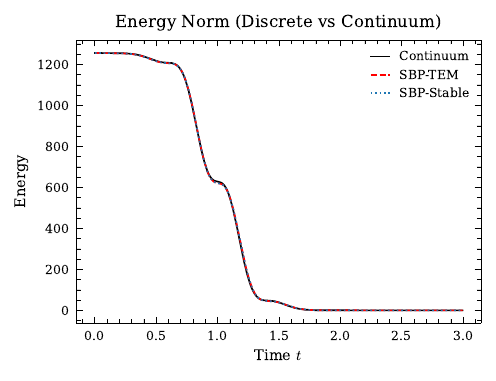}
\caption{Comparison of the continuum and discrete energies as a function of time. The discrete energies are computed from the numerical data, both in~SBP-TEM and~SBP-Stable schemes. Both of these norms remain non-increasing and in close track with the continuum energy, computed from a closed-form solution.}
\label{fig:energy_norm_disc_cont}
\end{figure}

\begin{figure*}[t]
\centering
\includegraphics[width=1.0\linewidth]{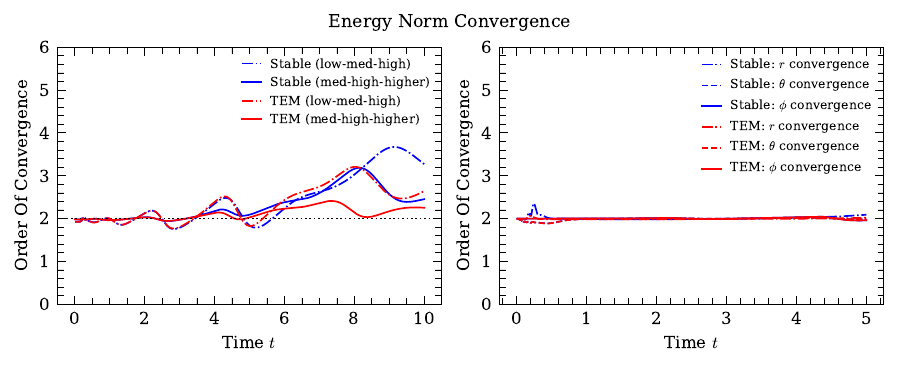}
\caption{Convergence order in the energy norm as a function of time for the scalar field satisfying the~LWE,~\cref{eq:LWEP} with~$F = 0$, and with the~ID given in~\eqref{eq:psi_ID_00_20_22}. \emph{Left panel}: Convergence order both in~SBP-TEM and~SBP-Stable discretizations using four successively refined uniform resolutions. The convergence order approaches~$2$ at higher resolutions. \emph{Right panel}: Directional convergence obtained by using three successively refined resolutions in each coordinate direction separately. In all cases, the convergence remains close to second order throughout the evolution.}
\label{fig:norm_conv_TEM_Stable}
\end{figure*}

\begin{figure*}[t]
\centering
\includegraphics[width=\textwidth]{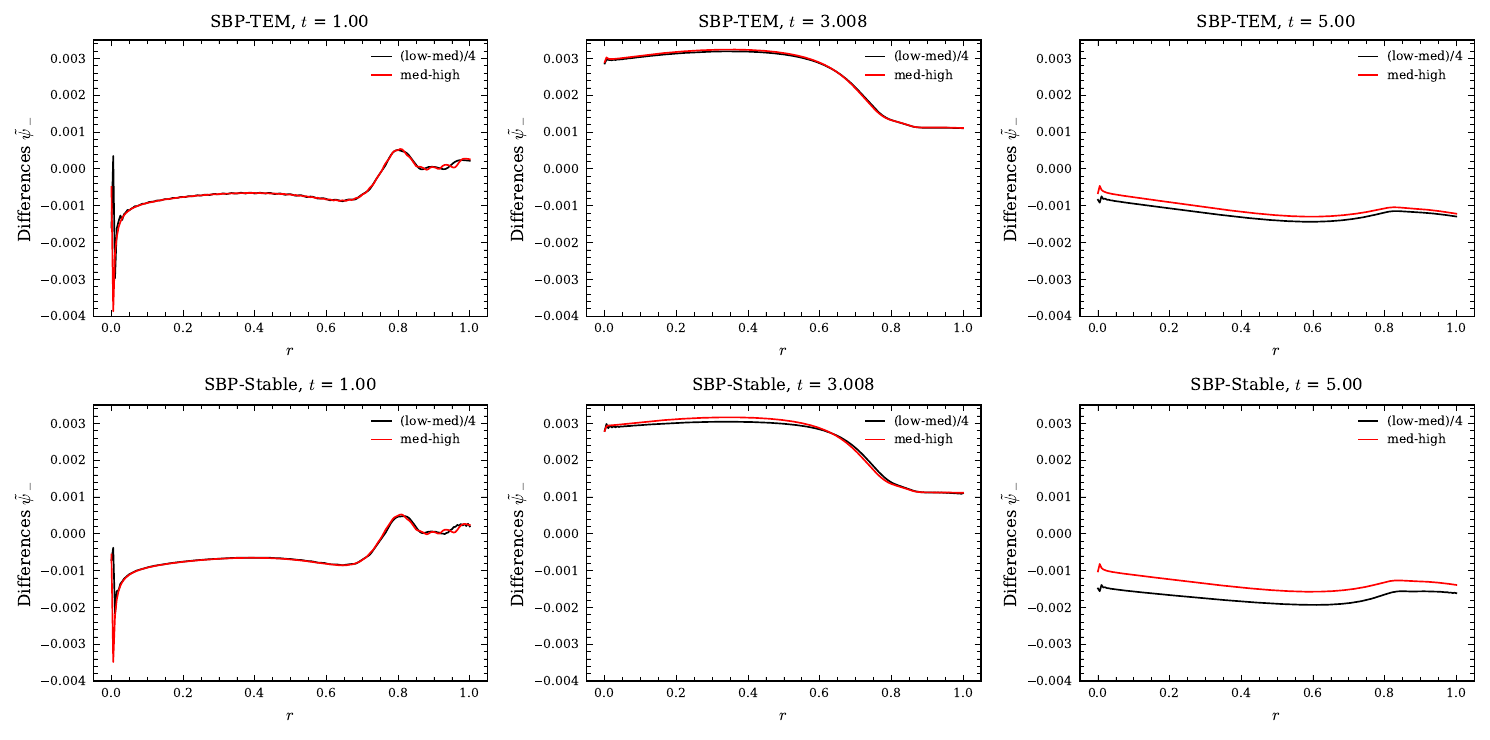}
\caption{Pointwise convergence in the radial direction for the outgoing mode $\tilde{\psi}_-$, integrated over the angles~$\theta$ and~$\phi$. These plots show a second-order convergence both in~SBP-TEM and~SBP-Stable schemes, at various times. Spikes at the origin arise from the coordinate singularity.}
\label{fig:r_convergence_psim}
\end{figure*}

\begin{figure*}[t]
\centering
\includegraphics[width=\textwidth]{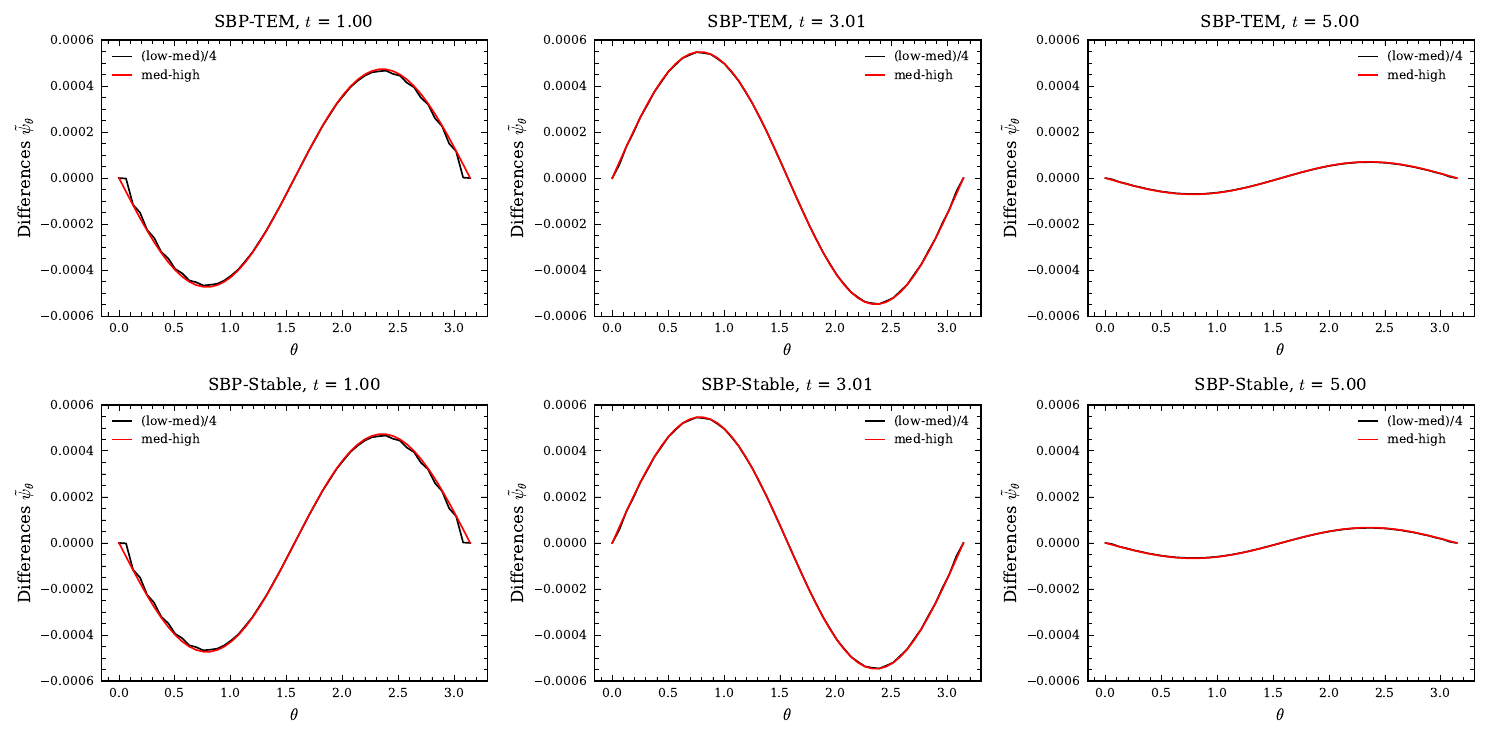}
\caption{Pointwise convergence for~$\tilde{\psi}_{\theta}$ along~$\hat{\theta}$, integrated over~$r$ and~$\phi$, both in~SBP-TEM and~SBP-Stable discretizations. The plots demonstrate a perfect second-order convergence across the domain, despite coordinate singularities on the polar axis.
}
\label{fig:theta_convergence_psitheta}
\end{figure*}

\begin{figure*}[t]
\centering
\includegraphics[width=\textwidth]{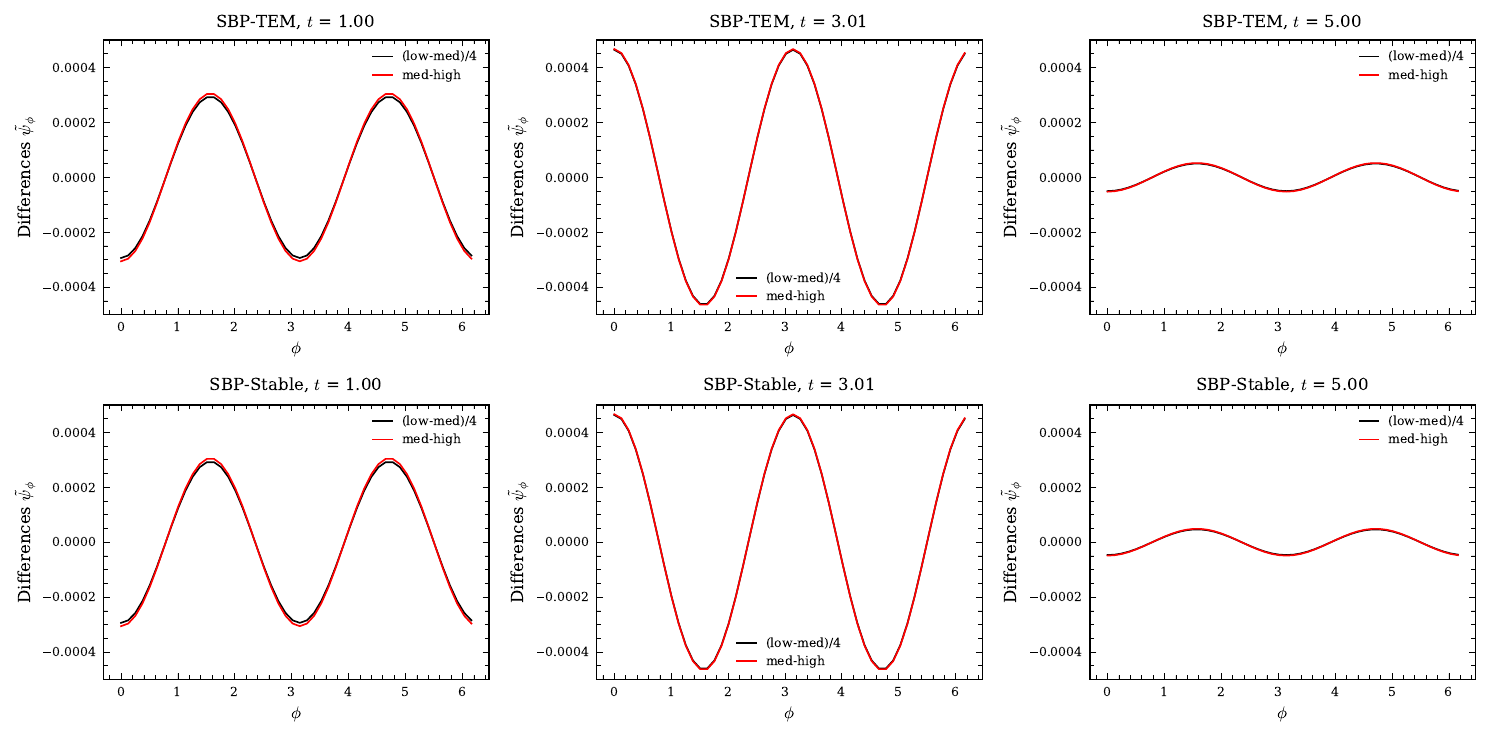}
\caption{Pointwise convergence of~$\tilde{\psi}_{\phi}$ along~$\hat{\phi}$, integrated over~$r$ and~$\theta$, both in~SBP-TEM and SBP-Stable schemes, demonstrating a perfect second-order convergence across the periodic domain~$\phi \in [0, 2\pi)$.}
\label{fig:phi_convergence_psiphi}
\end{figure*}

\begin{figure*}[t]
\centering
\includegraphics[width=1.0\linewidth]{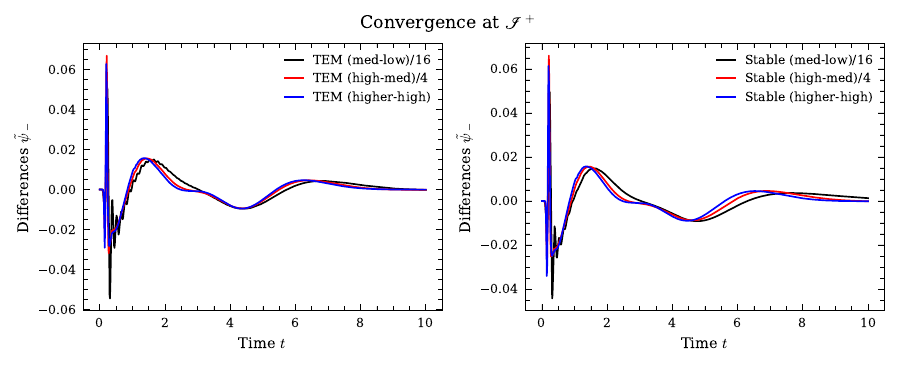}
\caption{Convergence at $\mathscr{I}^+$ of the outgoing mode $\tilde{\psi}_-$ integrated over the~$2-$sphere, both in the~SBP-TEM and~SBP-Stable discretizations. The results are exhibited as a function of time, demonstrating a second-order convergence with higher-order wiggles that die away with increasing resolution.}
\label{fig:scri_convergence_psiminus}
\end{figure*}

\subsection{Linear Wave Equation, $F=0$}\label{sec:F=0}

We begin by examining the basic behavior of the numerical solutions by visualizing how a narrow pulse in the~ID, defined by~\eqref{eq:psi_ID_00_20_22} with~$a = 1$ and~$\sigma = 10$, evolves over time. Fig.~\ref{fig:Contour_TEM} illustrates that this pulse splits into outgoing and ingoing components, which eventually reach~$\mathscr{I}^+$ in two bursts. This occurs because of the chosen time symmetry,~$\psi_T(T = 0) = 0$, in the~ID that keeps both outgoing and ingoing components on equal footing. The outgoing velocity of these pulses equals unity, which is consistent with our construction, cf. Sec.~\eqref{sec:Hyperboloidal_slices}.

The contour plot in Fig.~\ref{fig:Contour_TEM} shows the regions where the integral~$\int |\tilde{\psi}|^2 \sin\theta \, d\theta \, d\phi \geq 10^{-5}$, with color bands displaying the values of this integral in the~$(t,r)$-plane on a logarithmic scale. At the resolution of~$(N_r, N_\theta, N_\phi) = (200, 50, 50)$, the plot shows no numerical noise reflected from the outer boundary, the polar axis, or the origin.

To benchmark our numerical scheme, we compared our numerical solutions against the analytical ones for different choices of~ID. However, for demonstration purposes, we present here a solution composed of the spherical harmonics~$(l,m) = (0,0), (1,0)$ and~$(2,2)$, which, following the construction given in~\cite{Rin25}, has the general closed form
\begin{align}\label{eq:psi_00_10_22_analytical}
\psi(T,R,\theta,\phi) & = \frac{f(T+R) - f(T-R)}{R} + \bigg[ \frac{1}{R^2} [f(T+R) \nonumber \\
& - f(T-R)] - \frac{1}{R} [f'(T+R) \nonumber \\
& + f'(T-R)] \bigg] \cos\theta +  \bigg[ \frac{3}{R^3} [f(T+R) \nonumber \\
& - f(T-R)] - \frac{3}{R^2} [f'(T+R) + f'(T-R)] \nonumber \\
& + \frac{f''(T+R) - f''(T-R)}{R} \bigg] \sin^2\theta \sin 2\phi \, .
\end{align}
Here,~$f(x)$ is a general function defined and~$O(1)$ for all~$x \in \mathbb{R}$. For simplicity and elegance, we choose~$f(x) = e^{-9x^2}$. From this solution, one can compute the closed-form expressions of all the rescaled~FOR variables~$\tilde{\psi}_\pm$,~$\tilde{\psi}_\theta$ and~$\tilde{\psi}_\phi$. Introducing hyperboloidal coordinates, defined by \eqref{eq:Hyperboloidal_Coords}, \eqref{eq:Height}, \eqref{eq:Compactification} and \eqref{eq:Omega}, with~$n = 2$ and~$r_\mathscr{I} = 1$, once can then compute the total energy, defined by \eqref{eq:Energy_norm} and \eqref{eq:Energy_Density_Rescaled_Variables}, on these slices as a function of hyperboloidal time~$t$.

The above solution can also be used to generate the~ID for numerical evolutions, from which a similar energy norm is computed at every time step~$t$, to be compared against the analytical one. Fig.~\ref{fig:energy_norm_disc_cont} demonstrates an excellent agreement between the two norms for all time~$t$, both in the~SBP-TEM and~SBP-Stable discretizations, at the resolution of~$(N_r, N_\theta, N_\phi) = (200, 50, 50)$, confirming the reliability of the numerical results in both schemes.

The next task is to calibrate the accuracy of the numerical solutions for a given accuracy of the numerical scheme. As we confine ourselves to the second-order accurate~FD operators with the~RK4 time integrator, we expect a second-order accuracy of our numerical solutions. We examine this accuracy both locally (pointwise) and in the energy norm,~\eqref{eq:Discrete_Energy_Norm}-\eqref{eq:Discrete_Weights}.

To compute the convergence order in the energy norm, the system is evolved at four different resolutions, starting from~$(N_r,N_\theta,N_\phi) = (50, 8, 16)$ and successively doubling the resolution in all coordinate directions. We label these resolutions as \textit{`low', `med', `high'} and \textit{`higher'}, respectively. Considering the restriction of the variables in the state vector~$\mathbf{U}$ at all these resolutions to the grid points at the `\textit{low}' resolution, we take the differences~$\mathbf{U}_{low} - \mathbf{U}_{med}$,~$\mathbf{U}_{med} - \mathbf{U}_{high}$, and~$\mathbf{U}_{high} - \mathbf{U}_{higher}$, and then take the energy norm, defined by~\eqref{eq:Discrete_Energy_Norm}-\eqref{eq:Discrete_Weights}, of these differences, denoting them by~$E_{low}(t)$,~$E_{med}(t)$ and~$E_{high}(t)$, respectively. The convergence order in the energy norm as a function of time is then obtained by plotting
\begin{align}
n_1(t) \equiv \log_2 \sqrt{\frac{E_{low}(t)}{E_{med}(t)}} \, , \quad n_2(t) \equiv \log_2 \sqrt{\frac{E_{med}(t)}{E_{high}(t)}} \, ,
\end{align}
as shown in the left panel of Fig.~\ref{fig:norm_conv_TEM_Stable}. To include the contribution of~$\tilde{\psi}$ in convergence order, we take~$F = 1/\chi^2$ in this energy norm, despite~$F = 0$ in the~EOM.

The left panel in Fig.~\ref{fig:norm_conv_TEM_Stable} shows this convergence order, both in~SBP-TEM and~SBP-Stable discretizations, as a function of time. These plots oscillate around the convergence order~$n_{conv} = 2$ due to the higher-order contributions to the numerical errors. They appear dominant here because of the very low base resolution we chose due to limitations on computational resources and computation time. However, this plot also shows that these oscillations around convergence order~$n_{conv} = 2$ diminish with increasing resolution.

These results are obtained by eliminating information from most of the grid points at higher resolutions. To take all that data into account, we also study the above norm convergence in the following norms, defined using the `Ceiling function', denoted by~$\lceil \cdot \rceil$,
\begin{align}\label{eq:E1_norm}
E_1(t) & = \sum_{(I,J,K) = (0,0,0)}^{(2 N_r, 2 N_\theta, 2 N_\phi-1)} \frac{\Delta r_I \Delta \theta_J \Delta \phi_K}{8(1 + \delta_{I,N_r})(1 + \delta_{J,N_\theta})} W^\alpha_{I,J,K} \nonumber \\
& \bigg[ \frac{1}{8} \sum_{(I',J',K') = (I-1,J-1,K-1)}^{(I,J,K)} U^L_\alpha \left( \left\lceil \frac{I'}{2} \right\rceil, \left\lceil \frac{J'}{2} \right\rceil, \left\lceil \frac{K'}{2} \right\rceil \right) \nonumber \\
& - U^H_\alpha (I,J,K) \bigg]^2 \, ,
\end{align}
and
\begin{align}\label{eq:E2_norm}
E_2(t) & = \sum_{(I,J,K) = (0,0,0)}^{(2 N_r, 2 N_\theta, 2 N_\phi-1)} \frac{\Delta r_I \Delta \theta_J \Delta \phi_K}{8^2(1 + \delta_{I,N_r})(1 + \delta_{J,N_\theta})} W^\alpha_{I,J,K} \nonumber \\
& \sum_{(I',J',K') = (I-1,J-1,K-1)}^{(I,J,K)} \bigg[ U^L_\alpha \left( \left\lceil \frac{I'}{2} \right\rceil, \left\lceil \frac{J'}{2} \right\rceil, \left\lceil \frac{K'}{2} \right\rceil \right) \nonumber \\
& - U^H_\alpha (I,J,K) \bigg]^2 \, ,
\end{align}
where,~$U_\alpha = \{ \Psi, \Psi_+, \Psi_-, \Psi_\theta, \Psi_\phi \}$,~$W^\alpha$ are the corresponding weights, and~$\delta_{I,J}$ are the Kronecker deltas. Here we defined the norm of errors obtained from the differences at `\textit{low}' and `\textit{med}' resolutions, consisting of~$(N_r, N_\theta, N_\phi)$ and~$(2 N_r, 2 N_\theta, 2 N_\phi)$ grid points, respectively. Norms of errors at higher resolutions can be obtained similarly. In these norms again, we observe the same behavior as above. However, the norm~$E_2$ is more sensitive to the higher-order contribution to the numerical errors.

In addition to the uniform refinement, we also examine the same norm convergence by increasing resolution only along a single direction, while keeping it unchanged along the remaining ones. The prime advantage of this approach is that we can now start with a significantly higher base resolution with the same computational resources. Moreover, it highlights the coordinate directions contributing more to the higher-order corrections in the convergence order.

\begin{figure*}[t]
\centering
\includegraphics[width=1.0\linewidth]{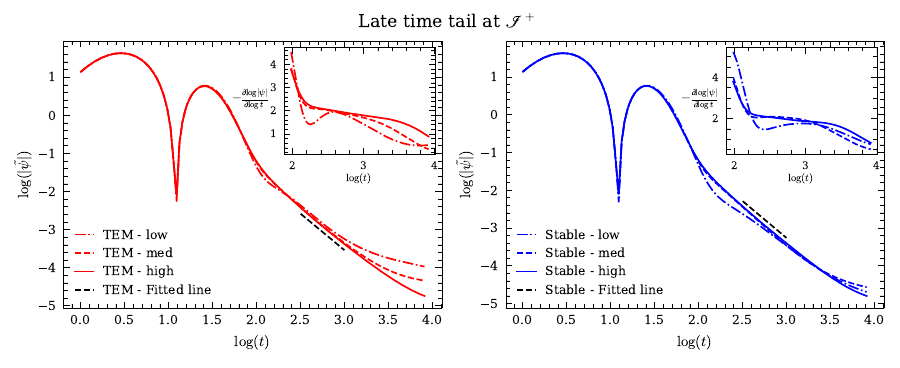}
\caption{Late-time behavior at~$\mathscr{I}^+$ of the scalar field satisfying the~LWE~\eqref{eq:LWEP} with potential~$F = 1/\chi^2$, shown at three successively refined resolutions. The data are shown for the rescaled field~$\tilde{\psi}$ in log--log scale, and the late-time signal exhibits a clear power-law decay with a least square fitted slope of approximately~$-1.91$ in the SBP-TEM and~$-1.95$ in the SBP-Stable schemes. The nested plots show that these slopes approach a constant limit with increasing resolution.}
\label{fig:late_time_tail_psi}
\end{figure*}

Here, the system is evolved at three different resolutions, starting from~$(N_r, N_\theta, N_\phi) = (200, 20, 20)$ and successively doubling it along the radial direction, keeping~$N_\theta$, and $N_\phi$ fixed, both in~SBP-TEM and~SBP-Stable discretizations. This exercise is then repeated by increasing the resolution along the polar and azimuthal directions separately, starting with the base resolutions of~$(N_r, N_\theta, N_\phi) = (100, 50, 20)$ and~$(100, 20, 50)$, respectively. The results are shown in the right panel of Fig.~\ref{fig:norm_conv_TEM_Stable}. Interestingly, the convergence order at these higher base resolutions is now predominantly~$2$, and the higher-order corrections at these base resolutions are now diminished.

We demonstrate pointwise convergence along the~$\hat{r}$,~$\hat{\theta}$ and~$\hat{\phi}$ directions in Figs.~\ref{fig:r_convergence_psim},~\ref{fig:theta_convergence_psitheta} and~\ref{fig:phi_convergence_psiphi}, respectively. The radial convergence is obtained from three different evolutions with base resolution of~$(N_r, N_\theta, N_\phi) = (200, 20, 20)$, and then doubling~$N_r$ successively, keeping~$N_\theta$ and~$N_\phi$ fixed. The rescaled errors are then integrated along the angular directions and then plotted with~$r$ at different times. We obtain a reasonable second-order convergence in all variables and for all times, and Fig.~\ref{fig:r_convergence_psim} demonstrates it in~$\tilde{\psi}_-$ in three different snapshots, both in~SBP-TEM and~SBP-Stable schemes.

Pointwise convergence along~$\hat{\theta}$ and~$\hat{\phi}$ directions are obtained similarly, with the base resolutions of~$(100,50,20)$ and~$(100,20,50)$, and integrating over~$r$ and~$\phi$, and~$r$ and~$\theta$, respectively. In all cases, we obtain an excellent second-order convergence, as demonstrated in Figs.~\ref{fig:theta_convergence_psitheta} and \ref{fig:phi_convergence_psiphi}, respectively. Both~SBP-TEM and~SBP-Stable schemes exhibit indistinguishable convergence behavior along these directions, as expected, since they differ only in their treatment along the radial direction near~$\mathscr{I}^+$, keeping it along the angular discretizations identical.

Lastly, we examine the convergence at~$\mathscr{I}^+$. We use the same evolutions as the ones used in the left panel of Fig.~\ref{fig:norm_conv_TEM_Stable}, and consider the numerical data corresponding to the grid points at~$\mathscr{I}^+$, and study their pointwise convergence there, both in the~SBP-TEM and~SBP-Stable discretizations, integrated over the angles, as a function of time. We again observe reasonable second-order convergence with a small higher-order contribution, due to a very low base resolution, with~$(N_r, N_\theta, N_\phi) = (50, 8, 16)$, that converges away with increasing resolution. The results are illustrated in Fig.~\ref{fig:scri_convergence_psiminus} for the outgoing mode~$\tilde{\psi}_-$.

\subsection{Scattering Potential, $F=1/\chi^2$}\label{sec:F=1/chi^2}

All the convergence properties here are similar to those in the~$F = 0$ case, and so, we do not display them here again. However, the presence of a non-zero scattering potential here changes the propagation properties and leads to the formation of late-time tails at~$\mathscr{I}^+$. These tails are expected to decay as a power law in time, as in Price's law~\cite{Pri72}.

Fig.~\ref{fig:late_time_tail_psi} shows this characteristic late-time power-law behavior in the numerical solution at~$\mathscr{I}^+$. Results are shown both in the SBP–TEM and SBP–Stable schemes at three successively refined resolutions, starting from~$(N_r, N_\theta, N_\phi) = (50, 8, 16)$, in both cases, and then doubling it in each coordinate direction at each refinement level. In both of these cases, the solutions exhibit a late-time behavior at all resolutions, with the higher-resolution evolutions displaying an increasingly linear behavior at late times in the~$\log–\log$ scale.

\begin{figure*}[t]
\centering
\includegraphics[width=\textwidth,
trim = {4.2cm 2.8cm 3cm 3.325cm}, clip]
{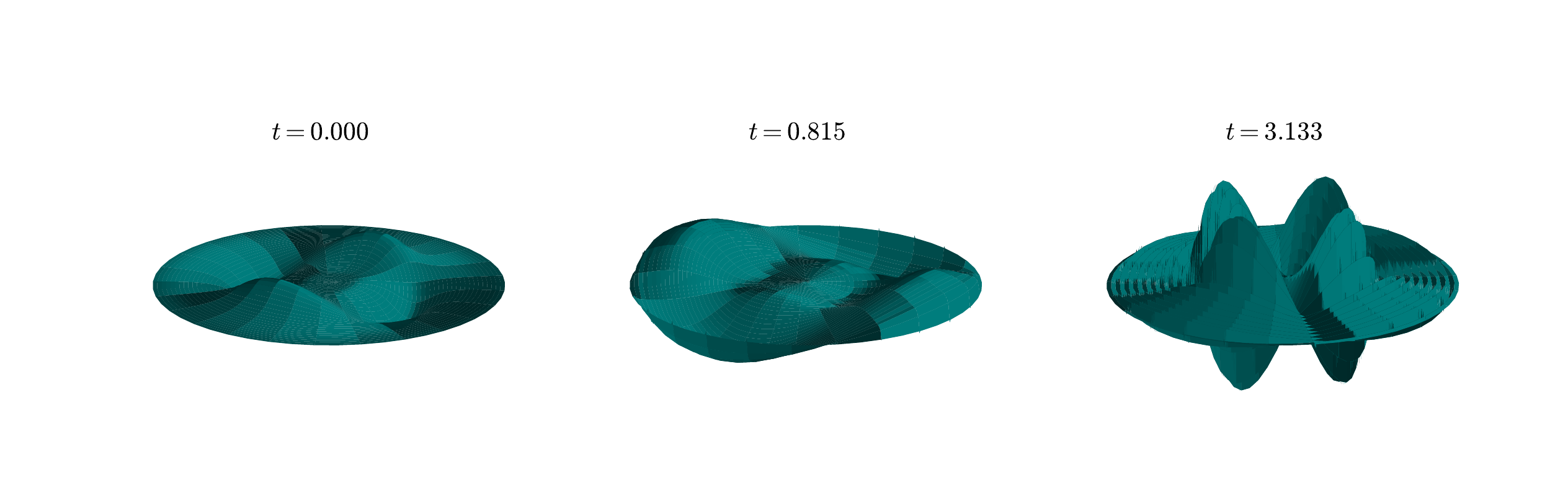}
\caption{Propagation of the numerical solution to the massive Klein-Gordon equation in hyperboloidal slices. The snapshots demonstrate how the solution keeps bouncing back and forth along the radial direction keeping the total energy conserved for all times.}
\label{fig:m2_evolution}
\end{figure*}

To quantify this decay rate, we focus on the late-time interval~$2.5 \leq \log(t) \leq 3.0$, as the highest resolution curve in this interval is almost linear, and perform a linear fit in the~$\log-\log-$ scale to the highest-resolution curve in each scheme. The resulting fits in both of the panels of Fig.~\ref{fig:late_time_tail_psi}, shown in black-dashed line segments, yield a slope of~$-1.91$ in the SBP-TEM discretization, and~$-1.95$ in the SBP-Stable one. The nested plots show how these slopes change with time. We expect these slopes to asymptote to~$-2$ if we work with sufficient resolution and wait for a long enough time~$t$.

\begin{figure}[t]
\centering
\includegraphics[width=1.0\linewidth]{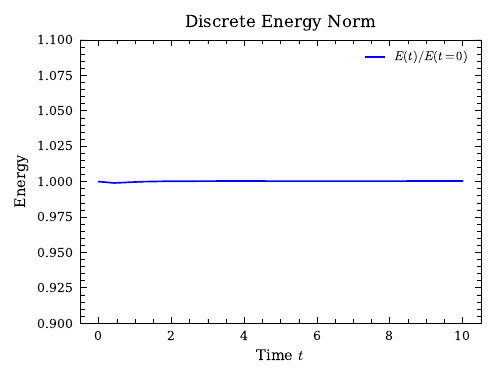}
\caption{Total energy with time for the massive fields on hyperboloidal slices from the numerical evolution using the SBP–Stable scheme without artificial dissipation. The horizontal straight line shows that the energy remains conserved even at the discrete level, as there is no flux at~$\mathscr{I}^+$.}
\label{fig:m2_energy_norm_zero_dissipation}
\end{figure}

\begin{figure*}[t]
\centering
\includegraphics[width=1.0\linewidth]{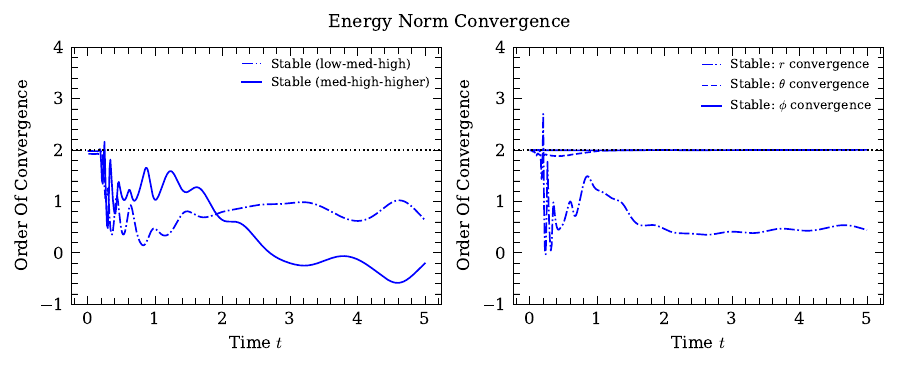}
\caption{Convergence order in the energy norm in the massive case~$F=m^2$. \emph{Left panel}: Convergence with uniform refinement along all directions. \emph{Right panel}: Convergence obtained by refining one coordinate direction at a time. While the angular directions exhibit a perfect second-order convergence, convergence in the radial direction is lost due to the singular behavior at~$\mathscr{I}^+$.}
\label{fig:m2_energy_norm_convergence_Stable}
\end{figure*}

\begin{figure*}[t]
\centering
\includegraphics[width=1.0\linewidth]{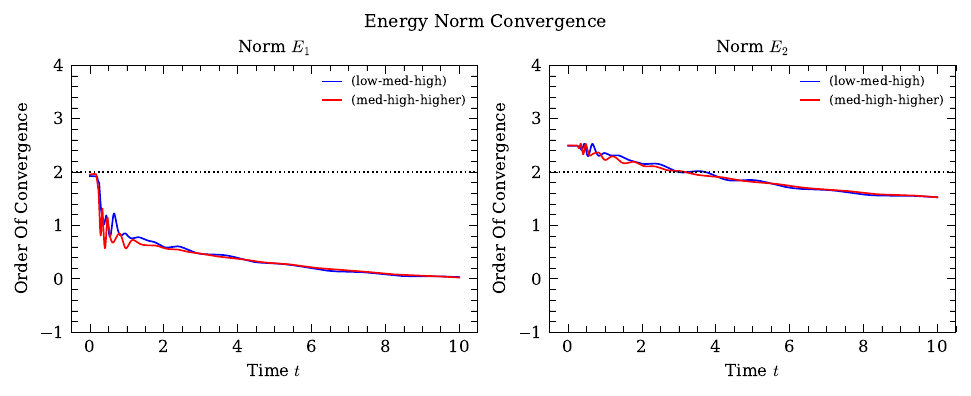}
\caption{Convergence order in the energy norm for~$F=m^2$, with norms given in~\eqref{eq:E1_norm} and~\eqref{eq:E2_norm}.}
\label{fig:m2_energy_norm_convergence_Stable_E1_E2}
\end{figure*}

\subsection{Massive Klein-Gordon Equation, $F=m^2$}\label{sec:F=m^2}

Evolving the massive fields, corresponding to the case~$F = m^2$ in~\eqref{eq:LWEP}, on hyperboloidal slices suffers from the problem that the mass terms in the equations and the energy norm become singular at~$\mathscr{I}^+$. However, it is expected that the solutions with a sufficiently decaying~ID, such that the mass terms in the~EOMs and the energy norm vanish at~$\mathscr{I}^+$, keep a similar fall-off behavior towards~$\mathscr{I}^+$ for all times~\cite{Kla93, Win88, GauVanHil21}, and, therefore, propagate within the light cone. As a consequence, no radiation reaches~$\mathscr{I}^+$ and the flux through $\mathscr{I}^+$ remains zero forever. We can, therefore, use this property to rewrite the equations and the energy norm at~$\mathscr{I}^+$.

The~ID~\eqref{eq:psi_ID_00_20_22} is one such example. With this choice, we will, therefore, take~$\varepsilon = \dot{E}(t) = 0$ at~$\mathscr{I}^+$, and erase the~$F$-terms from the~EOMs~\eqref{eq:hyp_LWEP_rescaled_FOR_null_Tilded_scri}-\eqref{eq:hyp_LWEP_rescaled_FOR_null_Tilded_z-axis_scri}, or, equivalently,~\eqref{eq:Discrete_EOM_scri}-\eqref{eq:Discrete_EOM_z-axis_scri} in our numerical implementation, giving the following~EOMs at~$\mathscr{I}^+$
\begin{align}\label{eq:Discrete_MKG_EOM_scri}
& \dot{\tilde{\Psi}} = 0 \, , \, \dot{\tilde{\Psi}}_+ = 0 \, , \, \dot{\tilde{\Psi}}_{\theta} = 0 \, , \, \dot{\tilde{\Psi}}_{\phi} = 0 \, , \non \\
& \dot{\tilde{\Psi}}_- = -{\Upsilon^{-1}_{r}}\left(\frac{\tilde{D}_{r}+D_r}{2}\right)\tilde{\Psi}_{-}+ {\Upsilon^{-1}_{r}}\left(\frac{\tilde{D}_{r}-D_r}{2}\right)\frac{\tilde{\Psi}_{+}}{\chi} \, ,
\end{align}
We shall evolve this system only in the~SBP-Stable scheme, as the other one does not guarantee energy conservation in the numerical solution, with a modified~$D_r$, and hence~$\tilde{D}_r$, at~$\mathscr{I}^+$
\begin{align}\label{eq:MKG_Dr_Scri}
& (D_r f)_{N_r, J, K} \equiv -\frac{f_{N_r-1, J, K}}{2} \, , \quad \textrm{and,} \non \\
 \quad & (\tilde{D}_r f)_{N_r, J, K} \equiv - \frac{(\tilde{W}_-)_{N_r-1} f_{N_r-1, , K}}{2 (\tilde{W}_-)_{N_r}} \, .
\end{align}
These modifications in the~EOMs will give a vanishing boundary term,~$\tilde{\Psi}_-^T (B+B^T) \tilde{\Psi}_-$, and, thereby, impose the vanishing flux at~$\mathscr{I}^+$ without breaking the~SBP scheme.

Additionally, as the mass term creates a potential well in the compactified coordinates, growing like~$R' \, F$, we expect the shorter wavelength modes to travel farther. These modes will be further generated due to the coupling between the incoming and outgoing modes introduced by the mass term, with the coupling coefficient~$R' \, F$ blowing up at~$\mathscr{I}^+$. Therefore, the high-frequency modes generated in the numerical solution here will be mostly physical, and introducing artificial dissipation will actually damp these modes, along with the unphysical ones. To avoid this situation, we first evolve our system without using dissipation and obtain numerical results.

The first result is that, despite the singular behavior at~$\mathscr{I}^+$, not only does the system evolve stably, but also the total energy of the system remains conserved for all times, even at the discrete level. This result is completely an artefact of the~SBP-Stable scheme, and completely demonstrates its strength even in such extreme cases. Fig.~\ref{fig:m2_energy_norm_zero_dissipation} demonstrates this result for the~ID~\eqref{eq:psi_ID_00_20_22}, at the resolution of~$(N_r, N_\theta, N_\phi) = (50, 8, 16)$ until~$t = 10$. Total energy at time~$t$ is normalised here with the initial energy, at~$t = 0$, and it remains close to unity throughout the evolution, consistent with the continuum results. Fig.~\ref{fig:m2_evolution} shows the propagation of this system on different time slices, in the~$\theta = \pi/2$ plane.

In any case, we need to introduce artificial dissipation to eliminate the high-frequency modes that the grid cannot resolve; otherwise, the numerical solution will appear noisy. However, increasing the radial resolution will eventually begin to resolve some of the physical modes that were not resolved at a lower resolution. Consequently, the solution will start showing non-convergence from the moment these intermediate-frequency modes are generated in the physical solution. As even higher-frequency modes emerge over time, this non-convergence time will take higher values for higher base resolutions. This fact is demonstrated in the left panel of Fig.~\ref{fig:m2_energy_norm_convergence_Stable}.

Since the singularity lies only along the radial direction, the convergence properties in the angular directions remain consistent with those observed in previous cases. The right panel of Fig.~\ref{fig:m2_energy_norm_convergence_Stable} shows the norm convergence when the resolution is increased separately along the~$\hat{r}$,~$\hat{\theta}$ and~$\hat{\phi}$ directions, confirming this behavior, and consistency with the Lax-Equivalence theorem~\cite{Tho95}. In all these convergence tests, we used the same sequence of resolutions employed in the~$F = 0$ case. Fig.~\ref{fig:m2_evolution} also shows numerical noise only along the radial direction, thereby confirming this behavior.

The left plot in Fig.~\ref{fig:m2_energy_norm_convergence_Stable} also shows a loss of correlation between the convergence orders at the lower and higher resolutions after some time. This happens because, as pointed out earlier, these convergence orders are computed by eliminating data from most of the grid points available at higher resolutions. However, we do not observe this correlation-loss in the convergence orders obtained from the~$E_1$ and~$E_2$ norms defined in~\eqref{eq:E1_norm} and~\eqref{eq:E2_norm}, respectively, as shown in Fig.~\ref{fig:m2_energy_norm_convergence_Stable_E1_E2}. As these norms contain information from all grid points from all resolutions, Fig.~\ref{fig:m2_energy_norm_convergence_Stable_E1_E2} demonstrates the strengths of these norms over the ones computed on the coarser grid.

\section{Conclusions}\label{sec:conclusions}

This work extends the~SBP framework developed by~\cite{GunGarGar10, GauVanHil21} to full~$3$D, and to the nontrivial foliations of Minkowski spacetime via spacelike hypersurfaces, with a focus on hyperboloidal slices with~$\mathscr{I}^+$-fixing coordinates. Starting with the behavior of the linear wave equation~(LWE) at the points of coordinate singularities, namely the~$z$-axis and the origin, we redefined the equation in those regions. We then introduced the first-order reduction~(FOR) in a covariant way, which was followed by introducing the compactified hyperboloidal coordinates and rescaling. We then arrived at a natural rescaling factor that is associated with the compactification, given by~\eqref{eq:Chi_natural}. Interestingly, this rescaling not only reduces to the conformal rescaling in the case of conformal compactification, but also greatly simplifies the regularized covariant divergence operators, given by~\eqref{eq:Tilded_Operators} and~\eqref{eq:Tilded_Operator_simplified}.

The~FOR system is further modified by adding constraint damping terms. It turns out that the resulting system remains symmetric-hyperbolic provided these constraints are added to the spatial part of the~FOR system via vector addition, which implies that the damping parameters~$\zeta$'s are equal to each other. Moreover, the temporal part of the system is modified by adding artificial dissipation, giving a geometric form to the overall modification.

The overall~SBP scheme is then derived by discretizing the partial derivatives and regularized covariant divergence operators, and then by imposing the energy conservation~\eqref{eq:E-dot_rescaled} and~\eqref{eq:Stokes_diff_form_rescaled}. Suitable dissipation operators are also introduced in Sec.~\ref{sec:Dissipation} in spherical-polar coordinates, which are defined everywhere, including the boundary points, and satisfy the dissipative property~(DP) in the energy norms. Notably, the definition of these operators varies depending on the~FOR variable they act upon, specifically because these variables correspond to the components of a~$4$-vector. In the overall scheme, time is treated to be continuous.

We derive two types of~SBP schemes, namely~SBP-TEM and~SBP-Stable, that differ only in their treatment of the outer boundary. Both of them have their strengths and weaknesses, as the former one keeps the accuracy of the numerical scheme intact throughout the domain, while the latter guarantees stability and negative-definiteness of the energy-flux at~$\mathscr{I}^+$, for~$\tilde{\zeta} = 0$. We demonstrate these properties through a series of numerical tests. Although the~SBP-TEM scheme exhibits better convergence properties, the~SBP-Stable discretization can effectively evolve even the systems with singularities, such as the massive Klein-Gordon fields, evolved in Sec.~\ref{sec:F=m^2}. However, in the massive case, it is clear that we lose accuracy over time, implying that more work needed in this direction. Both schemes excel in handling coordinate singularities, particularly in curvilinear coordinates—spherical polar coordinates in our case—while maintaining convergence properties comparable to fully regular systems. This is one of the remarkable properties of this scheme.

We confined ourselves here to finite-difference~(FD) methods with second-order accuracy, but the scheme in its abstract form, given in Sec.~\ref{sec:SBP_Overview} and~\ref{sec:SBP_Scheme}, can also be adapted for higher-order accurate~FD schemes and pseudo-spectral discretization. Even at very low resolutions and limited accuracy, this scheme has effectively captured essential physical properties of the corresponding continuum systems. For instance, it captured the wave propagation to~$\mathscr{I}^+$ for the case~$F = 0$, cf. Fig.~\ref{fig:Contour_TEM}, late time tails at~$\mathscr{I}^+$ for systems with scattering potentials that decay sufficiently fast asymptotically, cf. Fig.~\ref{fig:late_time_tail_psi}, and energy conservation in the massive Klein-Gordon case, cf. Fig.~\ref{fig:m2_evolution}.

We also propose new norm convergence tests, given by~\eqref{eq:E1_norm} and~\eqref{eq:E2_norm}, that include data from all grid points at all resolutions. Not only do they display the higher-order contributions to the numerical errors more effectively, but they also deliver better convergence results, even in situations where traditional norm convergence tests—based on data restricted to coarser grids—fail. These results are demonstrated in Fig.~\ref{fig:m2_energy_norm_convergence_Stable} and~\ref{fig:m2_energy_norm_convergence_Stable_E1_E2}.

In the following works, we will generalize this scheme to curved and dynamical spacetimes, with the ultimate goal of evolving the fully nonlinear systems,  such as those described by the Einstein field equations~(EFEs), in general relativity.

%

\section{Acknowledgements}\label{sec:Acknowledgements}

The authors would like to express their gratitude to Shing-Tung Yau, Lars Andersson, Piotr Chru{\'s}ciel, Istvan R{\'a}cz and Puskar Mondal for their helpful discussions. They also thank An{\i}l Zengino{\u{g}}lu, Alex Va{\~n}{\'o}-Vi{\~n}uales and David Hilditch for their critical feedback on the manuscript. This work was partially conducted at the ``Extremal Black Holes and the Third Law of Black Hole Thermodynamics" workshop at the Institute for Computational and Experimental Research in Mathematics (ICERM) at Brown University, as well as at the ``Hyperboloidal Foliations and their Application" workshop at the Erwin Schr{\"o}dinger International Institute for Mathematics and Physics~(ESI) at the University of Vienna. All the simulations were performed on the Sonic cluster at the ICTS. PK and AR acknowledge support of the Department of Atomic Energy, Government of India, under projects no. RTI4019 and RTI14013. SG's research was partially supported by the Beijing National Science Foundation~(BJNSF) under the International Scientist Project~(ISP), Grant No.: IS25026. AR's research was supported by the ICTS-SN Bhatt Memorial Excellence Fellowship Program for 2024 and the ICTS Long-Term Visiting Students Program (LTVSP) for 2025-26.  PK also acknowledges support by the Ashok and Gita Vaish Early Career Faculty Fellowship
at the International Centre for Theoretical Sciences.


\bibliography{Refs}

\appendix

\begin{widetext}
\begin{align}\label{eq:Source_Matrix}
\mathbf{S} = \left( \begin{array}{ccccc}
0 & \frac{1}{2\chi} & \frac{1}{2} & 0 & 0 \\
- \frac{\chi \, R' \, F + \zeta_R \chi'}{(2R'-1)} & \frac{1}{2R'-1} \left( \frac{R'}{R} - \frac{2 \, \chi'}{\chi} \right) - \frac{\zeta_R}{2} & \frac{\chi}{(2R'-1)} \left( -\frac{R'}{R} + \frac{\zeta_R}{2} \right) & \frac{\chi \, R' \, \cot\theta}{R^2 \, (2R'-1)} & 0 \\
-R' \, F + \frac{\zeta_R \, \chi'}{\chi} & \frac{R'}{R \, \chi} + \frac{\zeta_R \, (2R'-1)}{2 \, \chi} & - \frac{R'}{R} + \frac{\chi'}{\chi} - \frac{\zeta_R}{2} & \frac{R' \, \cot\theta}{R^2} & 0 \\
0 & 0 & 0 & - \zeta_\theta & 0 \\
0 & 0 & 0 & 0 & - \zeta_\phi \\
\end{array} \right) \, .
\end{align}
\end{widetext}

\end{document}